**The information and its observer: external and internal information processes, information cooperation, and the origin of the observer intellect**

**Vladimir S. Lerner, USA, lernervs@gmail.com**

In observing interactive processes, each observation is act changing the observing process, running to conversion of observed uncertainty to certainty. That natural phenomenon of interaction creates Yes-No actions of information Bits in its information observer. The information observer emerges from interacting random field of Kolmogorov probabilities, which link Kolmogorov 0-1 law's probabilities and Bayesian probabilities observing Markov diffusion process by its probabilistic 0-1 impulses. Each such No-0 action cuts maximum of minimal entropy of the impulse correlation, while following Yes-1 action transfers this maxmin between the impulses performing the minimax principle of converting the process entropy to information. Merging Yes-No actions generate microprocess within the bordered impulse, producing information Bit with free information when the microprocess probability approaches 1. The free information follows from the cutting correlation connecting the Markov process impulses. Each impulse' free information attracts the interacting bits. The borderer impulse' attracting interaction captures energy of a certain interaction which memorizes the Bit. The multiple Bits, connected by the free information, move a macroprocess which self-joins its bits in triplet macrounits. Each memorized information binds the reversible microprocess within impulses with the irreversible information macroprocess along the multi-dimensional process.

The observation automatically converts cutting entropy to information conveying the observer process inner causality, logic, and complexity. The triple macrounits logically self-organize information networks (IN) encoding the units in information geometrical structures enclosing triplet's code. The encoded units select objective or subjective information observer depending on the units' number. The IN geometry self-forms the observer dynamical and geometrical hierarchical structures with a limited boundary. The IN time-space distributed structure self-renews and cooperates information which decreasing the IN cooperative complexity.

The triplet units, built through attraction and resonance, have a limited stability, leading to a finite triplet structure which each IN ending triplet bit encloses. The observing process self-builds multiple IN with finite triplet number by the IN-ending bit free information. After each IN potentially loses stability, evolving in a chaos, it possesses ability of self-restoration by a cooperating triplet. Multiple cooperative triplets bind their IN whose ending triplets enclose the subjective observer information enables encoding its cognition. The observer's cognition assembles through the common units multiple attractions in resonance loops which accept only the units recognizing each IN node in the IN-triplet hierarchy. The ended triplet of observer hierarchical informational networks measures level of the observer intelligence. Maximal number of the accepted triplet levels in a multiple IN measures the observer maximum information intelligence comparative to other intelligent observers. The intelligent observer recognizes and encodes digital images in message transmission, being self-reflective enables understanding the message meaning. The entropy of the impulses and time interval of observations integrates entropy functional (EF) along the observing process.

The converted impulse information measures information path functional (IPF).

The minimax variation principle, applying to the EF-IPF, determines Equations of Integrating Interactive Information Dynamics formally describing the micro-macro processes, the IN, the selective objective and subjective information observers, and invariant conditions of observer's self-organization and self-replication.

Trajectory of observation process carries wave function, both probabilistic and certain, which self-organizes the hierarchical structures. These functional regularities create united information mechanism, whose integral logic self-operates this mechanism, transforming multiple interacting uncertainties to physical reality-matter, human information. The applications and practical implementations confirm the formalism theoretical concepts and results.

**CONTENTS**

## 1. Introduction. The Approach Essence

Revealing information nature of various interactive processes, including multiple physical interactions, human observations, Human-machine communications, biological, social, economic, other interactive systems, integrated in information observer, becomes important scientific task.

One of them is tracing random interactions of different objects, physical particles, others, up to interacting people by observing their probabilities of occurrences. Like an astronomer traces unobserving planet tracing its probabilistic trajectory until it become most probable and informative.

Physical approach to an observer, developed in Copenhagen interpretation of quantum mechanics [1-5], requires an act of observation, as a physical carrier of the observer knowledge. But this observer' role not describes the formalism of quantum mechanics.

According to D. Bohm ontological interpretation of quantum physics [6-8]: physical processes are determined by information, which "is a difference of form that makes a difference of content, i.e., meaning", while "meaning unfolds in intention, intention into actions"; and "intention arises from previous perception of meaning or significance of a certain total situation." That observer entails mental processes.

J.C. Eccles's quantum approach [9] "is to find a way for the "self" to control its brain".

A. Wheeler's physical theory [10-14] of information-theoretic origin of an observer, introduces doctrine "It from Bit". Wheeler hypothesized that the Bit participates in creating the origin of all physical processes. However, Wheeler's theory remained unproven, and the theory does not include how the Bit self-creates itself.

Before existence of J.A. Wheeler's theory, many physical scientists [1-14], including Einstein [15], Penrose [16], other, define the observer only physical origin.

S. Weinberg, focusing on probability on quantum mechanics [17], gets the trouble from this probability natural origin.

As it follows from physical quantum field [18], with vacuum quantum fluctuations, a natural probability of this fluctuation originates physical particles. Actually, instead of identifying particles, it measures their probability.

J.A. Weller has included the observer in wave function [14] according to standard paradigm: Quantum Mechanics is Natural.

Nonetheless, "Quantum Bayesianism" [19], which combines quantum theory with probability theory, states: 'the wave function does not exists in the world - rather it merely reflects an individual's mental state.'

Since information initially originates in quantum process with conjugated probabilities, its study should focus not on physics of observing process' interacting particles but on its information-theoretical essence.

A Kolmogorov [20] establishes probability theory as foundation of information theory and logics.

C.E. Shannon's entropy [21] is a measure of probability of signal including errors in communication channels, which contemporary science automatically applied to random process states as a measure of information process.

Kullback–Leibler's divergence [22] between the probabilities distributions measures relative "information" connections the states of the observed process.

E.T. Jaynes [23] applies Bayesian probabilities to propose a plausible reasoning mechanism [24] whose rules of deductive logic connect maximum Bayes entropy to a human's mental activities, as a subjective observer.

These references, along with many others, make no difference between Shannon entropy and information which up to now, as well as "information process" has not had a scientifically conclusive definition.

Moreover, notion of Observer assumes it observes physical events which build his physical information structure revealing the observation certainty.

The multiple references, studying information mechanisms in intelligence, explain these through various physical phenomena, whose specifics are still mostly unknown.

So, considering an observer, it's not clear what it actually observes as physical information: entropy of random signals maximizing a capacity in communication channel or a path to a certain reality?

Science knows that interactions have built structure of Universe as its fundamental phenomena.

Reality is built from emerging interactions.

There have been many studies of these interactions specifics; however, no one approach has unified the study of all their common information origins, regularities, and differentiation.

The first approach unifying these studies is published in [26, 27], and extends results in [28-34] which we review.

The approach focuses on observations as interactions producing the Observer itself.

The aim is formal principles explaining how information and information process originate and create Information Observer enables self-creating Information in interactive observations.

We study creation of Information Observer from zero information. At beginning, the only random interactive process emerging from probability field can be observed, while ability of its observation generates a virtual observer which is also emerging during the interactive observation. The emerging information during interactive probabilistic observation builds the Information Observer with its information structure. In the process of building, the arising information structures request needed information for their further development in the observer building.

This unified approach shows how an information observer emerges from the interacting observations of a random process running to conversion of observed uncertainty to certainty. This certainty *is* information.

Any single certain inter-action is a "Yes-No" action known as a Bit, the elementary unit of information, directly connecting information with interactions.

Multiple observations generate the Bits flow joined in information process which moves informational dynamics.

Information and the Information Process emerge as phenomena of natural interactions. Multiplicity of these interactions enfolds a probabilistic field conserving the sets of possible events, the sets of actual events, and their probability function. The observing process, emerging from the field, possesses ability of observation the actual events by getting their information covered in multiple elementary interactions of the observing process.

The approach initial points are:

1. The notions of object, particle, and image are replaced with the probability of these events observation.

This probability virtually observes the events occurrence during interactive observation. The events occurrence reveals the certain observation.

2. The interactions are abstract “Yes-No” actions of an impulse, probabilistic or real. The field of probability is source of information and physics.

3. Multiple interactions create random process whose interactive impulses model Markov process. The objective probabilities of the Kolmogorov 1-0 law’s axiomatic probabilities interact with the Markov process. The sequence of the probabilistic 0-1 (No-Yes) impulses initiates the Bayes probabilities within the Markov process, which self-observe the running Markov process, generating correlation of observing process events. Particular probability observes specific set of events whose entropy of correlation holds uncertainty. When the entropy is erased, uncertainty disappears, instead appearing as an equal certainty.

Removing the entropy produces information which emerges from particular set of the observing probabilistic events that create specific information observer.

Uncovering certainty from uncertainty is a scientific path to determine facts of reality.

4. In observing interactive processes, conversion of observed uncertainty to observer certainty is natural phenomenon•

The observing objective Yes-No probabilities measure the idealized (virtual) process’ impulses as a *virtual observer.*

Such formal mathematical approach to emerging information and its observer from a probability field is similar to some Quantum field approaches, and extends formalism of Wheeler’s *“It from Bit”* origin of the Observer.

The approach formalism models the interactions of Kolmogorov law probability events with Markov process events, generating Bayes probability virtually observing the Markov diffusion. The interacting events, the objective probabilities, and the Markov process emerge from the formal probabilistic field.

Such mathematical modeling is a probabilistic; it reveals a possibility of formal emergence, first, a virtual and then information Observer during the interactive observations.

This formal model leads to the following features.

1. The emerging Bits organize themselves in information process and structures triplets, which logically self-organize and assemble an informational network (IN).
2. In the process of network assembling, the triplets merge and interact with each other. Each triple interaction is memorized and becomes a node of the IN.
3. The sequence of logically organized nodes defines a code of the network. The code encloses all the information about the network.
4. The ending triplets of multiple cooperating INs enclose an observer information code enables encoding its cognition.
5. The observer’s cognition assembles hierarchical informational networks whose ending triplet encodes and measures level of the observer intelligence.
6. This code integrates all prior observations and *is* an immerged information observer.

7. The multi-dimensional observations, starting from a maximal probable uncertainty, identify a path to certainty of the maximal probable observations, which measure integrating information of the emerging Observer.

The informational observer emerges from probabilistic observation without any preexisting physical law.

From the formalism follows generation of the Observer with its functional properties as a natural phenomenon, which will be found in any intelligent Observer system being natural or artificial.

Below are brief explanations of the initial points.

An observer, processing random interactions, generates its virtual probability measurement of the random process uncertainty in an observable process of the virtual observer. (In this probabilistic model, a potential observer of hidden (random) bits models Markov observable process located in the surrounded random field and therefore affected by the field's random impulses-frequencies). With growing probabilities of virtual impulses, the observing process' correlations increase, and real impulses emerge. The observable Markov process correlation virtually cuts the probabilistic impulses and their hidden impulses release.

The observing multiple interactions integrate its growing correlation in the process entropy.

The impulses, cutting (removing) entropy of the correlations, create real observing information, which moves and self-organizes the information process.

The integral measure of Kolmogorov-Bayes objective probabilities, connecting the interactive observations, starts from virtual probabilistic observation and virtual observer, and evolves from objective interaction to a subjective real. It integrates the observing impulses, enable self-observation, and unities the observing information in Information Observer.

Merging Yes-No actions of the Markov probabilistic impulse generates microprocess within the bordered impulse.

The microprocesss produces information Bit with free information when its probability approaches one.

A transitive gap, separating the micro-macroprocess on an edge of reality [33], overcomes the injection of the needed energy which allows memorizing the bit and free information.

The free information, converted from the entropy connecting the Markov process impulses, holds the correlation links.

Each impulse' free information attracts the interacting bits.

The multiple Bits, connected by the free information, move a macroprocess of the information process where the microprocess randomly appears. The macroprocess bits attract other bits in duplets through its free information. Free information from one bit out of the pair gets spent on the assembling the duplet. Free information from the duplet bits attracts a third forming bit, assembling all three in a triplet-basic elementary structure of macrounits. The triplet third bit's free information, attracting another duplet of bits, creates two bound triplets, which self-joining encloses another triplet, and so on. This continuing process creates the nested levels of the bound triplets which self-organize an informational network with the levels of hierarchical structure.

The last-ending triplet in the information network (IN) collects and encloses the entire network's information.

Each bit's memorized information binds the reversible microprocess within each impulse in the irreversible information macroprocess along the multi-dimensional observing process.

The triplets, emerging from this process, cooperates the IN.

The Bayes sequential probabilities generate probabilistic logic in the observing process, while the information process transforms it to certain logic. The observation consecutively and automatically converts cutting entropy to information conveying the process information causality, certain logic, and both the bit logical complexity and the bound (cooperative) complexity of the IN. Each triplet's three segments (of the information macroprocess) generates three symbols and one impulse-code from the free information, composing a minimal *logical code* that encodes the macroprocess in a physical information process. The triplet macrounits logically self-organize information networks (IN) encoding the units in the information geometrical structures enclosing triplet's code, which select objective or subjective information observer depending on the encoding units. The IN geometry self-forms the observer dynamical and geometrical hierarchical structures with a limited boundary. The IN time-space distributed structure self-renews and cooperates information, decreasing the IN complexity through the attraction of their triplets. The IN ended triplet contains maximum amount of free information, which enables self-built other INs through the attraction, creating the multiple networks (domain). The informational networks cannot connect to each other when free information of the third bit in a triplet is not sufficient to attract another bound duplet. This triplet becomes the IN ended triplet, and the network completes as a finite network. The process of self-building network stops, the IN loses stability creating chaos of bits. (In a DNA, the ended triplet's code forms telomerase which controls the DNA life cycle).

However, in the chaos, there could be a pair of bits that bind creating a duplet, having enough of free information to attract another bit. Then the attracting bits create a triplet, and building the network may continue.

Hence, the finite IN chaotic process could self-restore the bound triplets creating next generation of the INs-domains, assembling the observer hierarchy. The observing probabilistic and certain logic, assembling the triplets, networks, and domains through free information logic, build the observer' cognitive logic and intelligence code.

Every observer has different amount of the observed logic bits needed for building specific and individual cognitive logic and intelligence code.

(This process is analogous to human brain cognition through the neuron yes-no actions modeling a bit. Moreover, all described self-creation of the duplet-triplets, the finite informational networks, multiple networks, as well as self-restoration, performs the human brain' information machine in the process of probabilistic observation.)

Each attracting free logical bit, while forming a triplet, selects the bits with equal speed creating a resonance. The bits cohere in the resonance which assembles their common logical loop. In the loop, only the bits involved in the resonance can recognizes each other. *Cognition has ability of recognizing the binding bits in a resonance process.*

If the attracting logical bits are not sufficient for performing the cognitive actions, cognition does not emerge. It cannot build any duplets and triplets. At the enough observed information, cognition arises at all levels of the informational networks. Along distributes units of the IN hierarchy, each IN node accepts only units that node concentrates and recognizes. The self-organized hierarchy of distributed logical loops builds a chain of self-assembles multiple logical hierarchical units. The resonance frequencies, self-organizing the cognitive logic of the units, provide interactive actions attracting the impulses with an external energy.

The attracting actions carry free logic of the assembling logical unit, which opens a switch-interaction with an external impulse carrying the Landauer's energy that starts erasing entropy and memorizing the information in its bit. The unit logic is memorized and encoding in triple code.

The IN hierarchy generates the multiple triplet codes, which the observer logical code integrates.

The observation process carries wave functions- probabilistic and real with the spinning space –time trajectory.

The wave function frequencies self-organize the observer space-distributed hierarchical structures during movement along this trajectory. The observing process, generating the triplets' IN time-space logic, composes a double spiral space (DSS) integral triple logic code which memorizes the DSS information helix structure.

The DSS helix structure rotates, spinning physical wave function frequencies, self-organizing the multiple local bits coding units, encodes the observer triplet coding structure which we call *observer intelligence.*

The logical switching actions perform *intelligence functions*, which generate each local code.

The logical switching of the free information at all hierarchical level performs the *intelligence functions*, which generate each local code. These functions are distributed hierarchically along the assembling logical units of the cognitive chain. The distributed intelligence coding actions at each hierarchical level control entrance the needed external physical processes. The DSS encodes the triplet units' information macrodynamic process in related physical irreversible thermodynamic process. The thermodynamic process' forces and flows-speeds determine a power to physically implement the encoding actions. The following observation of performance of these actions provides feedback to observer self-controlling the performance.

The approach results describe the emerging the observer's information regularities and intelligence, satisfying a simple natural law during conversion uncertainty to certainty in the observer interactive probabilistic observation of environment. Natural (real) interactions converts this entropy to information as the interactions' phenomena.

The approach formalism of the information law comes from Feynman concepts [24A] that a variation principle for the process integral with the problem solution mathematically formulates the physical law regularities for this process.

The variation problem for the integral measures of observing process' entropy functional and the bits' information path integral [29-30] formalizes the minimax law, which describes all regularities of the observing processes.

Solving the problem [26,27,33,34], mathematically defines the micro-macro processes, the IN, selective objective and subjective information observers, and invariant conditions of observer's self-organization and self-replication.

These functional regularities create united information mechanism, whose integral logic self-operates this mechanism, transforming the multiple interacting uncertainties to physical reality-matter, human information and cognition, self-originating the observer information intellect. This logic holds invariance of information and physical regularities, following from the minimax information law. The approach focuses on formal information mechanisms in an observer, without reference to the specific physical processes which could originate these mechanisms. The formally described information regularities contribute to basic theory of brain function and information mechanisms of Human-Machine interactions that allow finding information structure of artificial designed observer toward artificial brain.

The information formalism describes a self -building information machine which creates Humans and Nature.

**2. Essence of the formalism. Main components**

Forewords

Interactions are natural fundamental phenomena of multiple events in common environment of Universe.

Interactions have built Nature.

Elementary natural interaction consists of action and reaction, which represents abstract symbols: Yes-No or↓↑ actions of an impulse modeling a Bit.

In physical examples, a sequence of opposite interactive actions models rubber ball hitting ground, the reversible micro-fluctuations, produced within irreversible macroprocess in physical and biological processes.

Here the Yes-No physical actions are connected naturally.

How does the bit and their logical sequence originate?

The information structure of such logic is a basis of DNA, brain information mechanism, and forms many other physical micro and macro-processes.

We show that probabilistic interactions, instead of interacting particles, create information, physical processes, and an observer of this information.

Probability measures only multiple events. Therefore, the probability measures the interacting event-interactions.

The probability of interactive actions can predict both real interaction and the particles.

Interaction of the objects or particles is primary indicator of their origin, which measure probabilities.

The field of probability is source of information and physics.

The approach aim is the formal principles and methodology explaining the procedure of emerging interacting observer self-creating information. The aim derives from unifying the different interactions independent of origin, and focusing on observation as the interactive observer.

The objective Yes-No probabilities measure the idealized (virtual) impulses in observation in a *virtual observation.*

The observation correlates random interactive action. The observing interactive random process of multiple interactions evaluates probabilities, measured by equivalent entropy of the correlation.

Cutting correlation in the impulse high probable observation removes entropy or uncertainty producing certainty (information).

Integrating the cutting correlations currying information finally produces the information observer.

Natural (real) interactions innately convert this entropy to information and the information observer.

Information observers may reproduce a brain as its memorized copy-image built during observation

Below we outline the main steps creating the Information Observer .

2.1. The starting model

Multiple interactive actions are random events in a surrounding random field.

The interacting random events formally describe probabilities in Kolmogorov Theory of Probabilities [35].

The probability field defines mathematical triple: $\Phi=(\Omega, F, P)$, where $\Omega$ is sets of all possible events $\omega$, $F$ is Borel's $\sigma$-algebra subsets from sets $\Omega$, and probability $\mathrm{P}$ is a non-negative function of the sets, defined on $F$ at condition $\mathrm{P}(\Omega)=1$.This triple formally connects the sets of possible events, the sets of actual events, and their probability function. The actual events of the field conserve their appearance with a related real energy.

Each abstract axiomatic Kolmogorov probability predicts probability measurement on the experiment whose probability distributions, tested by relative frequencies of occurrences of events, satisfy condition of *symmetry* of the equal probable events. Some of them form a multiple infinite sequence of independent events satisfying Kolmogorov 0-1 law.

In the random field, sequence random events, collected in independent series, forms a random process, including Markov diffusion process [36] modeling multiple interactions.

The Markov diffusion process describes multi-dimensional probability distributions in the random field.

The events, satisfying the Kolmogorov law 0-1 probabilities, affect a Markov diffusion process', distributed in this field, which randomly switch the drifts (speeds) of the Markov process.

The switching Markov speeds sequentially change the process current a priori-a posteriori Bayes probabilities, whose ratio determines probability density of random No-Yes impulses 0-1 or 1-0, as a part of the Markov process. That links the Kolmogorov's 0-1 probabilities, the Markov process' Bayes probabilities, and the Markov No-Yes impulses in common Markov diffusion process. Within the Markov process, the Bayes probabilities densities randomly observe the impulse process, composing the observing process of a virtual observation. The Kolmogorov law probabilities do not observe, but initiate discrete probabilities actions on Bayes probabilities which do observe. The observing process No-Yes probability hold random impulse of 0-1, or 1-0 random actions.

The multiple random actions describe some probability distributions of the observing sequence of specific set of events, emanating from the probability field, which formally define the observing triple above.

Random probabilistic impulses of multiple interactions represent an immanent randomness in a stochastic process in a surrounding random field with some random time arrow.

So, how do the observing random probabilistic interactions become a logical sequence encoding the bits?

## 2.2. Observation. Virtual observer. Uncertainty, Information, Certainty

These Kolmogorov objective 0-1 probabilities quantify idealized (virtual) impulses whose Yes-No actions represent an act of a virtual observation where each observation measures a probability of the possible events for a potential observer.

Multiple virtual observations transit these probabilities along an interactive random process, generated by a virtual probability measurement, which models an observable process of a potential (virtual) observer.

In the impulse's virtual Yes-No reversible actions, each second (No) through recursion [37] affects the predecessor (Yes) connecting them in a weak correlation, if there was not any of that.

The arising correlation connection memorizes this action, indicating start of observation with following No-Yes impulse. The correlation encloses a mean time interval [36] which begins a time of observation.

The observing Bayes a priori-a posteriori probabilities determine arrow of the time course of an observation process with continuous correlations.

Uncertainty of observation measures conditional entropy of Bayesian a priori—a posteriori probabilities.

Maximal uncertainty measures non-correlating a priori-a posteriori probabilities, when their measure approaches zero. Such theoretical uncertainty has infinite entropy measures, whose conditional entropy and time do not exist.

The finite uncertainty measure has a nonzero correlating finite a priori – a posteriori probability of the interactive events with a finite time interval and following finite conditional entropy.

Example of such finite uncertainty' process is "white noise".

That allows measuring the observable process' uncertainty relative to uncertainty-entropy of the white noise.

The most common formal model of a natural random non-stationary process is Markov diffusion process, which includes the considered interactive observable process.

(The same way as an astronomer observes a probabilistic trace of unknown object, the Bayes probabilities formally observe the observable Markov process.)

If an elementary Dirac' delta-impulse increases each Bayes a posteriori probability, it concurrently increases probability of such virtual impulse (up to real impulse with probability 1), and decreases the related uncertainty.

Information, as notion of certainty–opposed to uncertainty, measures a reduction of uncertainty to maximal posteriori probability 1, which, we assume, evaluates an observing probabilistic fact. Actually, it's shown [30,32], that each of the delta-impulse No-0 action cuts maximum of the impulse minimal entropy, while following Yes-1 action transfers this maxmin between the impulses, decreasing the following entropy of the process.

Such impulse observation imposes the minimax principle increasing each posteriori probability.

*Simple example.* When a rubber ball hits ground, energy of this interaction partially dissipates that increases entropy of the interaction, while the ball's following reverse movement holds less entropy (as a part of the dissipated), leading to max-min entropy of the bouncing ball. Adding periodically small energy, compensating for the interactive dissipation, supports the continuing bouncing. •

The observation under Kronicker' [0-1] impulse-discrete analog of Dirac' delta-function,also imposes the minimax principle automatically.

The impulse minimax principle, imposed on the sequential Bayesian probabilities, leads to growing each posteriori probability, correlations, and reducing the process' entropy along the observing process.

The correlation freezes entropy and Bayes probability hidden within correlation, which, connecting the hidden process correlations, conveys probabilistic causality along the process.

The correlation connects the Bayesian a priori-a posteriori probabilities in a temporal memory that does not store virtual connection, but renews, where any other virtual events (actions) are observed.

If the observing process is self–supporting through automatic renewal virtual inter-actions, it calls a Virtual Observer, which acts until these actions resume.

Such virtual observer belongs to a self–observing process, whose Yes-action virtually starts next impulse No- action, and so on, sending the self-observing probing impulses. Both process and observer are temporal, ending with stopping the observation. Starting of virtual self-observation limits the identified threshold [33].

Each new virtually observing event–action temporary memorizes whole pre-history events from the starting observation, including the summarized (integrated) maxmin-minimax entropy of virtually cutting correlations.

The memory of a last of the current correlation connection automatically holds the integrated entropy of the correlations.

The memory temporary holds the difference of the probabilities actions, as a virtual measure of a distance between the impulses' No-Yes actions and a probabilistic accuracy of measuring correlation.

The measuring, beginning from the starting observation, identifies an interval from the start, which is also virtual, disappearing with each new connection, identifying a next interval memorized in that connection.

The random process' impulses hold virtually observing random time intervals with the hidden entropy and the events.

Collecting and measuring the uncertainty along the random process integrate entropy functional (EF) [31].

The minimax of interacting impulses of whole process carries the minimax variation principle (VP) imposed on the EF, which brings invariant measure for the running time intervals.

The correlation indicates appearance of an invariant time interval of the impulse–observation.

The difference of the probabilities actions that temporary holds memory of the correlation identifies a virtual measure of an *adjacent distance* between the impulse No-Yes actions.

That probable distance originates a space shift, quantified by a related probable curved time.

The space displacement shifts the virtual observation from the source of random field to a self-observing space –time process, initiating the probabilistic emergence of *time-space coordinate system* and gradient of entropy-an entropy force depending on the entropy density and space coordinates.

The displaced self-observing process with the space-time priori-posteriori actions continues requesting virtual observation with time-space probing impulses, which intends to preserve both these probabilities and invariant impulses. The observations under these impulses enclose the reduced entropy of the space-time movement forming a volume. The movement increases the correlation connections and the observer's space–time entropy, starting self-collecting virtual space-time observation in a *shape* of space-time correlation of the *virtual Observer volume.*

The space-time entropy' force rotates the curved time-space coordinate system (within the volume) developing the rotating Coriolis force and a momentum. Both are moving a space trajectory along the coordinate system, which depends on the gradient and velocity of running movement.

The gradient entropy along the rotating interval of the trajectory could engage each next impulse in rotating action, which increases the correlation, temporally memorizing the time-space observation.

The memory temporary holds a difference of the starting space-time correlation as accuracy of its closeness, which determines the time-space observer location with its shape. The evolving shape gradually confines the running rotating movement which *self-supports* formation of both the shape and Observer.

The virtual observer, emerging from the displacement virtual process, sends the discrete time-space impulses as virtual probes to test the preservation of Kolmogorov probability measure of the observer process with probes' frequencies. Such test checks the abstract probability via a symmetry condition indicating both the probability correctness and the observing time-space structural location of the Observer.

The increasing frequencies of the observer's self-supporting probes check the growing probabilities and their symmetry.

The virtual observer self-develops its space-time virtual geometrical structure during virtual observation, which gains its real form with transforming the integrated entropy of the correlated events to the equivalent specific information observer.

With no real physics affecting the virtual observation, the virtual probing impulses replicate the information impulses and start a probabilistic path from maximal entropy (uncertainty) to information-as a maximal real certainty.

2.3. The cutting entropy functions within Markov process, the related impulse, and the cutting EF satisfying the VP

Since the Yes-No discrete actions, forming virtual or real controls, cut the EF within the Markov process, they should preserve its additive and multiplicative properties.

That requirement limits the admissible controls' class by two real and two complex functions.

Applying these control functions identify the cutting invariant impulse.

The VP extreme, imposed on the process' impulse cutting by the EF, proves that each three invariant virtual impulses of the process time intervals enable generating a single invariant information impulse. Instead of these three emerges the last.

Information delivered by the impulse step-up control, being transferred to the nearest impulse, keeps information connection between the impulses, providing persistence of the impulse sequence.

This condenses each two of the previous impulses intervals and entropies in the following information impulse interval.

It also proves that action, starting the information impulse, captures the Markov multiplicative entropy increments.

Such impulse includes three parts:

1. First is delivered by multiplicative action by capturing entropy of random process;
2. Second is delivered by the impulse step-down cut of the process entropy;
3. Third is information delivered by the impulse step-up control, and then transferred to nearest impulse. It keeps information connection between the impulses and provides persistence continuation of the impulse sequence during the process time.

Each of the three parts holds its invariant portion within the impulse measure.

Since each cutting impulse preserves it invariant information measure, each third of the sequential cutting impulses triples and condenses information density.

It implies that a final IPF impulse condenses all previous IPF impulses, while the final time interval is limited, depending on process time course.

Because the final time interval evaluates the density of this impulse information or entropy, both are also limited, as well as the total IPF.

Conclusively, the IPF finite maximal information density limits a finite minimal physical time interval in accessible time course. The finite three times intervals within the invariant impulse parts allow identifying the related discreet correlation functions and its cutting increments.

The results [33] verify the estimated entropy contributions in all parts of the impulse and the following information increments. The increment measures the memorized correlation of the impulse *probability* events in the impulse *time interval's invariant measure* $|1|_M$ [34]. With growing probability and correlations, the intensity of entropy per the interval (as entropy density) increases on each following interval, indicating a shift between virtual actions, measured in the time interval' unit $|1|_M$. With growing the density of the opposite actions, they *merge* in a jumping impulse whose cutting action *curves* an emerging ½ time units of the starting impulse time interval, while a following rotating curved time-jump action initiates a *displacement* within the impulse opposite rotating Yes-No actions.

The rotating displacement's space shift quantifies the curved time. When the displacement shift of two space units replaces the curved ½ time units, a transitional impulse within the same impulses $|1|_M$ rises.

The *transitional* time-space impulse preserves probability measure $|2\times 1/2|_M = |1|_M$ of the initial time impulse, holding the impulse probability for two space units (as a counterpart to the curved time, Fig.1).

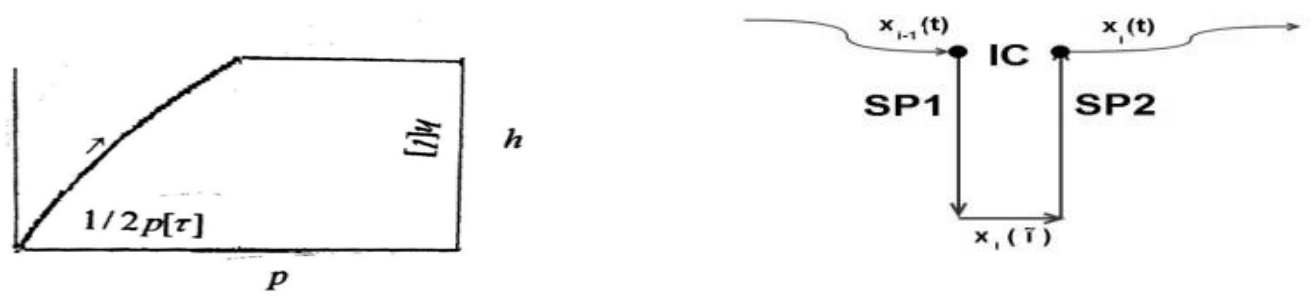

**Fig.1. Illustration of origin the impulse space coordinate measure $h[l]$ at curving time $\tau$ coordinate measure $1/2p[\tau]$ in transitional movement. On the right is an initial impulse with step-down (SP1) and step-up (SP2) actions before the jump at SP1 locality.**

From obvious relation $2\pi h[l]/4 = 1/2p[\tau]$, their coordinate/time ratio $h/p$ leads to the ratio of measuring units: $[\tau]/[l] = \pi/2$ with elementary space curvature $K_s$ equals to inverse radius: $h[l]^{-1} = K_s$.

Since each transitional time-space of the observed impulses holds equivalent probability, it allows appearance of both simultaneously in the observation.

2.4.Microprocess within the impulse. Emerging Information Observer

The merging impulse-jump on the impulse border, cutting the curving time, spots a “needle curve pleat” at transition from the cutting time to a finite space unit.

Thus, the growing Bayes a posteriori probability, along observations merges neighbor impulses, generating interactive jump on each on the impulse border.

The jump brings the extreme discrete displacement, which rotates the jump opposite actions in the anti-symmetric entropy increments, the starting *microprocess* within the bordering impulse [34].

The merging anti-symmetric entropy increments relate to the impulse’ actions superposition.

The jump initiates the inner transitional “mini” impulse within the microprocess.

The jump-wise displacement preserves the Yes-No probabilities of the transitional impulse, and the emerging time-space movement conserves these probability’s measures in *discrete time-space form of the impulse* between the probabilities. The time interval the interactive jump-curving impulse estimates the impulse curvature’ entropy measure and its time-space invariant measure equal to π [34].

At satisfaction of the symmetry condition, the impulses’ axiomatic probability begins transforming to the microprocess ‘quantum’ probability with pairs of conjugated entropies and their correlated movements.

The rotating conjugate movement starts *discrete time-space* micro-intervals forming the transitional mini-impulse at rotating angle $\pm\pi/4$. The opposite rotation with the discrete intervals makes symmetrical mirror copies of the observation within the transitional impulse.

The growing correlation adjoins the conjugated symmetric entropy fractions, uniting their entropies into a running pair *entanglement,* which confines an entropy volume of the pair superposition in the transitional impulse at angle of rotation $\pi/2$.

The entangled pair of anti-symmetric entropy fractions appears simultaneously with the starting space interval. The correlation, binging this couple with a maximal probability, is extremely tangible. The pair of correlated conjugated entropies of the *virtual* impulse is not separable with any *real* action *between* them.

The entangled increments, captured in rotation with forming volumes, adjoin the entropy volumes in a stable entanglement, when the conjugated entropies reach equalization and anti-symmetric correlations cohere. Arising correlated entanglement of the opposite rotating conjugated entropy increments condenses the correlating entropies in the entropy *volumes* of the microprocess.

The stable entanglement *minimizes* a quantum uncertainty of the entangled virtual impulses and increases their Bayes probability.

As maximal a priori probability approaches $P_a \to 1$, both the entropy volume and rotating moment grow. Still, between the maximal a priori probability of virtual process $P_a < 1$ and a posteriori probability of real process $P_p \to 1$ is a small microprocess' gap, associated with time-space probabilistic transitive movement, separating the entropy and appearance of its information.

The gap implies a distinction of statistical possibilities with the entropies of uncertain reality from the information-certainty of reality. The Bayes probabilities measure may overcome this transitive gap.

The gap holds a hidden real locality within the hidden correlation.

The rotating momentum, growing with increased volume, intensifies the time–space volume transition over the gap, acquiring physical property near the gap end at the rising probability.

When a last posterior probability approaching $P_p = 1$ overcomes a last prior virtual probability, the curving momentum may physically cut the transferred entropy volume.

Growing a posteriori probability of the virtual impulse successively brings a reality to its posteriori action, which injects energy, capturing through real interaction (like a bouncing ball) that cuts (erases) the entropy-uncertainty hidden in the correlations.

When the conjugated pair of the correlated entropies is cut, this action transforms the adjoin entropy increments to real information which binds them in real Bit of the *entangled couple* where changing one acts to other.

The step-down (No) control's cut *kills* a total entropy' volume during *finite* time-space rotation and *memorizes* dynamically this cutting entropy as the *equivalent information with its asymmetric geometry*.

Ability of an observer to overcome its gap depends on the amount of entropy volume, enclosing the observed events collected during virtual probes. The probes entropy force and momentum spin the rotating momentum for transition over the gap, and the real control jump adds energy covering the transition.

Cutting the curved volume places a real needle pleat with time–space interacting impulse.

The cutting action, killing the process entropy near $P_p \to 1$, produces an interactive impact between the impulse No and Yes actions, which *requires the impulse access of energy* to overcome the gap.

The impact emerges when virtual Yes-action, ending the preceding imaginary microprocess, follows real No-action which delivers equivalent information compensating for this impact, while the virtual cuts avoid it. Thus, the information impulse appears with energy within it.

Transition maximal probability of observation through the gap, up to killing the resulting entropy, runs a *physical* microprocess with both local and nonlocal entangled information units and real time-space, which preempt memorizing information.

When the posteriori probability is closed to reality, the impulse positive curvature of step-up action, interacting with an external impulse' negative curvatures of step-down action, transits a real interactive energy, which the opposite asymmetrical curvatures actions cover.

During the curved interaction, the asymmetrical curvature of step-up action compensates the asymmetrical curvature of the step-down real impulse, and that real asymmetry is memorized through the erasure by the supplied external energy [38].

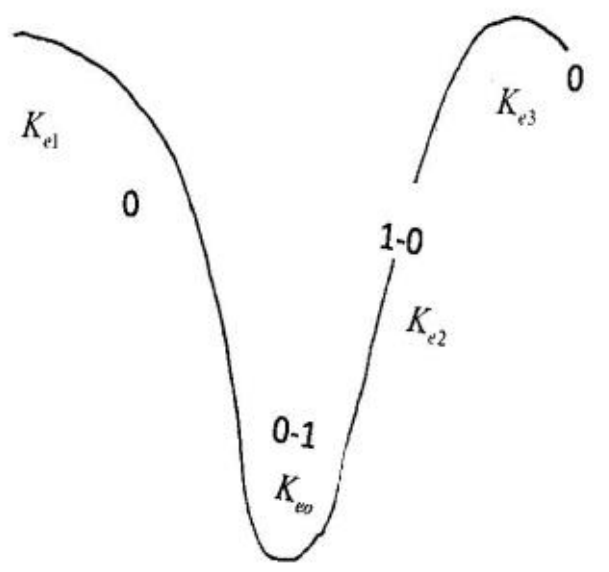

Fig. 2A

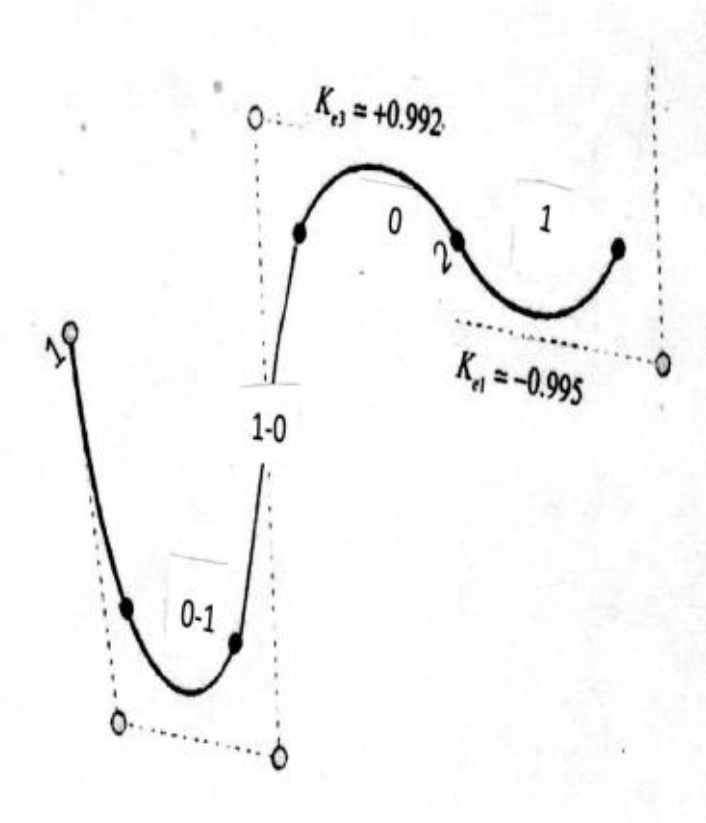

Fig. 2B

**A virtual impulse (Fig.2A) starts step-down action with probability 0 of its potential cutting part; the impulse middle part has a transitional impulse with transitive logical 0-1; the step-up action changes it to 1-0 holding by the end interacting part 0, which, after the inter-active step-down cut, transforms the impulse entropy to information bit.**

**In Fig. 2B, the impulse Fig. 2A, starting from instance 1 with probability 0, transits at instance 2 during interaction to the interacting impulse with negative curvature $-K_{e1}$ of this impulse step-down action, which is opposite to curvature $+K_{e3}$ of ending the step-up action ($-K_{e1}$ is analogous to that at beginning the impulse Fig.2A). The opposite curved interaction provides a time–space difference (a barrier) between 0 and 1 actions, necessary for creating the Bit.**

**When the interactive process provides Landauer's energy with maximal probability (certainty) 1, the interactive impulse' step-down ending state memorizes the Bit. Such certain interaction injects the energy overcoming the transitive gap, including the barrier, toward creation the Bit.**

The step-up action of an external (natural) process' curvature $+K_{e3}$ is equivalent of potential entropy $e_o = 0.01847 Nat$ which carries entropy $\ln 2$ of the impulse total entropy 1 Nat.

The interacting step-down part of internal process impulse' invariant entropy 1 Nat has potential entropy $1 - \ln 2 = e_1$.

Here the interacting curvature, enclosing entropy density, lowers the initial energy and the related temperatures in the above ratio. From that follow *conditions creating a bit in interacting curved impulse* [34].

1. The opposite curving impulses in the interactive transition require keeping entropy ratio 1/ln2.

2. The interacting process should possess the Landauer energy by the moment ending the interaction.

1.3. The interacting impulse hold invariant measure M=[1] of entropy 1 Nat whose the topological metric preserves the impulse curvatures.

The last follows from the impulse' max-min mini-max law under its step-down and step-up actions, which generate the invariant [1]-Nat time-space measure with topological metric π (1/2circle) preserving opposite curvatures.

Theoretically, a pure probability predictability in “idealized measurement” challenges Kolmogorov's probability measure at *quantum mechanics*’ entanglement when both additive and symmetry probability for mutual exchangeable events vanish.

The observing Markov probabilities’ additive and multiplicative properties are changing within the microprocess. Specifically, the merging jump violates regular properties of Markov process leading to a sub-Markovian process [39].

From the starting jump-up to the following jump-down, launching space interval of transitional impulse, the microprocess holds additive properties for both probability and related conjugated entropy.

The transitional impulse, confining the entanglement, ends with jump-down, initiating only the multiplicative property for the entangled entropy and the probability.

The microprocess within the transitional impulse is reversible.

The microprocess within whole impulse is reversible until the impulse ending action cuts its entropy.

It concurs with property of quantum wave function before and after interactive measurement.

For each random impulse, proceeding between a temporary fixed the correlated random No-Yes actions, such microprocess is multiple whose manifold decreases with growing the probability measure. At a maximal probability, only a pair of additive entropy flows with symmetric probabilities, which contains symmetrical-exchangeable states advance in the superposition.

The multiple impulses initiate a manifold of virtual Observers with random space-time shape in a collective probabilistic movement.

With maximal observing probabilities, the manifold of the virtual observers also decreases.

The microprocess is different from that in quantum mechanics (QM), since it rises as a virtual inside probing impulse of growing probability under jumping action, and then evolves to the real under final physical No-actions. Its superposing rotating anti-symmetric entropy flows have additive time–space *complex amplitudes* correlated in time-space entanglement, which do not carry and bind energy, just connects the entropy in joint correlation, whose cuts models elementary interaction with no physics.

The QM probabilistic particles carry the analogous conjugated probability amplitudes correlated in time-space entanglement.

Virtual microprocess does not dissipate but its integral entropy decreases along No-Yes reversible probes in the observation.

The real microprocess builds each information unit-Bit within the cutting impulse in real time, becoming irreversible after the cut (erasure). These operations, creating Bit from the impulse, reveal structure of Weller Bit, which memorizes the Yes-No logic of virtual actions, while the Bit free information participates in getting the multiple Bits information.

Such Bit-Participator is primary information observer formed *without a priori physical law.*

Whereas *the observing probable triple in the field specifies each information observer.*

The information observer starts with real impulse cutting off the observing process and extracting hidden information Bit. That identifies the information observer as an extractor and holder this information emerging in observation. Killing physical action converts entropy of virtual Observer to equivalent information of real Information Observer.

Killing the distinct volumes densities converts them in the Bits distinguished by information density and curvature.

The curving impulse of each Bit accumulates the impulse complimentary–opposite actions carrying *free information*, which initiates the Bits' attraction.

During the curved interaction, a primary virtual asymmetry, measured by equivalent entropy, compensates the asymmetrical curvature of a real external impulse. The real asymmetry is memorized as information through the entropy erasure by the supplied external Landauer's energy.

The entropy' cutting interaction, curving asymmetry and producing information, performs function of Maxwell demon, which emerges with the curving asymmetry.

Between different Bits rises information gradient of attraction minimizing the free information which finally binds Bits. That connects the Observer's collected information Bit in units of an information process, which finally *builds Observer information structure.*

The maxmin-minimax principle, rising in the impulse observation, leads to the following attributes of emerging information Bit:

-it information is delivered by capturing and cutting entropy of virtual observing correlated impulses;

-its free information is transferring to the nearest impulse that keeps persistence continuation of the impulse sequence via the attracting Bits;

-the persistent Bits sequentially and automatically convert entropy to information, holding the cutoff information of random process correlation, which connects the Bits sequences;

-the Bit holds time–space geometry following from the geometrical form of discrete entropy impulse;

-the information, memorized in the Bit, cuts the symmetry of virtual process;

-the free information, rising between the Bits cutting from random process, is spending on binding the attracted Bits;

-each three Bits' free information allows binding the triple bits in a new bit, which is different from the primary Bit cutoff from random process;

-both primary and attracting Bits' persistence continuation integrate the time-space real information process, which is composing the elementary information units in space-time information structure of Information Observer.

While the Bit preserves origin of its information, the growing information, condensed in the integrated Bit's finite impulse size (limited by the speed of light), increases the Bit information density. The increasing density conserves growing energy being equivalent of interacting physical particles-objects.

The optimal estimation of a lower limit on the increment of probability events evaluates the probabilities and limitations on process of observations [33].

Evaluation the information constraints, limitation on the microprocess, and conditions of the observer start predicts the probability path from virtual probes to the probability approaching a real cutoff.

On the path of observing uncertainty to certainty, emerge causality, information, and complexity, which the information Bit's geometry encloses in the impulse time, curvature, and space coordinates.

An edge of reality, within the entropy-information gap, evaluates Plank's fine structural constant in a sub-plank region of uncertainty [34]. A minimal displacement within this region estimates number of probing impulses to reach that region, which concurs with virtual probes approaching the real cutoff.

## 2.5.Information Macroprocess

The rotating movement (Fig.3) of the information Bits binds them in information macroprocess, which describes the extremals of the EF variation problem (VP) that was solved.

The macroprocess free information integrates the multiple Bits in information path functional (IPF) which encloses the Bits time-space geometry in the process' information structure.

Estimation of extremal process shows that information, collected from the diffusion process by the IPF, approaches the EF entropy functional. The IPF formalizes the EF extreme integral for the cutting interactive impulse, whose information approaches EF at the bit number $n \to \infty$, [32].

The IPF integrates flows of Bits units with finite distances and sizes.

The IPF maximum, integrating unlimited number of Bits, limits the total information that carries the process' Bits and intensifies the Bit information density, running it to infinite process dimension.

At infinitive dimension of the macroprocess, it describes both the EF and IPF extremals.

The limited number of the process units, which free information assembles, leads to limited free information transforming the impulse microprocess to the macroprocess with a restricted dimension.

The randomly applied deterministic (real) impulses, cutting all process correlations, transforms the initial random process to a limited sequence of independent states.

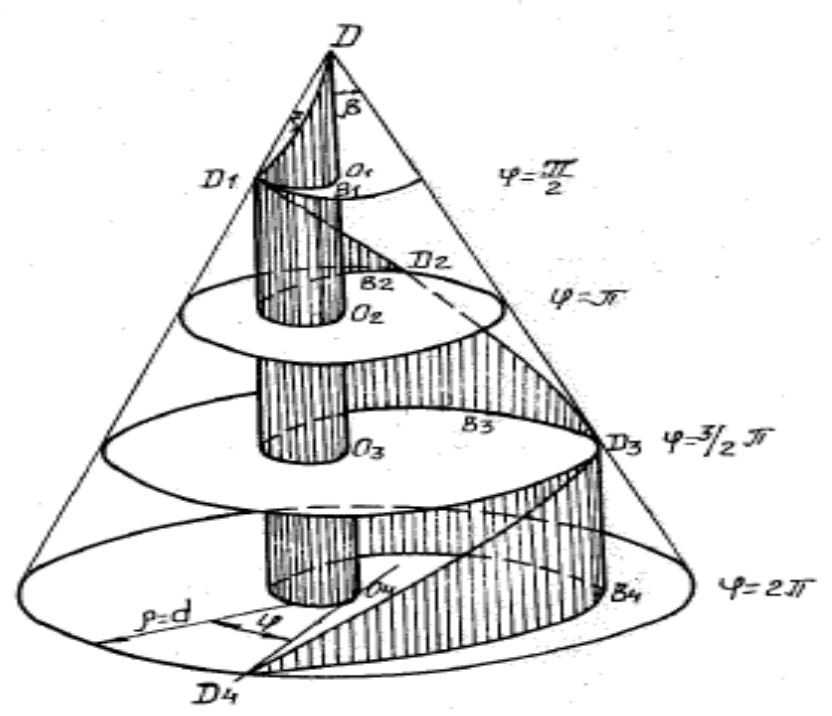

.

**Fig.3. Forming a space -time spiral trajectory with current radius $\rho = b\sin(\varphi\sin\beta)$ on conic surface at points D, D1, D2, D3, D4 of spatial discrete interval DD1=$\mu$, which corresponds to angle $\varphi = \pi k / 2$, $k = 1,2,...$ of radius vector's $\rho(\varphi,\mu)$ projection of on the cone's base (O1, O2, O3, O4) with the vertex angle $\beta = \psi^o$.**

The ratio of primary a priori- a posteriori probabilities beginning the probabilistic observation identifies the initial conditions for the EF and its extremals. The initial conditions determine the entropy function starting virtual observations at a moment which depends on minimal entropy (uncertainty) arising in the observations.

This entropy allows finding unknown posteriori entropy starting virtual Observer.

The initial conditions bring complex (real and imaginary) entropies for starting conjugated processes in a virtual observer' microprocess. This process' minimal interactive entropy becomes threshold for starting information microprocess, beginning the real observation and information Observer.

Starting the extreme Hamiltonian processes evaluate two pairs of real states for the conjugated rotating dynamic process [27-31]. That allows finding the equations unifying description of virtual and real observation, microprocess, and the dynamic macroprocess.

Applying the macrodynamic equations [30, 34] to traditional form of dynamic model with unknown control function solves the initial VP which determines the optimal control for the traditional model. These controls, formed by feedback function of the macrostates, bring step-up and step-down actions which sequentially start and terminate the VP constraint, imposed on the model. That allows extracting the cutoff hidden information from the localities joining the extremal segments of the observing trajectory. The extracted information feeds the observer macrodynamics with the recent information from the current observations.

This feedback' process concurrently renovates the observing Markov correlations connecting the macrostates and the controls The correlations identify the Markov drift function transferred to the equations of the information macrodynamics [29,34].

Within each extremal segment, the information dynamics is reversible; the irreversibility rises at termination of each VP constraint between the segments.

The feedback identifies cutting correlations and automatically transforms the EF to IPF. That reveals the integrated information hidden in the observing process randomness.

While the sum of cutting correlation functions identifies the IPF integrant.

The EF-IPF Lagrangian integrates both the impulses and constraint information on its time-space intervals.

The identified VP constraint leads to invariant relations for each impulse information, interval, and the intervals between information impulses holding three invariant entropies of virtual impulses, as well as invariant conjugate vector on the extremal. These invariants estimate each segment information on the macrotrajectory, the locality between segments, predicting where the potential feeding information should transfer the feedback or measuring.

The invariants allow encoding the *observing process* using Shannon's formula for an average optimal code-word length of the code alphabet letters.

The information analog of Plank constant evaluates maximal information speed of the observing process, which estimates a time interval and entropy equivalent of the gap separating *the micro- and macroprocess.*

Shifting this time to real time course automatically converts its entropy to information, working as Maxwell's Demon, which enables compensating for the transitive gap.

*The equations of observation finalize both math description of the micro-macro-processes and validate them numerically.*

2.5.1. The information process' basic structural unit-triplet.

The cutoff attracting bits, start collecting each three of them in a primary basic triplet unit at equal information speeds, which resonates and coheres joining in triplet units $UP_o$ of information process.

Specifically, each three bits, processing the joint attractive movement, cooperate in the created new attracting bit, which composes a primary basic triple unit and then composes secondary triple unit, and so on.

That cooperates a nested sequence of the enclosed triplet unit emerging during cooperative rotation, when each following triplet unit enfolds all information of the cooperating units in its composite final bit.

The $UP_o$ size limits the unit starting maximal and ending minimal information speeds, attracting and forming new triplet unit by its free information along the macroprocess. The process movement selects automatically each $UP_o$ during the attracting minimax movement, by joining two cutoff Bits with a third Bit, which delivers free information to next cutting Bit, forming next cooperative triple unit $UP_i$ (Figs.4,5). Forming the multiple triplet trajectory follows the same procedure, (Fig.5,6)

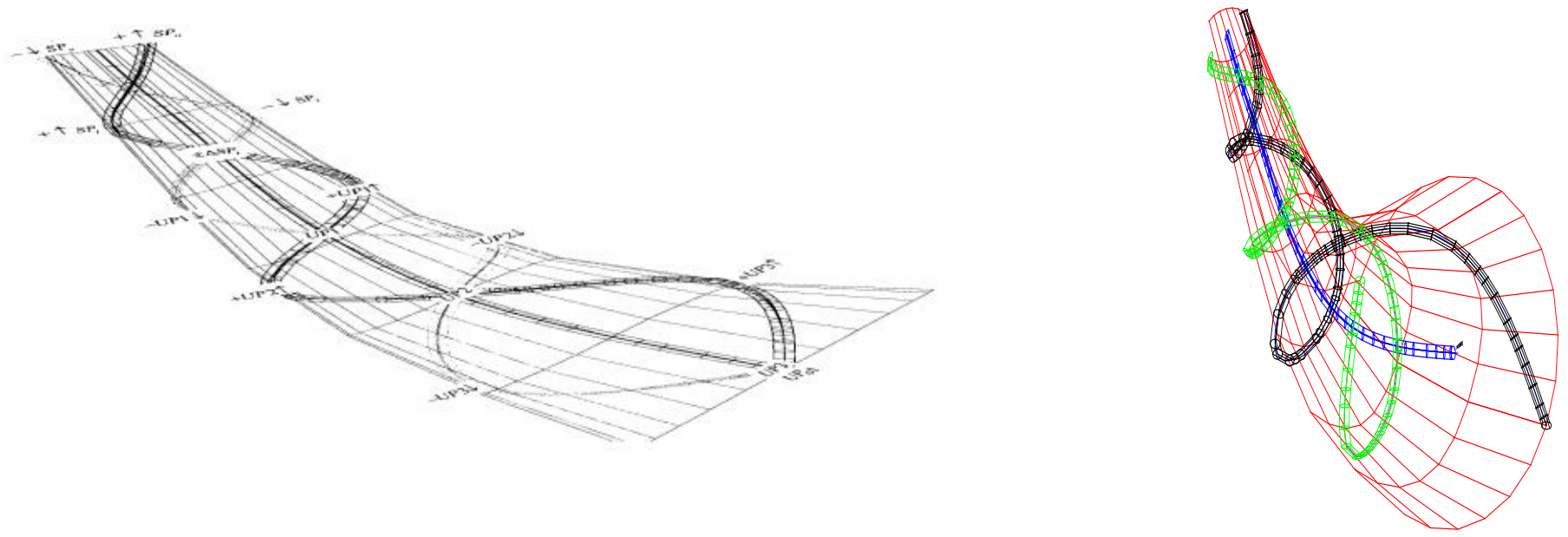

**Fig. 4. Time-space opposite directional-complimentary conjugated trajectories $+\uparrow SP_o$ and $-\downarrow SP_o$, form spirals located on conic surfaces (analogous to Fig.2). Trajectory connected bridges $\pm\vartriangle SP_i$ binds the contributions of process information unit $\pm UP_i$ through the impulse joint No-Yes actions, which model a line of switching controls.**
**Two opposite space helixes and middle curve are shown on the right.**

Each self-forming triplet joins two segments of the macro-trajectory with positive eigenvalues by reversing their unstable eigenvalues and attracting a third segment with negative eigenvalues. The third segment' rotating trajectory moves the two opposite rotating eigenvectors and cooperates all three information segments in a triplet's knot (Fig. 5). Each triplet unit generates three symbols from three segments of information dynamics and one impulse-code from the control, composing a minimal *logical code* that encodes this elementary physical information process.

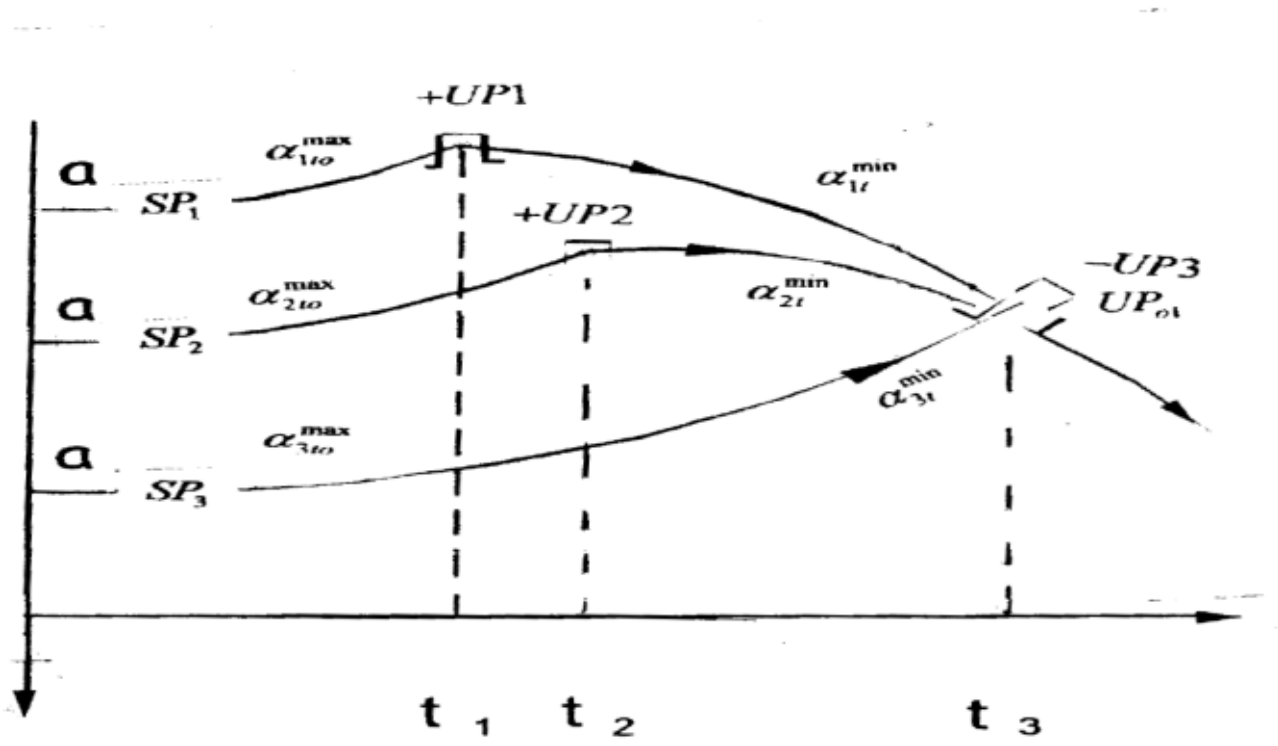

**Fig.5. Illustrative dynamics of assembling units $+UP1, +UP2, -UP3$ on the space-time trajectory and adjoining them to $UP_{o1}$ knot along the sections of space-time trajectory $SP_1, SP_2\ SP_3$ (Fig.4) at changing information speeds from $\alpha_{1to}^{\max}$, $\alpha_{2to}^{\max}, \alpha_{3to}^{\max}$ to $\alpha_{1t}^{\min}$, $\alpha_{2t}^{\min}, \alpha_{3t}^{\min}$ accordingly; a is dynamic information invariant of an impulse.**

Fig.5. shows how a minimum three self-connected Bits assemble optimal $UP_o$-basic *triplet* whose free information requests and binds new triplet, which joins and binds three basic triplets in the ending knot, accumulating and memorizing information of all trees. Each current unit $UP_k, k=1,2.3$ composes its bits in triplet code that encodes and connects the units.

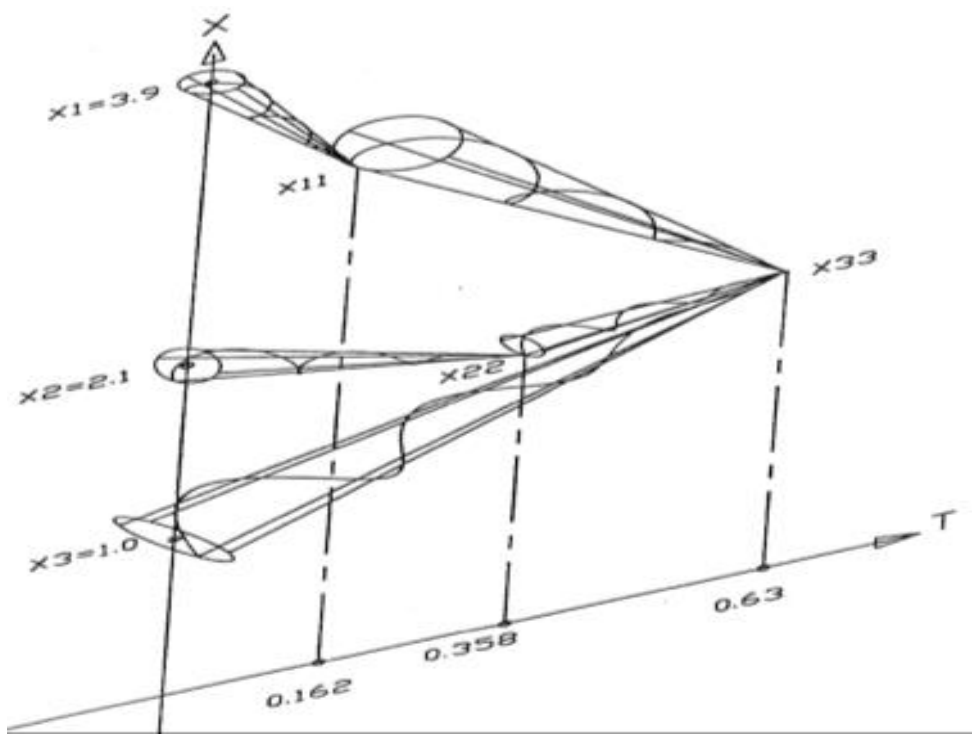

**Fig. 6. Simulation of forming a triplet's space structure on a knot node, composed from time-space spirals Fig.3 (according to Fig.4). The Fig.6 indicated time intervals measure real times $\Delta t_{13}, \Delta t_{13}, t_3$ during the simulation shown on cones diameters.**

The pair of opposite directional rotating units equalizes their eigenvalues with the rotating third one, attracts and binds them in the triple by the starting attracting force.

2.5.2. Assembling the information network (IN).

During the macro-movement, the joint triple unit' free information transfers the triple code to next forming triplet, which assembles a network that is building by the forming triplet' code.

The multiple triples sequentially adjoin time-space hierarchical network (IN) whose free information requests from observation and attaches new triplet unit at its higher level of the knot-node, concurrently encoding the IN triple logic (Fig.7).

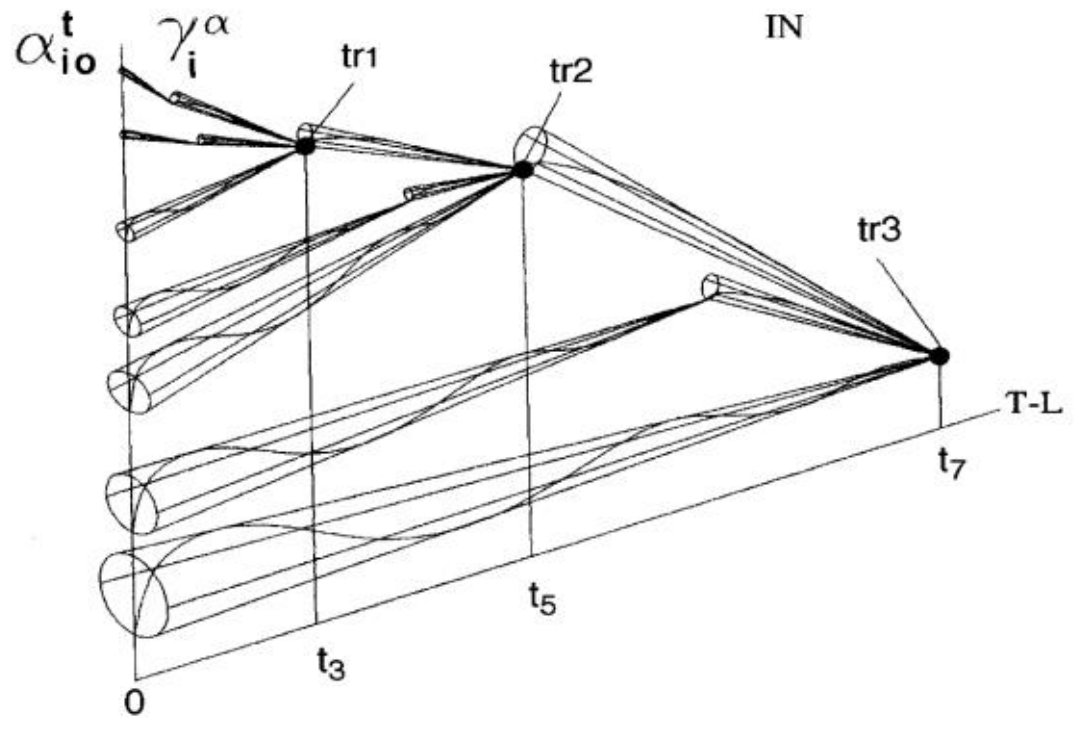

**Fig.7. The IN information geometrical structure of hierarchy of the spiral space-time dynamics (Figs.3-6) of the triplet nodes (tr1, tr2, tr3, ..); where $\{\alpha_{io}^{t}\}$ are a ranged string of the initial eigenvalues, cooperating on $(t_1, t_2, t_3)$ locations of T-L time-space, where $\gamma_i^{\alpha} = \alpha_{io}^{k} / \alpha_{io}^{k+1}, k$ is number of a nearest IN triplets.**

Each triplet has unique space-time position in the IN hierarchy, which defines exact location of each code logical structure. The IN node hierarchical level classifies *quality* of assembled information.

The currently ending IN node integrates information enfolding all IN levels in this node knot (Fig.7).

The IN is automatically feeding novel information, which it concurrently requests by a deficit of needed information.

The deficit creates equal entropy-uncertainty, requesting new information, which initiates probing impulses and the frequency of the observing entropy impulses.

The information probing impulses interact with the observing cutting their entropy.

That provides the IN-feedback information, which verifies the IN's nodes' requesting information.

New information for the IN delivers the requested node interactive impulses, whose impact on the probing impulses, with observing frequency, through the cutoff, memorizes the entropy of observing data-events. Appearing new quality of the information triplet currently builds the IN temporary hierarchy, whose high level enfolds the information triplet logic that requests new information for the running observer's IN, attaches it, and extends the logic code. The emergence of observer current IN level indicates observer's information surprise measured by the attaching new information.

The IN time-space logic encodes in double spiral space (DSS) triple code that rotates, encircling the conic structures (Fig.3,4,7), which multiple INs logic extends.

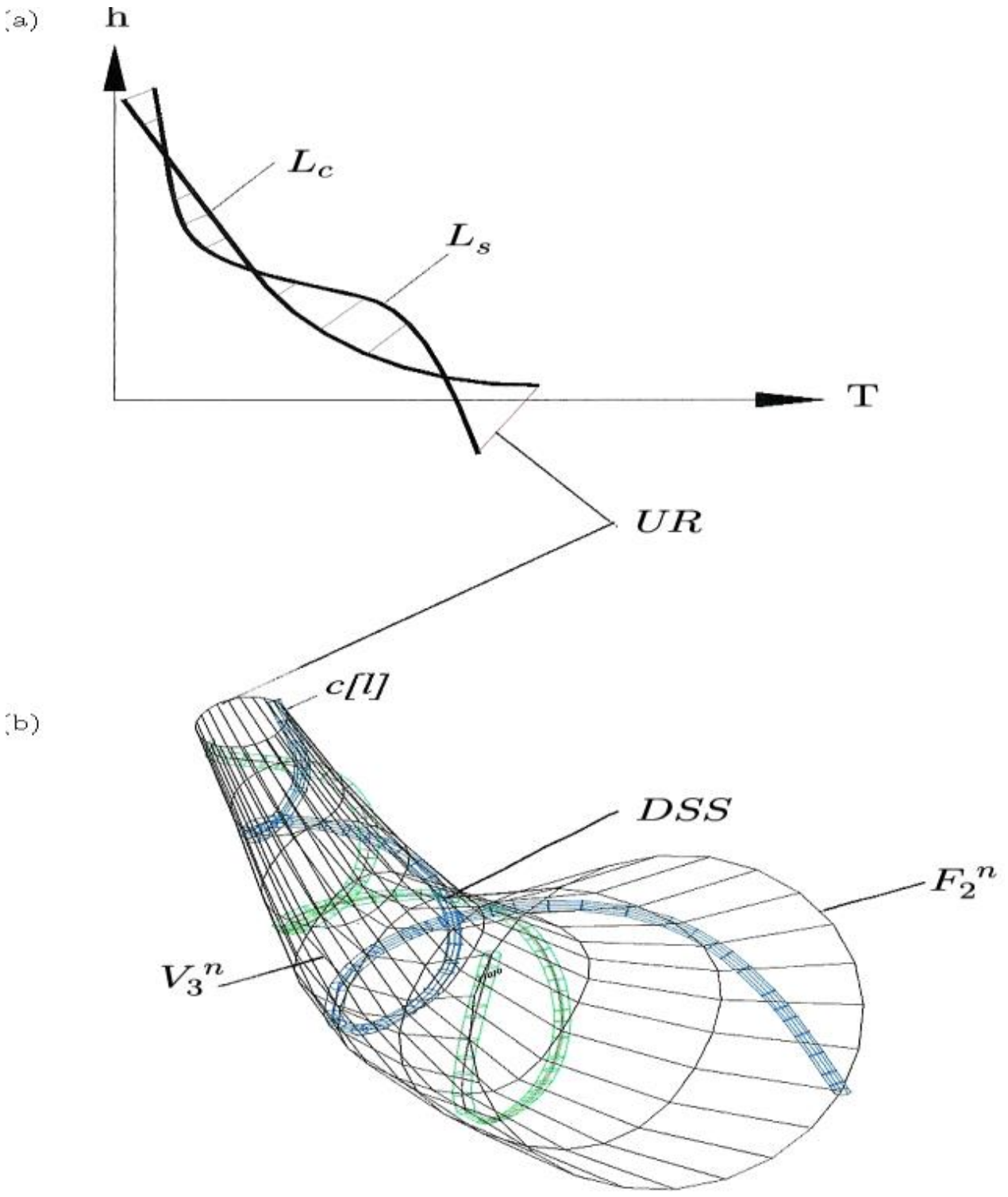

**Fig.8. Simulation of forming double spiral cone's structure (*DSS*) with cells (c[l]), arising along switching control line hyperbola $L_c$ (shown in (Fig.3)) and uncertainty zone (UR) geometry, surrounding the $L_c$ .**

**The curved macro trajectory models line $L_s$ whose rotation around $L_c$ forms UR, enfolding the geometry volume $V_3^n$ with space surface $F_2^n$ This space structure encloses each of the spiral models on Fig. 3, 5, 6.**

The ending knots of a higher IN's level assemble its three units $\pm UP_{i0}$ in the next IN level's triplet which starts it, Fig.8.

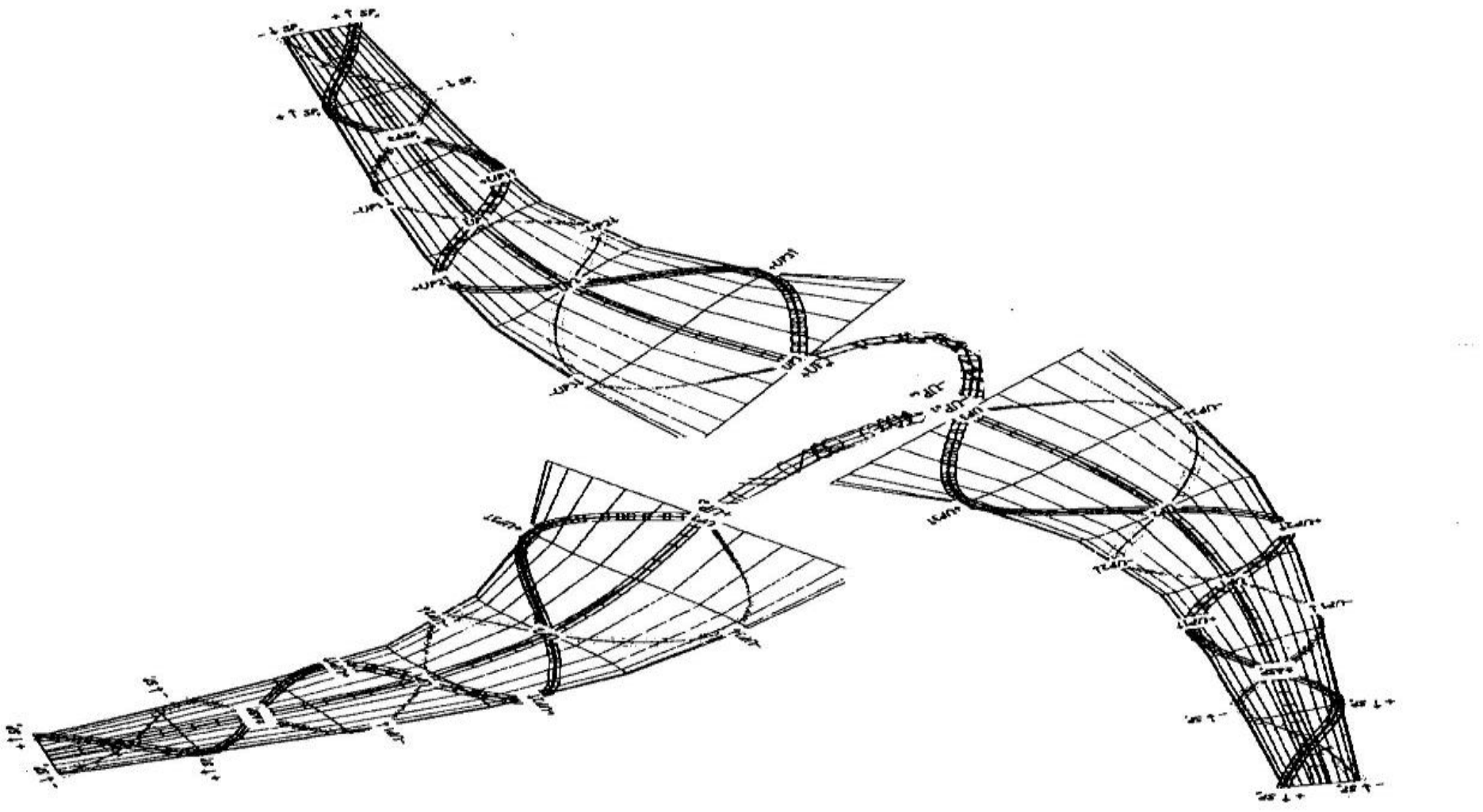

**Fig.9. Assembling three formed units $\pm UP_{i0}$ at a higher triplet's level connecting the units' equal speeds (Fig.4).**

The attracting process of assembling knots forms a rotating loop shown on Fig.7 at forming each following cooperative triplet tr1, tr2,…. (The scales of the curves on Figs 8 distinct from the interacting knots' curves on Fig. 7, since these units, at reaching equal speeds, resonates, which increases size of the curves on Fig. 9).

The rotating process in the loop of the harmonized speeds-information frequencies at different levels is analogous to the Efimoff scenario [40]. The loop could be temporal until the new formed IN triplet is memorized. The loop includes Borromean knot and ring, which was early proposed in Borromean Universal three-body relation.

The multiple IN time-space information geometry shapes the Observer *asymmetrical structure* of cellular geometry (Fig.10) of the DSS triple code, (Fig.9).

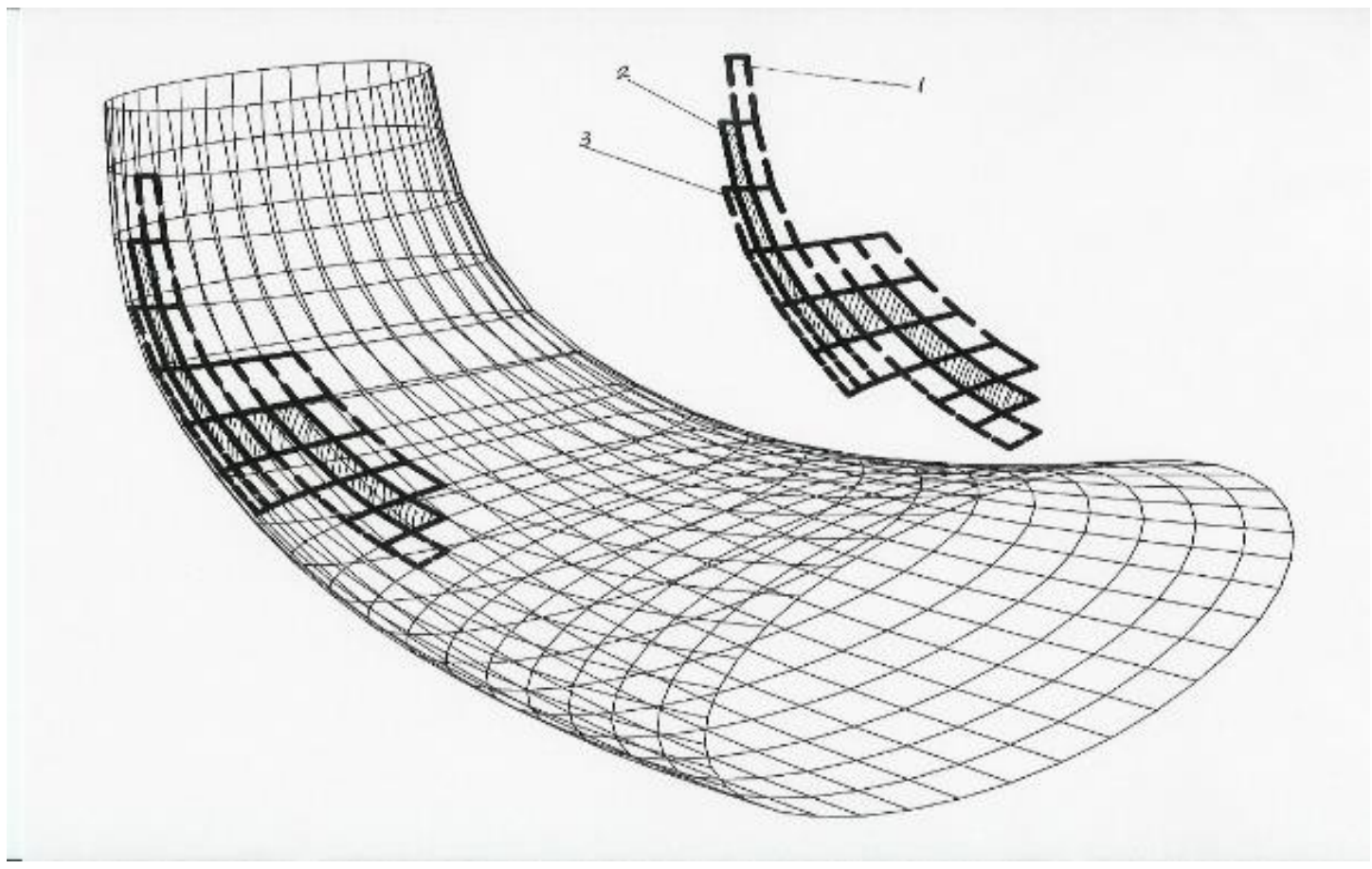

**Fig. 10. Structure of the cellular geometry, formed by the cells of the DSS triplet's code, with a portion of the surface cells (1-2-3), illustrating the space formation. This structure geometry integrates information contributions modelling on Figs.3, 7.**

The macroprocess integrates both imaginary entropy of the merging impulses micro processes and the cutoff information of real impulses, which sequentially convert the collected entropy in information physical process during the macro-movement.

The observation process, its entropy-information, and micro-macro processes are Observer-dependent, information of one Observer distinct of the information of others Observers.

However, the invariant information minimax law leads to the invariant information *regularities* for different Observers.

Observing the same process with different probability field triple, each Observer gets specific information. The information requests its current IN during its optimal time-space information dynamics. Optimal information process determines the extremal trajectories of the entropy functional, solving the minimax variation problem for the observing process.

The information macrodynamic equations [29-34] describe information macroprocess, which averages all observing microprocesses and holds regularity of observations under the maxmin-minimax impulses. These equations *predict optimal information path* from the starting virtual observation up to information process, and the physical macrodynamics.

## 2.6. Forming structure of information Observer and its regularities

The macro-movement in rotating time–space coordinate system forms Observer's information structure confining its multiple INs that determine the *Observer time of inner communication* with *self –scaling* of both requesting and accumulating information. Each Observer *owns the inner time of information processing* and the time *scale* of the required information (on the micro and macrolevels), depending on density of the IN nodes information.

(a). The current information *cooperative force*, initiated by free information, evaluates the observer's *selective* actions attracting new high-quality information. *S*uch quality delivers a high density-frequency of the related observing information through the IN selective mechanism of the requested information.

These actions engage acceleration of the observer's information processing, coordinated with new selection, quick memorizing and encoding each IN node information with its logic and space-time structure, which minimizes the spending information. It determines observer's self-organized feedback loop.

The optimal criterion is growing quality of the observing information, collected by the IN, which selects the needed information that the observer acquires.

(b). The observer optimal *multiple choices limit and* implement the minimax self-directed strategy, which evaluates the amount of the information emanated from the IN integrated node *that identifies the attracting cooperative force*.

(c). The IN nested structure holds cooperative complexity [41] measuring *origin* of complexity in the interactive dynamic *process* cooperating doublet-triplets. Their free information anticipates new information, requests it, and automatically builds hierarchical IN with the DSS that decreases complexity of not cooperating yet information units.

(d). The information structure, self-built under the self-synchronized feedback, drives self-organization of the IN and the *evolution macrodynamics* with ability of its self-creation.
(e). The observer cognition emerges from the evolution process, as evolving intentional ability of requesting, integrating, and predicting the observer needed information that builds the observer growing networks.

The evolving free information builds the Observer specific time–space information *logical* structure that conserves its "*cognition*".

Such logical structure possesses both virtual probabilistic and real information causality and complexity, whose information measures the cognitive intentional actions.

The rotating cognitive movement connects the impulse microprocess with the bits in macroprocess, composing the elementary macrounit-triplet, and then, through the growing IN's levels quality information, integrates multiple nested IN's information logic in an information domains.

The observer's cognition assembles the common units through the multiple resonances at forming the IN-triplet hierarchy, which accept only units that each IN node concentrates and recognizes.

The cognitive movement, at forming each nodes and level, processes a *temporary loop* (Fig.6) which might disappear after the new formed IN triplet is memorized. Along the IN hierarchy runs the distributed resonance frequencies spreading a chain of the loops.

The chain rotates the thermodynamic process (cognitive thermodynamics) with minimal Landauer energy, which performs natural memorizing of each bit on each evolution level.

The cognitive actions model the correlated inter-actions and feed-backs between the IN levels, which controls the highest domain level. Both cognitive process and cognitive actions emerge from the evolving observations, which maintain the cognitive functions' emerging properties and encodes the cognitive logic information language.

## 2.7. The emerging information Intelligence.

The observer intelligence emerges on the path from staring observation, virtual observer, creation microprocess, bits, information macroprocesses, and nested networks (IN) with growing quality of information and the nested logic.

These self-generate the observer selective actions, ability of their prediction, the IN concurrent renovation, and extension to complex self-built IN domains enclosing maximal quality of condense information and its logic.

The self-built structure, under self-synchronized feed-backs, drives self-organization of the IN and evolution macrodynamics with ability of its self-creation.

Within the evolving processes, integrating by the IN space-time coherent structure, emerges the observer cognition, which starts with creation elementary units of virtual observer holding a memory at microlevel.

The coordinated selection, involving verification, synchronization, and concentration of the observed information, necessary to build its logical structure of growing maximum of accumulated information, unites the observer's self-organized cognitive actions performing *functional organization.*

The functional organization of these intelligent actions spent on this action evaluate the memorized amount of quality information at each IN' ending hierarchical level. This functional organization integrates the interacting observers' IN levels and domains in the observer IN highest (ending) hierarchical level. Maximal level measures maximal cooperative complexity enfolding maximal number of the nested INs structures, which memorize the ending node of the highest IN. The DSS helix structure (Fig.8) self-organizing the multiple local bits coding units, which encodes the observer triplet coding structure (Fig. 10), we call *observer intelligence.*

The intelligence has an ability of the observer to build the informational networks and domains, which includes the cognitions. The ended triplet of observer hierarchical informational networks and domains measures level of the observer intelligence. All observers have different levels of intelligence which classify observer by these levels.

The quality of information *memorized in the ending node of each observer IN highest levels* measures the observer information Intelligence. The cognitive process at each triplet level preempts the memorizing.

An observer that builds maximum number of the hierarchical informational networks and domains has maximum intelligence. This observes have an imbedded ability to control other observers.

The *Observer Intelligence* holds ability to uncover causal relationships enclosed in evolving observer networks and self- extends the growing quality information and the cognitive logic on building *collective observer intellect.*

The intelligence of a multiple interactive observers integrates their joint IN's ending node.

The Observer *Cognition* emerges in two forms: a virtual, rotating movement processing temporal memory, and the following real information mechanisms, rotating the double helix geometrical structure and memorizing it.

The DSS concurrently places and organizes the observing information bits in the IN nodes, whose sequential knots memorize information causality and logic.

The self-directed strategy develops multiple logical operations of a self-programming computation which enhances collective logic, knowledge, and organization of diverse intelligent observers.

The EF-IPF measure allows evaluating information necessary to build a minimal intelligent observer.

The increasing INs hierarchy enfolds rising information density which accelerates grow the intelligence that concurrently memorizes and transmits itself over the time course in an observing time scale. The intelligence, growing with its time interval, increases the observer life span.

The self-organized evolving IN's time-space distributed information structure models *artificial intellect*, which we detail in Sec.8, where the specifics of cognition and intelligence are described.

## 2.8. The Observer individuality specifics:

-The probability field' triple observing specific set of probabilistic events that arise in emerging particular information observer. The asymmetry of final bit will conserve this specific;

-Time of observation, measuring quantity and density of information of the delivered bits;

-Cooperation the observing information in a limited number of the IN-triplet nodes and limited number of the observer's IN, which depends on individual observer selective actions [34].

-The selective actions define the cooperative information forces, which depends on the number of the IN nodes. The minimal cooperative force, forming very first triplet, defines minimal selective observer.

The individual ability for selection classifies information observers by levels of the IN hierarchy, time-space geometrical structure, and inner time scale whose feedback holds *admissible* information spectrum of observation.

The individual observers INs determine its explicit ability of self-creation.

-These specifics classify the observers also by level of cognition and intelligence.

The information mechanism of building all observers is invariant, which describes the invariant equation of information dynamics following from minimax variation principle.

**3. Information structure of artificial designed observer: the basic stages toward artificial brain**

Observers are everywhere, including people, animals, other species, multiple particles, and objects communicating and *interacting* between each other, accepting, transforming and exchanging information.

To design an artificial observer, which models AI brain information logic, cognition and intellect, we need to unify the interacting properties of various observers by a common structure of Information Observer.

Various interactions simplify inter-active Yes-No (↓↑) actions-an impulse known as a Bit, the elementary unit of information. Since the multiple interactions generates a manifold of such Yes-No actions, they can model interactive Information Observer.

If such unified Observer is feasible, it possess the common information regularities with invariant observer structure, considered above, which creates an interactive impulse *observation*.

Creation the artificial information Observer under such observation requires some common steps described below.

Since the multiple interactive actions are random, their multiplicity generates a random multi-dimensional process, and the interacting result evaluates its probability of occurrences. (No material particles or 'agents 'participates in such observation.)

More probable occurrence allows selecting the needed result estimating its certainty as observed information.

A simple example is observing some uncertain planet moving around a star. When probability of its observation increases, making a shot-film which brings the copy of that most probable observation. Such copy removes the observer uncertainty, while spending energy on the shot, which the copying film consumes. Then the film exposes a certain observation which, enclosing the spending energy brings the current information of the observing planet.

Therefore, the probabilities that the observation process generates, are sources of a random uncertainty which, being removed, brings an equivalent certainty, or information about the observation process.

Removing the uncertainty during natural interaction creates information as natural phenomenon of interaction.
The observation, searching the information, interacts with the current observing probabilities sending multiple Yes-No probabilistic impulses to decreases the uncertainty of the observing probabilities until getting certain Yes-No impulse of elementary bit of the information. Such active observation builds its observation process through decreasing the random uncertainty by the probabilistic impulses, which observe the probabilities of random uncertainty. The active observation models a 'virtual' observer whose increasing probabilistic impulses generate multiple bits of certain information observer.
Our aim is a conceptual design the invariant structure of information observer emerging in observation with individual properties of self-encoding and integrating information codes in self-creating space-time structure, which models specific AI brain information logic, cognition and intellect.

3.1. Probabilistic observation measured by discrete probabilistic impulses of the observing process

Since the probabilistic observation runs the multiple interactive impulses, they are the probability impulses probing a random environment, which formally describes a random field of multiple interactions.
The field of axiomatic probabilities defines mathematical triad: the sets of possible events, the sets of actual events, and their probability function. Each of this triad specifies the particular observer whose probability, belonging to the field, observes the current events. The multiple triads extend the impulses' probabilistic observation toward artificial brain emerging in the observations.
According to formal theory of probabilities, the axiomatic probabilities are objective by definition, formalizing multiple experimental occurrences of each random event.
In artificial observation, the probabilistic impulses can produce a random generator with the probabilities of chosen triad.
Formally, a source of such discrete axiomatic probabilistic impulses is Kolmogorov 0-1 law probabilities, which generates the sequence of *mutually independent* variables of the events from the field triad.
These no-yes probabilistic impulses interact with a probabilistic observing process in the field, whose reflecting actions are multiple yes-no impulses probabilities: a priory probability (before yes-action) and a posteriori probability (after yes-action) as the Bayes probabilities connecting a sequence of the impulses in the *observing* random process.
Classical process, connecting multiple interacting impulses is Markov diffusion process, widely used in many physical, chemical, biological and many other observations and applications,
It allows assigning Markov process as the observable random process.
The sequence of the probabilistic no-yes impulses of the random field initiates the Bayes probabilities within the Markov process, which self-observe the evolving Markov process inner probabilities generating the observing process.
The sequential Bayes probabilities provide a *virtual* observation of the observable process as a virtual (potential) observer. These, also axiomatic probabilities have analogy with standard Bayes probabilities that evaluate the probability of a hypothesis by a priori probability, which is then updated through a posteriori probability of the evidence.

Since the impulse probabilities start with specific triad within the random field, the initiated Markov process probabilities begin the observation process with the same triad.

Multiple observing Markov processes start with multiple triads.

In this probabilistic model, a potential observer of hidden (random) bits models Markov observing process located in the surrounded random field and therefore affected by the field's random impulses-frequencies.

Since our Universe build multiple interacting processes, their running multiplicity is actual source of random field.

In the formal probabilistic model, the axiomatic artificial probabilities objectively measure the observation process and build the observing process.

The observing process starts observation with the specific field triad which spots the current observing environment enclosing the observing events.

In the artificial observer, designing during the observation, these probabilities identify the multiple interacting events in the observable spot. The artificial designed observer generates the random yes-no impulses self-creating the observable process of its current environment, which self-builds the observing process of the hidden bits.

The process probabilities identify the multiple interacting events in the observable spot, which finally measures the probability revealing the bits.

The uncovered bits provide a future real code which generates that information observer.

3.2. Reduction of the observable process entropy under probing impulse and rising probabilistic logic

Uncertainty of random events in the observable process evaluates logarithmic measure of the event's relative probabilities, which defines relative entropy. The events connect correlation along the observable process.

Uncertainty of random events between the impulse yes-no probabilities identifies the relative probabilities' logarithmic measure which is a relative entropy of the probabilistic impulse.

This entropy is immanent to probabilistic probing impulses interactive actions $\downarrow\uparrow$ in the observing process.

The observable process correlation evaluates its time interval [36].

The correlation connection between the interactive actions $\downarrow\uparrow$ equivalent to time interval between these actions, which is the impulse' unit $\bar{u}_k$ measure $M[\bar{u}_k] = |1|_M$ .

Each No-probability of probing impulse virtually cuts this entropy and correlation in observable process.

Within each interacting impulse, its step-down No-cutting action maximizes the cutting entropy, while its step-up Yes reaction, spending an entropy on this action' creation, minimizes the cutting entropy.

That provides max-min principle for each impulse, and mini-max along the multiple observing impulses.

Between the No-Yes impulse actions, the maximal cutting No action minimizes absolute entropy that conveys Yes action (rising its probability), which leads to a max-min of relational entropy transferring the probabilities.

In simple example with bouncing rubber ball, when the ball hits ground, the energy of this interaction partially dissipates that increases the interaction's total entropy, while the ball's following the reverse movement holds less entropy (as a part of the dissipated), leading to max-min entropy of the bouncing ball. Adding periodically small energy, compensating for the interactive dissipation, supports the continuing bouncing.

As soon as the initial impulse 0-1 actions involve, the minimax principle is imposed along all impulse observation.

This leads to reduction the observable process entropy under the multiple probing impulses that increase a posteriori probability in the observing process.

The sequential a priori-a posteriori probabilities determine probabilistic causality along the observing process, which carries its probabilistic logic. (The hidden bit, covering this yes-no probabilistic causality, hides this probabilistic logic). The logarithmic ratio of those probabilities defines relational entropy which measures this logic for each impulse in the sequence.

The observing process logic integrates entropy functional (EF) along trajectories of the Markov process, which measures the process integral entropy.

The minimax principle formalizes its variation problem (VP), which, applying to the EF, allows analytically describing the observing process minimax trajectory as the extreme VP solution [27].

The solution brings invariant entropy increment for each discrete impulse, preserving its probability measure, and the impulse time interval measure $|1|_M \to 1Nat$ which is entropy equivalent of Nat.

The EF connects the entropy with the process total time, which synchronizes and adjoins the local impulse time measure along the process in an absolute time scale.

The impulse logical measure $1Nat$ includes *logical bit*, which measures $f_B = \ln 2 \cong 0.7Nat$, while the difference in $0.3Nat$ carries a wide of the impulse Yes action transferring $\cong 0.05Nat$ to next impulse.

That evaluates measure $f_l \cong 0.3 - 0.05 = 0.25Nat$ of the impulse *free logic*.

Each of this logic originates from the observing impulse's relational Bayes probabilities reducing the entropy along the observation process and growing the correlation dependencies of the impulses in the observing process.

The reduction of the observable process entropy under the observing probabilistic impulses determines the growing logical dependency of the impulses during the observation, which involves a virtual impulse' correlational attraction.

The free logic, logically connecting the impulses, measures the virtual attraction, the growing correlation, and the increasing a posteriori probability logic of the observing process.

Until this probability less than one, the probabilistic logic, even with growing this probability, is still uncertain (virtual). Closeness this uncertainty to certainty with the probability one measures particular relational probability and its entropy, which finalize the reduction of the uncertainty.

3.3. The emergence of space-time geometry in the impulse observing process

At growing Bayes a posteriori probability along observations, neighbor impulses ↓↑ and ↓↑ may merge, generating interactive jump ↑↓ on the impulses border.

The merge converges action with reaction, superimposing cause and effect and their probabilities.

The merge squeezes to a micro-minimum the inter-action time interval, which initiates a microprocess on the bordered impulses.

The jump increases the Markov process drift (speed) up to infinity which generates a high entropy density curving the correlation between these actions up to its dissolving-cut which virtually cuts the time measure of this correlation in an emerging ½ units of the bordered impulse.

The dissolve correlation measures the orthogonal correlations.

This jump, curving the time up to the related orthogonal correlation, rotates the impulse reaction↑ of ½ unit measure to its orthogonal *displacement* (Fig.1).

The time measured action ↑ of ½ units measure is the orthogonal *displaced to* second ½ unit $\bar{u}_k$ within the impulse, becoming its orthogonal space impulse measure.

This originates the curved space shifts, quantified by the impulse discrete probability' measure $\mathrm{p}[\bar{u}_k]$(1 or 0), which measures *two impulse parts (units)* including a counterpart to the primary curved time, Fig.1.

Thus, the displacement within the impulse $\bar{u}_k$ changes the impulse second ½ time to such discrete space shift which *preserves* its measure $M[\bar{u}_k]=|1/2\times 2|\xrightarrow{\mathrm{p}[\bar{u}_k]}|1|_M$ in the emerging time-space coordinate system.

The measure is conserved in following time-space correlated movement transferring the minimax.

*Beginning the space more precisely starts on* $1/2\pi \cong 0.159155$ part of the impulse invariant measure $\pi$ [33].

The curved impulse is measured by its curvature [34].(Therefore, the birth of space, created by time, provides invariant impulse measure $\pi$ which generates invariant free information that keeps invariant triplet.)

The virtual observer, being displaced from the initial virtual process, possesses the discrete *time-space* impulse which is sent as virtual probe. The observing probabilities define the probe' frequencies which *self-test* the preservation of axiomatic probability measure. (For example, by Bernoulli's formulas).

The minimax observer is *self-supporting* the probes increasing frequencies, which are checking the probability grows.

Such test checks this probability also via symmetry condition (imposed on the axiomatic probability) indicating the probability correctness (by closeness to the axiomatic probability).

A temporary memorized current correlation encloses a difference of the starting and current space-time correlation, identifying the time-space of current location of the virtual observer.

Sequence of Bayes priori-posteriori probabilities transfers their events' causal relationship along observation process in probabilistic logic.

The impulses, cutting entropy, temporary memorize in their correlation the logic as a code of virtual observer.

The rising time-space correlations memorize a time-space shape of the evolving observer.

The evolving shape gradually confines the running rotating movement, which *self-supports* developing both the curved shape and the observer time space geometrical structure.

Such virtual Observer *self-develops* its space-time virtual geometrical structure during virtual observation, which gains its real form with sequential transforming the integrated entropy to equivalent information.

3.4. The impulse transformation of the observing process entropy to information

This transformations emerge during the interactive impulse observation, when a posteriori probability of a last probing No action of this impulse, cutting the final minimal entropy of the observation process, follows the Yes action of that impulse, which brings the posterior probability one, or certainty.

The growing correlation of the observing interactive impulses interaction connects the impulse ending Yes action with the following opposite No action of the next interacting impulse in the virtual attraction.

Interaction of the curving space-time impulses brings opposite topological curvatures: positive for Yes pushing action and negative for No opposite action, which creates topological anti-symmetrical transitive transformation. The interaction $\downarrow\uparrow$ of the impulse opposite curvatures creates topological anti-symmetrical transitive transformation. That transformation brings logical asymmetry which evaluates a part of logical bit entropy $\cong 0.0105Nat$.

Transferring probabilistic logic to certain logic covers a transitive barrier between entropy and information [120].

The opposite curved interaction of observing impulse with an external impulse brings the asymmetrical entropy which could overcome the barrier.

The opposite curving impulses in the interactive transition require entropy ratio 1/ln2 to transfer entropy equivalent to the forming logical bit.

The opposite curved interaction decreases difference in the entropies ratio on relative amount $r_0 = (2\ln 2 - 1)/\ln 2 \cong 0.5573$ (where $2\ln 2$ counts entropies for both impulse potential bits in the transformation, from which deduct the entropy measure I Nat of the transforming impulse).

In process of the impulse interaction, part of the logical impulse $i_1 \cong (1.44 - \ln 2) \times 0.5573 \cong 0.2452bit$ with ratio $r_0$ identifies moment $t_1$ following appearance of potential (previously hidden) logical bit. Since the entropy measure $i_1$ evaluates the free logic in the curved interaction at moment $t_1 = 0.2452/1.44 \cong 0.17$ relative to the impulse invariant measure $|1|_M = 1.44bit = 1Nat$. Interval $\Delta t_f = 0.17Nat$ measures the part of the impulse delivering free logic which creates asymmetrical logic in information bit.

After interval $\Delta t_f$ follows interval $\Delta t_B = \ln 2/1.44 \cong 0.481352Nat$ of appearance the logical bit which indicates a logical certainty of the observing hidden bit. This logical bit of the certain logic appears with the certain free logic, which carries the certain logical attraction. The logical entropy bit becomes information bit through memorizing the anti-symmetrical logic in information $i_{13} \cong 0.0105 \times 1.44 = 0.015 bit$

The certain free logic evaluates $i_f \cong 0.2452 - 0.015 \cong 0.23 bit \cong 1/3 bit$ which appears on time interval $\Delta t_{f1} = 0.23/1.44 \cong 0.1595$ measured relatively to the impulse time interval in bit.

Thus, from begging of the impulse, time interval $t_{\Sigma} = 0.17 + 0.481352 + 0.1595 = 0.811$ appears that is evaluated relatively to the impulse interval measure $|1|_M$. Interval $t_{\Sigma}$ deducts from the external impulse' measure $|1|_M$ the relative time interval $t_{en} = 0.889 \cong 0.19$ which estimates the interval needed for memorizing the logical bit.

The observation *process*, reducing uncertainty of the random interactive process, concurrently integrates it in the maximal growing correlation which holding entropy equivalent of interval $t_{en} = 0.19$ temporary memorizes this logical bit.

The certain logical bit becomes physical information bit through erasure the entropy of this logic, which allows replacing the logic by memorizing its bit. (The logical bit conceals hidden entropy which was carrying energy in initial real interactive process before it had covered by the entropy of the multiple random interactions.). Injecting energy equivalent to the integrated uncertainty-entropy removes this entropy on interval $\Delta t_B$ of creation the asymmetrical logical bit.

Memorizing the asymmetrical logical bit erases the logical bit entropy.

The memory of physical bit must be stored or placed in some physical entity, which performs encoding of the memorized bit by extracting it initial position.

The physical reality of revealing the bit brings also its energy for potential erasing this entropy.

Acquiring certainty-information from the entropy, satisfying the second law in an irreversible process, requires cost of the energy equivalent to that in Demon Maxwell.

According to Landauer principle [38], any logically irreversible manipulation with information, such as encoding leads to erasure the information in a dissipative irreversible process.

Erasure of information Bit requires spending minimal energy $W = k_B \theta \ln 2$ ($k_B$ is Boltzmann constant, $\theta$ absolute temperature) which should be delivered outside by an environment.

Therefore, the transformation of observing impulse entropy to information includes: getting the anti-symmetric logic, its memorizing through erasure, and encoding the memorized bit by storing its position in an environmental process impulse.

This process needs energy for both getting the asymmetrical logical bit, memorizing it, and ending with the encoding bit.

The observing virtual process ending with the minimal entropy is reversible and symmetrical, while both logical bit and physical bit is anti-symmetrical. The transitive opposite curving anti-symmetric interaction curries asymmetrical logic in its correlation with no actual cost.
But, since this interaction reduces the anti-symmetrical entropy, getting the information anti-symmetric logic requires less energy for erasure than that memorizing bit.
Below we estimate the components of this process.
The information logic covers time of encoding $i_{en} = 1.44 - 1.23 = 0.21 bit$, where $0.21 bit \times 1.44 = 0.3 Nat$ is transferred to a next interacting impulse, which is the equivalent to entropy $e_1 = 1 - \ln 2 = 0.3 Nat$.
Both the free information and that needed for encoding carry information logic which we call encoding logic.
Since Landaurer energy allows memorizing only 1 bit, the free information currying the attraction and information cost of encoding are not becoming physical until the external impulse brings additional to Landaurer energy $W = k_B \theta 0.44$.
That adds cost $w_{\Delta 1} \cong 0.63478$ of Landaurer' energy, whereas the encoding costs is $w_{\Delta 2} \cong 0.2736$; total additional cost is $w_{\Delta} \cong 0.9$ at the same temperature.
This energy should be delivered on time interval $t_{\Sigma} - 0.17 = 0.64$ whose start identifies the impulse interacting time transferring entropy $e_1$.
The moment $t_1$ ends the interval of the curved opposite interaction which adds interval~0.02 bringing total interval 0.19.
Hence the moment of delivering total energy identifies a potential encoding time interval ending the internal impulse, which interacts with the external impulse through the opposite curvatures.
Since the external process' movement within its impulse ends at the impulse step-up stopping states, the thermodynamic process delivering this energy should stop in that state.
Hence, the erased impulse entropy memorizes the equivalent physical information $1.44 bit$ in the impulse ending state, where the encoding stores this information.
Such impulse starts producing the physical bit after the moment ending transitive logic and ends producing it by the moment ending delivering Landaurer energy.
This time interval estimates ln2.
The time intervals of memorizing free information and the encoding estimate accordingly $\Delta t_{f1} \cong 0.1597$ and $t_{en} = 0.19$.
The free information can attract a next logical bit during which the first memorized bit is encoding.
The invariant free information, being retrieved from cutting correlation's hidden information, preserves its invariant adaptive attracting properties which generate invariant encoding and decoding.

Such an attracting encoding can automatically produces the curving interaction of the next logical impulse with a next real environmental process since time interval of encoding equals to time interval of transitive curving.

Such interaction starts producing the next physical bit after the moment ending the transitive logic, and ends producing it by the moment ending delivering Landaurer energy.

That time interval also estimates ln2, while the time intervals of memorizing this bit's free information and the encoding estimate the same $\Delta t_{f1} \cong 0.1597$ and $t_{en} = 0.19$.

From that follows necessity of proper concurrence of the time curse of the impulse inner and external impulses and coordination the sequence of the moment appearance bit, its memory, and encoding.

These will allow consecutive integration of the process entropy and its transformation to process information.

3.5. Transformation of the integrated process entropy to the process integral information. Growing information density of the process impulses

The EF functional integrates the sequential Bayes probabilities of the observing impulses, which virtually observe and cut the observable process. The EF measures total process entropy as the process potential-prognosis integral information. The minimax impulses process the observing process' time trajectories with emerging space-time trajectories. The VP for the EF analytically describes the minimax time-space extreme trajectory of the observing process. After the impulses of the EF integrated multi-dimensional process reach certainty, the impulses cut with maximal probabilities transforming each impulse entropy invariant measure to the equivalent information invariant measure. These multiple information impulses integrate information path functional (IPF) on a path through the discrete impulses' time–space information measures.

The observing process probabilities approaching to its certainty measure automatically transform the EF virtual observing probabilistic multi-trajectories to the IPF information trajectories.

Each curved impulse invariant time-space measure $\pi$ encloses information measure $1Nat$ which includes bit, free information, and information needed for encoding.

This free information between a nearest impulse bits attracts them with intensity~1/3 bit per impulse.

The IPF integrates the bits with free information connecting the bits sequence.

The minimax principle, applied along trajectory of the IPF information process, maximizes information enclosed in each current impulse, squeezes its time interval, while its growing attracting free information minimizes the time interval between the nearest impulses proportionally to 1/3bit. For each third impulse that interval of information distance becomes proportional to 1 bit, which is the information of invariant impulse with time-space measure $\pi$.

Hence, each invariant impulse along the IPF extreme trajectory squeezes proceeding impulses distance to $h_i = \pi$.

The IPF extreme trajectory sequentially condenses the increasing information of each third following impulse that grows the information density of the invariant impulse. Free information of such impulse increases that intensifies information attraction between the invariant impulses in above proportion. By the end of the IPF integration, all integrated information is concentrated in a last impulse, whose information density approach maximal limit.

Since free information encloses information logic, the multiple bits with triple growing density raise the process information logic.

The IPF integral information with its logic is condensed in the last integrated impulse time –space interval volume.

For multiple information impulses, each third curved impulse having invariant measure $\pi$ appears in the information process with time frequency $f_i = k_i, k_i = 3,5,7,9,...$.

Information time-space density $D_i^I = k_i Nat / v_i^s$, concentrating $k_i$ Nat for each third impulse, increases with growing $k_i$ while the time-space volume $v_i^S$ holds invariant measure. The invariant impulse time coordinate $\tau_i = \pi / \sqrt{2}$ , the flat surface space coordinate measure $l_i = \sqrt{2}$ and orthogonal to them space coordinate space coordinate measure $h_i = \pi$ determine impulse volume $v_i^S = \tau_i \times l_i \times h_i = \pi^2$ .That determines $D_i^I = k_i Nat / \pi^2$ .

So, the current impulse time–space geometry encloses information, density and frequency, concentrating information logic and information of all previous impulses along the path.

The EFextreme trajectories, starting from the multi-dimensional observing process, the EF-IPF transformation converts to the multi-dimensional orthogonal processes whose curved impulses hold the above information measures. The EF-IPF space-time extremal trajectories rotates forming spirals located on conic surfaces Fig.3, which starts from virtual (entropy) process and continues as the information process.

Since each bit of this trajectory creates the cutting entropy in the impulse observation, the trajectory consists of segments of information process dynamics and the between segments intervals delivering each following bit to the segment. On Fig.3 each segment starts on the cone vertex-point D and ends on the point D4 which connects to a vertex of the following cone. The observing bit is delivering at each cone vertex.

The segment includes the impulse process with its logical bit, intervals of free logic and correlation connecting the nearest segment which temporary memorizes the segment logic.

The logical and information dynamics describes the process of sequential logical interaction of such multiple impulses, rotating with information speed determined by the impulse density. The dynamics between the cone vertexes is reversible and symmetrical analogous to Hamiltonian dynamics.

The logical anti-symmetry brings the anti-symmetrical logical bit prior to interaction with the external impulse which starts delivering the external energy. This bit is supplying at each cone vertex.

After the external energy generates physical multiple bits, the physical information process starts. Starting this process follows the time interval $\Delta t_f \cong 0.17$ of the logical anti-symmetrical interaction.

Since each impulse' curved measure $\pi$ with its relative time interval $\Delta t_f$ appears in the information process with frequency $f_{10} = \pi$ , the frequency of appearance this interval is $f_1 = 0.17 / \pi \times \pi = 0.17$, which equals to frequency of encoding time interval $t_{en}$. The frequency of spectrum $\omega_1 = 2\pi f_1 = 1.068$ identifies the time of opening supply an external energy, equal to the spectrum frequency of the encoding time interval.
Time interval of memorizing the bit $\Delta t_B$ identifies the bit information measure $\ln 2$ which is equivalent of invariant impulse part $\Delta t_B = \ln 2 / 1.44$ or frequency of appearance that interval $f_B$ within the impulse. It determines frequency $\omega_2 = 2\pi f_B = 2\pi \ln 2 / 1.44 = 3.02 < \pi$ of spectrum $\{\omega_1, \omega_2, \omega_1\} = \omega_o, \omega_o = (1.068, 3.0.2, 1.068)$ necessary for delivering energy memorizing and encoding the bit. Or these frequencies present spectrum $\omega_o = \{1, 28277, 1\} \times 1.068$. Along the information process, this spectrum frequency is currying proportional to frequency $f_i = k_i$.
After supplying the external energy during these time intervals, whose sum equals to the invariant impulse time interval, the whole impulse becomes the segment of physical information process.
Therefore, physical dynamics describe the IPF extremal trajectory rotating on sequential cones (Fig.3). Each cone vertex encodes the bit memorized in previous impulse-segement with frequency $\omega_1$, each current segment repeats with frequency $\omega_2$, and transfers to next cone its vertex with frequency $\omega_1$ of encoding the current impulse bit.

Hence, each physical information impulse carries spectrum $\{\omega_1 \omega_2 \omega_1\} = \omega_o$, while their sequential pair on the trajectory carries spectrum impulses $\{\omega_1, \omega_2, [\omega_1 = \omega_1], \omega_2, [\omega_1 = ...]\} = \omega_\Sigma$ .where $[\omega_1 = \omega_1]$ is the resonance frequency for two impulses whose distance is shortening on 1/3.
That allows closely connecting the impulses in the resonance. Along the trajectory, each of these pairs appears with the growing frequency of the impulses appearance $f_{io} = 1/3k_i$.

Since these fixed time intervals $\Delta t_f, \Delta t_B, t_{en}$ are relative to the invariant impulse measure, they are repeating for each invariant impulse with the increasing information density and with growing frequency. Thus, along the extreme trajectory, each third impulse will deliver triple $\{\omega_1, \omega_2, [\omega_1 = \omega_1]_{\Delta t_{10}}, \omega_2, [\omega_1 = \omega_1]_{\Delta t_{20}}, \omega_2, [\omega_1 = \omega_1]_{\Delta t_{30}}, \} = \omega_{\Sigma 10}$ with time intervals $| \Delta t_{10}, \Delta t_{20}, \Delta t_{30} |$. These time intervals, being sequentially proportional to the impulse distance measuring in 1/3 bit proportion: $1/k_i$ to the invariant time measure of the impulse, are shortening accordingly $\Delta t_{10} \sim 1/3 \sim \pi / 3\sqrt{2}, \Delta t_{20} \sim 1/5 \sim \pi / 5\sqrt{2}, \Delta t_{30} \sim 1/7 \sim \pi / 7\sqrt{2}$ are.
It sequentially shortens the distance between the impulses on the extreme trajectory assembling each such three impulses in a triple of the resonance frequencies.

### 3.6. Self-forming triplet logical structures and their self-cooperation in information network (IN) hierarchical logic

In the multi-dimensional observing process, minimum of three logical bits with free logics can appear, which, attracting each other, would cooperate in a logical triple.

Multiple probabilities of interacting impulses in this multi-dimensional process produce the numerous frequencies. Some of those, minimum of three, can generate the attractive resonance cooperating the triple.

This triple logic starts temporary memorizing two sequential pair cross-correlations during in their time of correlation. That memorizes the asymmetrical logic in a locally asymmetric cross-correlation during the time of this process. When this process is ending, the triple correlations temporary memorize the triple logical bits.

According to [91] the minimal entropy of cross correlation ln2 can be memorized at cost of the equivalent minimal energy of logical bit.

This is information cost of memorizing the triple logical bit which includes additional free logic.

The attracting free logic of the emerging three logical bits starts the bits self-cooperation in the following sequence.

The free logic of the emerging logical bit holding frequency $\omega_2$ attracts next logical bits of toward a resonance with the equal frequency of next bit's free logic, assembling the two in joint resonance.

This resonance process links these bits in duplets.

The free logic from one bit out of the pair gets spent on the binding of the duplet. The free logic from the duplet' bit attracts the third bit and binds all three in a knot bit creating the triplet logical structure.

The knot bit still has free information and it is used to attract a different bound pair of emerging bits, creating two bound triplets. This process continues creating nested layers of bound triplets, three triplets and more (Figs. 4-6).

Hence the triplet logical structure creates the resonance frequencies of the attracting logic joining the triple bits.

The free logic attraction toward the triple resonance of their equal frequencies is core information mechanism structuring an elementary triplet.

The trajectory of forming triplet describes the rotating segments of their cones (Fig.5), whose vertexes join the knot starting the base of the following cone. The knot frequency joins the cone vertexes in resonance along the cone base when the next spiral segment starts. It connects next triplet in the resonance and so on, creating the nested layers of logical space-time information network (IN), where the layers' knots hierarchy identifies the nested nodes of the IN hierarchy.

Triplets are the basic elements that form a nested informational time-space network with a hierarchical structure.

Each triplet unit generates three symbols from three segments of information dynamics and one when the segment attracting triple logic is binding in the logical triplet knot. These symbols can produce triplet code, while the knot logic symbol binds the triple code for a potential encoding all triple. The knot encoding will release its free information logic which transfers this triple code to next triplet node.

Thus, the nodes logically organize themselves in IN code.

The attracting free information in resonance of three bits frequencies creates the triplet information logical structure which carries the unbound free information logic including the encoding logic with related frequencies. The frequency of free information logic determines the moment of time when the external energy starts memorizing the bit. The frequency of the triple encoding logic determines the moment of time when the external energy starts encoding the knot triple code.

The IN emerging logical structure carries the triple code on each node space time-hierarchy, and the last triplet in the network collects and encloses the entire network's information.

The network, built through the resonance, has limited stability and therefore each IN encloses a finite structure.

That's why the observing process self-builds multiple limited INs through free information of its ending nods.

The final triplet in every network contains the maximum amount of free information.

Because of this, the networks are self-connected through the attraction of their *ended triplets.*

Even after each IN potentially loses stability evolving in a chaos, it possesses ability of self-restoration[33].

The multiple INs self-cooperate in hierarchical domain, starting with the cooperation of each tree ended triplets' free information in a knot which joining this INs' triples in resonance. This IN ending knot's free information resonates with other three INs' ending free information,, forming triplet structure analogously to the core triplet. This high level triplet joins these three INs structuring in a next IN of the domain hierarchy. The hierarchical logical trajectory describes the space-time spiral structure (Figs.7,9), which also presents trajectory of information process evolving in observations. This hierarchy enables generating sequential triple code locating on the rotating trajectory of the cone vertexes, which are distributed at the different hierarchical levels of the multiple IN and the domain hierarchy. Such space-time code integrates the observing process in space-time information geometry of self-organizing observer.

### 3.7. Self-forming hierarchical distributed logical structure of cognition. The observer wave function

The multiple moving INs, sequentially equalizing the speeds-frequencies of the nodes attracting information logic in resonance, assemble total observer logic.

This logic consists of the mutual attracting free information, which, sequentially interacting, self-organizes the cooperative logical rotating spiral loops enclosing all observing information.

We call it observer cognitive logic, which encloses both probabilistic and information causalities distributed along all observer hierarchy. The logical functions of the self-equalizing free information in the resonance perform the cognitive functions, which are distributed along hierarchy of assembling units: triplets, IN nested nodes, and the IN ending nodes. These local functions self-organize the observer cognition.

Assembling runs the distributed resonance frequencies spreading along this hierarchy.

Since each unit, ending high level structure encloses all levels information logic, the unit' impulse invariant time-space interval, containing this information, concentrates more information density than the unit of lower level hierarchy.

The attracting resonance frequencies of the attracting free information hold the cognitive logic loop, which includes local loops structuring local information units and self-creating the unit hierarchy.

Thus, the frequency delivering spectrum $\{\omega_1, \omega_2, [\omega_1 = \omega_1]_{\Delta t_{10}}, \omega_2, [\omega_1 = \omega_1]_{\Delta t_{20}}, \omega_2, [\omega_1 = \omega_1]_{\Delta t_{30}}, \} = \omega_{\Sigma 10}$ is increasing growing in the triple sequentially shortening intervals $|\Delta t_{10}, \Delta t_{20}, \Delta t_{30}|$ for each $i$ trajectory segment.

More details are in propositions which proved at the following conditions.

The space-time spiral trajectory of the EF extremal (Fig.3) describes sequence of multi-dimensional curving rotating segments, representing interacting impulses of the observing process, which integrates the observing process' logic.

Each segment' impulse has invariant entropy measure $1Nat$ moving along the trajectory and rotating the curved impulse invariant measure $\pi$, which includes time coordinate measure $\tau_i = \pi / \sqrt{2}$, the flat surface space coordinate measure $l_i = \sqrt{2}$ and the orthogonal to them space coordinate space coordinate measure $h_i = \pi$; measure $1Nat$ includes the impulse logical bit $\ln 2$ and free logic $f_{li} = 1 - \ln 2 \cong 0.3Nat$, which holds asymmetric logic of such segment. The logic density per each third segment increases according to $D_i^I = k_i Nat / v_i^s$, where $v_i^S = \tau_i \times l_i \times h_i = \pi^2, k_i = 3,5,7,9$.

The asymmetric logic divides the sequential segments by barriers, which transfer the between segments anti-symmetrical interaction with logic interval $\Delta t_f$ following interval $\Delta t_B$ of memorizing bit and interval $t_{en}$ of the free information ending the segment.

Along the space-time trajectory, each sequential segment repeats this triple with invariant frequency spectrum $\{\omega_1, \omega_2, \omega_1\} = \omega_o, \omega_o \cong (1.068, 3.0.2, 1.068)$.

Ratio of alternating bridge-middle part-bridge sequences along the segments on the trajectory identify the frequencies of this spectrum.

*Proposition.*

Along each $i$-dimensional segment rotates *three dimensional space wave functions*, spinning like a top (Fig.A), with rotating speed around each spiral cross-section $\alpha_i^s = 1[square / radian]$,or $\alpha_i^{s_o} = \pi / radian$, and orthogonal to this rotation space speed $\alpha_i^h = 1[volume / radian]$,or $\alpha_i^{h_o} = \pi / radian$. The related frequencies of the orthogonal rotations are $\omega_i^s = \alpha_i^s / 2\pi, \omega_i^{s_o} = 1/2$ and $\omega_i^h = \alpha_i^h / 2\pi, \omega_i^{h_o} = 1/2$ accordingly.

Each $i$-dimensional segment cross-sectional rotation spreads the space rotation on space interval $\pi$ of the segments invariant measure.

The three-dimensional wave function distributes the space rotation along the segments trajectory with the above invariant speeds delivering the invariant spectrum $\{\omega_1, \omega_2, \omega_1\} = \omega_o, \omega_o \cong (1.068, 3.0.2, 1.068)$.

These spectrum frequencies identifies the alternating ratio of bridge-middle part-bridge along each sequential segment.

Finally the periodical wave function includes the sequence of repeating arguments along both orthogonal rotations:

$u_{sh} = u_s \times u_h, f_{ws} = \{f_{iws}\}, f_{wh}\{f_{iws}\}, i = 1,...n$, which performs the multiple three-dimensional movement with *three dimensional space wave functions* the like a top (Fig.A) .

The shape of the multiple wave functions describes the extreme multi-dimensional trajectory formalizing the minimax observation process which models rotating segments on cones (Fig.3). •

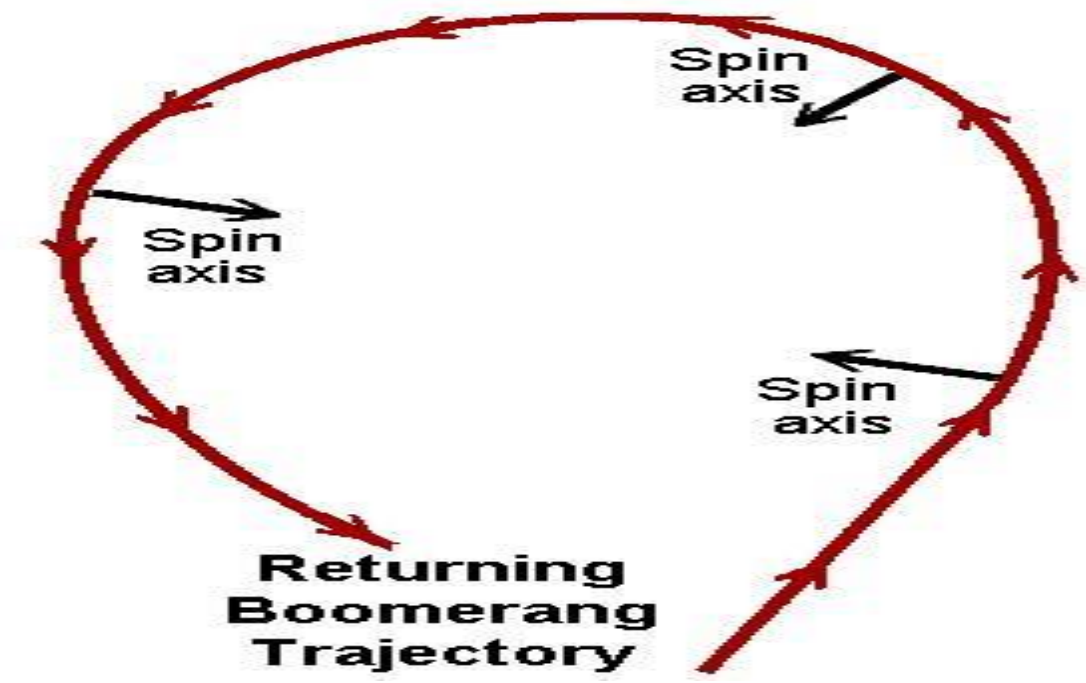

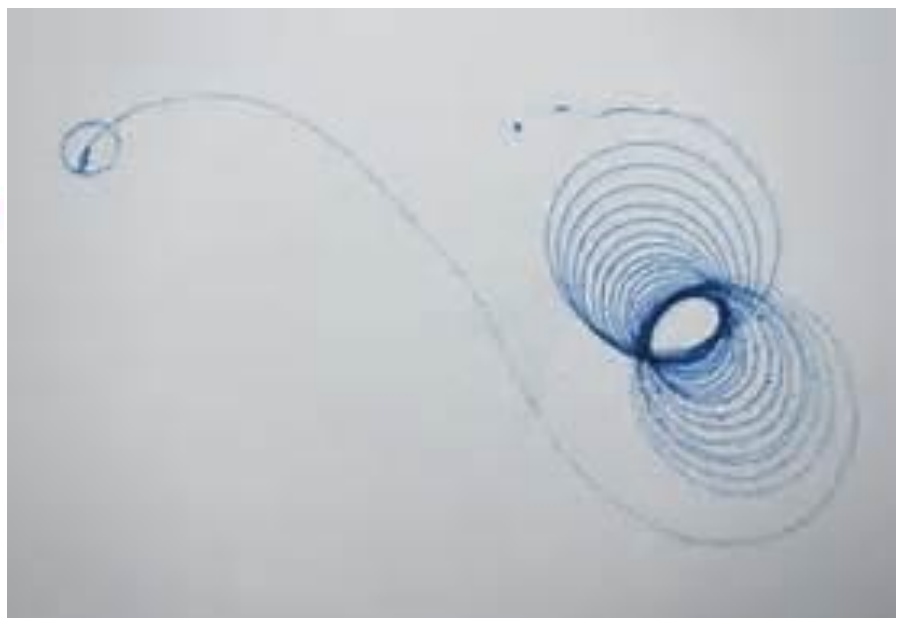

**Fig.A. Illustrative schematic of spinning top (from Google' top trajectory), where schematic bellow illustrates a rotating segment with bridge on the EF-IPF trajectory distributing the wave space –time movement.**

The wave function frequencies and properties

1).The wave function with above speeds and frequencies emerges during the observation process when a space interval appears within the impulse microprocess during reversible time interval of $\varepsilon_{ok} = 0.015625$ of the impulse invariant measure $\pi$ equivalent to $1Nat$. Before that, the observing trajectory has described the probabilistic time function whose probability $P_{\Delta}^{*} \cong 0.821214$ indicates the appearance of a space-time probabilistic wave.

During the probabilistic time observation, entropy of the Bayes priori-posteriori probabilities measures probabilistic symmetric logic of a sequence of these probabilities.

Thus, wave function starts emerging in probabilistic observation as a probability wave in probability field.

At beginning of the microprocess, the probabilistic wave measures only time of its propagation.

2). The asymmetrical logic emerges with appearance of free logic interval $\Delta t_{f1} \cong 0.1597 \cong 1/2\pi$ which, repeating with equal wave frequency $\omega_i^s$ indicates beginning of the interactive rotating asymmetry on a segment' bridge. From that, the observation logic on the trajectory becomes the asymmetric part of total free logic $f_{li} = 1 - \ln 2 \cong 0.3Nat$.

The asymmetric *logical wave* emerges. The asymmetric logic probability approaching $p_{\pm a} = \exp(-2h_{\alpha}^{o*1}) \cong 0.9866617771$ appears with certainty-reality of observing the previously hidden asymmetrical bit. Such logic temporary memorizes correlation with the probability which carries a logical bit of the certain logic. Such certain logical bit may carry energy in real interactive process, which had covered by the entropy of the multiple random interactions of the Markov process.

The path to creation of the certain bit includes an increment of the probability $\Delta P_{ie} = 0.981699525437 - 0.9855507502 = -0.1118$ starting injection of energy from an interacting impulse of the Markov process. Thus, the certain free impulse logic carries the certain logical attraction.

The wave function in the microprocess is probabilistic until the certain logical information bit appears.

The certain asymmetrical logical bit become physical bit through erasure the entropy of this logic, which allows replacing the logic by memorizing its bit.

3). The wave function starts on the observation process which the EF extreme trajectory prognosis, carrying the probabilistic wave transforming the observation process to certainty of real observation. The spinning movement of the space –time trajectory describes the invariant speed around the cross-section of its rotating impulses-segments, which spreads the invariant rotation space speed along the segments trajectory. The segments' invariant spectrum $\{\omega_1, \omega_2, [\omega_1 = \omega_1]_{\Delta t_{10}}, \omega_2, [\omega_1 = \omega_1]_{\Delta t_{20}}, \omega_2, [\omega_1 = \omega_1]_{\Delta t_{30}}, \} = \omega_{\Sigma 10}$ repeats the triple frequencies of these three time intervals between them. That shortens distance of the equal spectrum frequencies and assembles them in resonance creating joint logical structures-triplets units up to the IN hierarchies and domains. The frequency absolute maximum indicates a finite end of its creation. A minimal energy of the resonance supports the forming logical loop. •

3.7.1. Distribution of the space-time hierarchy

1).The hierarchy of self-cooperating triplet's units distributes the space rotation emerging along the EF segments of the time-space extreme trajectory, where each third impulse progressively increases the information density measure of its bit in triple. The time-space hierarchy of the units starts with emerging in observation the symmetrical logic at appearance of space interval in the microprocess. This logic self–forms a hierarchy of the logical unit structures through the impulse' mutual attracting free logic which, sequentially attracting the moving unit' speeds, equalizes their frequencies in resonance that assembles the observer logic along the units hierarchy.

2).The hierarchy of the logical cooperating units becomes asymmetrical with appearance of certain logical bit on the extreme trajectory. The repeating free logic interval indicates the wave frequency $\omega_1 = f_i^s = 1/2\pi$.

The EF rotating trajectory of three segments equalizes their information speeds joining in the resonance frequency during the space rotation, which cooperates each third logical bit's segment on the trajectory and logically composes each triplet structure in the unit space hierarchy.

3).The appearance of the asymmetrical logical bit on the extreme trajectory indicates entrance the IPF information measure on its path to forming logical bit-ln2. The path starts on relative time interval $\Delta t_f = 0.23/1.44 \cong 0.1597$ of the logical asymmetry, which identifies the segment bridge. During the triple impulses, the third time interval $\Delta t_{3r} = 3\Delta t_{1r} = 3/2\pi \cong 0.4775152$ indicates the end of the triple cooperative logic, starting build the triplet knot. Forming the triplet knot needs a time interval during which the triple free logic binds in the triplet bit. The time interval of creating the bit approaches $\Delta t_B = \ln 2/1.44 \cong 0.481352$. Difference $\Delta t_B - 3\Delta t_{3r} \cong 0.004$ evaluates the time of binding the triplet. Thus, the wave space interval delivers the logical bit with wave spectrum frequency $\omega_2 = 2\pi\Delta t_B = 2\pi \ln 2/1.44 = 3.02 < \pi$, while the triplet knot repeats with spectrum frequency $\omega_{20} = 2\pi 3/2\pi = 3$.

4). Delivering external energy for memorizing the logical bit identifies relative moment $t_1 = 0.2452/1.44 \cong 0.17$ ending the interval of the asymmetry. The resonance frequencies of asymmetrical logic by this moment have already created. Along the IPF path on the trajectory, this moment follows interval $\Delta t_B$ of creation logical bit, ending the emergence of the knot that binds the free logic. The interval of memorizing physical bit requires the same interval $\Delta t_B$ during which the entropy of logical bit erases. The needed external impulse, erasing asymmetric logical bit, starts with interval $\Delta t_B$ and ends with interval of encoding the bit $t_{en} = 0.19$. The external energy, supplied on time interval $t_{\Sigma b} = 0.481352 + 0.19 = 0.671352$, includes both erasure the logical bit and its encoding. Whereas interval of information free logic $\Delta t_f = 0.23/1.44 \cong 0.1597$ is left for attracting a new bit, adding interval~0.02 of the opposite asymmetrical interaction. It brings total $t_{\Sigma bo} = 0.671352 + 0.02 = 0.691352 \cong \ln 2$ for the interval of the external bit. When this bit' free information starts encoding on interval $t_{en} = 0.19$, interval 0.02 has been already spent. That re-identifies interval of this encoding $t_{eno} = 0.17$ -the same as the external impulse interval of asymmetry.

Therefore, the frequency spectrum, initiating the encoding, equals $\omega_1$ in sequence $\{\omega_1\omega_2\omega_1\}$. This triple sequence identifies the segments alternating on the trajectory with the repeating ratio of bridge-middle part-starting next bridge. The ratio holds on the bridge relative interval $t_{en} = 0.19 - 0.02 = 0.17$. Thus, the sequence of segments on the EF-IPF extreme trajectory carries its wave function' frequencies which self-structure the space-time unit of logical bits' hierarchy that self-assembles the observer logic. The logic controls memorizing and encoding physical bits and the hierarchical structure of these units space-time information geometry.

5).The segments-impulses on the EF spiral trajectory sequentially interact through the frequencies repeating on the bridges time–space locations, which connect the segments in trajectory. The segments' sequence on the EF-IPF extreme trajectory (Figs.3, 4) carries its wave function' frequencies which self-structure of the unit logical bit hierarchy that self-assembles total observer logic. This logic controls the memorizing and encoding both physical bits and their units' hierarchical structure. •

The observer cognitive logic, which encloses both, probabilistic and information causalities distributed along all observer hierarchy. The logical functions of the self-equalizing free information in the resonance perform the cognitive functions, which are distributed along hierarchy of assembling units: triplets, IN nested nodes, and the IN ending nodes.

These local functions self-organize the observer cognition. Assembling runs the resonance frequencies $[\omega_1 = \omega_1]$ spreading along this hierarchy. Since each unit, ending high level structure encloses all its information logic, the high unit' impulse invariant time-space interval, containing this information, increases more information density than the unit of lower level hierarchy.

The resonance frequencies of spectrum $\{\omega_1, \omega_2, \omega_1\} = \omega_o$ holding the cognitive logic loop-self creates the unit hierarchy.

3.7.2. The observer logical structure

1). The observer logical structure self–forms the attracting free information, which self –organizes hierarchy of the logical triplet units assembling in resonance frequencies. Each triplet logical structure models Borromean ring consisting of three topological circles linking by Brunnian link-loop. The spinning top space trajectory, FigA, as well as EF-IPF trajectory, includes a Borromean rings' chain modeling a distributed hierarchical logic.

2). The observer logical structure carries the wave along the trajectory' segments , where each third segment delivers triple logic of information spectrum $\{\omega_1,\omega_2,[\omega_1=\omega_1]_{\Delta t_{10}},\omega_2,[\omega_1=\omega_1]_{\Delta t_{20}},\omega_2,[\omega_1=\omega_1]_{\Delta t_{30}},\}=\omega_{\Sigma 10}$ with sequentially shortening intervals $|\Delta t_{10},\Delta t_{20},\Delta t_{30}|$,at $\Delta t_{io}/k_{io},k_{io}=3,5,7,9$ and the increasing segment's information density.

Two sequential segments synchronize resonance frequencies $[\omega_1=\omega_1]_{\Delta t_{10}}$ and $[\omega_1=\omega_1]_{\Delta t_{20}}$ while the triplet synchronizes resonance frequency $[\omega_1=\omega_1]_{\Delta t_{30}}$. This triple logic holds one bit in each observer's triplet logical structure unit.

The sequential triplets' attracting free logic conveys that resonance spectrum with progressively shortening time intervals and growing their frequencies, which cooperate the logical units in IN nested hierarchy. The needed spectrum with the increasing frequencies automatically carries each consecutive segment along the EF-IPF trajectory.

The emanating wave function delivers the frequencies cooperating a growing hierarchy of the logical units.

The self-built hierarchy of the logical structures self-integrates the observed logic which the structure encloses.

3). The hierarchy of distributed logical loops self-connects logical chain. The logical chain wide determines the invariant impulse' relative interval $0.17$ enclosing the assembled logical code. The growing density of consecutive impulses along the trajectory sequentially squeezes absolute value of this interval whose ratio preserves the invariant impulse.

The absolute time-space sizes of the logical chain are squeezing through the multi-level distributed hierarchy.

4).The logical chain rotation, carrying the frequencies of synchronized spectrum, requires a minimal energy to support the chain. This energy is equivalent to the logical bits code. The integrated chain logic holds this code.

The observer logic chain synchronizes the triple rhythms along the EF-IPF trajectory in a melody. Or the hierarchical space-time chain harmonizes the melody of the rhythms. $\bullet$

Therefore, the wave function frequencies, initiating self-forming the observer cognition, emerges along the EF-IPF extreme trajectory in form of probabilistic time wave in probability field.

The probabilistic impulse observation starts the microprocess, where entangled space rotation code develops the rotating space-time probability wave.

The emerging opposite asymmetrical topological interaction shapes the space–time wave function, becoming certain, as well as the observer' cognitive logic.

*These results conclusively and numerically determine the structure and functions of cognition.*

3.8. The multi-level self-encoding the hierarchical cognitive logic in intelligence code enclosing the Observer information geometrical structure

The self-organized hierarchy of distributed logical loops along the chain self-assembles multiple logical hierarchical units.

The resonance frequencies, self-organizing the cognitive logic of the units, provide interactive actions attracting the impulses with an external energy. The attracting actions carry free logic of the assembling logical unit, which opens a switch-interaction with an external impulse carrying the Landauer's energy that starts erasing entropy and memorizing the information in its bit . The bit free information is encoding the memorized bit.

Each memorized bits encodes its triplet space-time information structures cooperating in the units of INs hierarchy.

The multiple units' bits hierarchy, integrating the local units' codes with the encoded triplet structures up to the highest level, self-organizes the observer space–time information geometry which structures the double spiral triplet code (DSS).

The DSS integrates the observing process minimax impulses, the triplet codes' optimal structure' units, which enclose the structured units' bits energy.

The DSS physical rotating helix structure, spinning physical wave function frequencies, self-organizing the multiple local bits coding units, encodes the observer triplet coding structure which we call *observer intelligence.*

The logical switching of the free information at all hierarchical level performs the *intelligence functions*, which generate each local code. These functions are distributed hierarchically along the assembling logical units of the cognitive chain.

The DSS encodes the triplet dynamics in information macrodynamic process which implements the observer encoding logic. The cognitive movement, beginning in virtual observation, holds its imaginary form, composing entropy microprocess, until the memorized IN bit transfers it to an information macromovement. That brings *two forms* for the cognitive helix process: *imaginary reversible with a temporal memory, and real-information moving by the irreversible thermodynamics memorizing incoming information.*

These basics we detail below.

The memorizing and encoding physical information structure of the unit space-time hierarchy

1).To memorize and encode the unit logical hierarchy in physical information structure, the observer logic provides the sequence of frequencies $\omega_{\Sigma}$ and $\omega_{\Sigma 1o}$ along the chain according to growing the triple density for each current trajectory impulse. That identifies the sequence of the moments for entrance external information with energy for both memorizing the units hierarchy and its physical encoding.

This sequence of frequencies is the same that hold the observer logic which delivers the observer logic minimal energy.

The attracting free information of the memorized bit encodes the physical bits, connecting them, first, in the triplets, second, in the INs nested nodes, and then, in each IN ending triplet code.

Therefore, along the observing trajectory emerges the triplet code through the sequence of segments whose frequency brings logical triple following its physical encoding. The cognitive process at each triplet level preempts the memorizing.

2).The spinning three-dimensional space wave functions' frequencies, synchronizing the segments, synchronize the triplet code in rotating spiral structure.

The code, which the switching time clock synchronizes has rhythmical sequence of time intervals (windows) where each observer logical structural unit gets the needed external energy. The clock time course assigns the frequency through the repeating time intervals, which determine each local resonance frequency of assembling the structural unit. These frequencies-local rhythms identify the moments of ending interval of the free information at each unit level, or the interacting cognitive and intelligence local actions.

3).The code' free information holds frequency identifying moments of physical encoding the code multiple bits.

This frequency is following from the integral cognitive logic chain which connects all local cognitive loops identifying the moments off of physical encoding each of the unit bits code. The final frequency of encoding high level physical bit brings free information that integrates energy of free information from units of local code bits. The high level coding bit, by the moment of encoding gets all physical energy necessary for the encoding.

Thus each following unit frequency open widow for integral encoding all previous units' bits.

Therefore, cognitive logic logically encodes intelligence of this logic.

Each stable observer conserves its widows according to the variation law regularities.

4). Each bit, memorized in the conjugated interactive bridge (Fig.4, left), divides the trajectory on reversible process section, excluding the bit bridge, and the irreversible bridge between the reversible sections, on the triplet knot, located on the cone vertex. The information logical dynamics memorize as *information physical dynamics* in the double spiral structure (DSS). The observing process' impulse trajectory is realized as information dynamics which compose the triplet dynamics. The multiple triplet's build the DSS which memorizes the information dynamics. The EF-IPF optimal trajectory predicts the information dynamics and the observer optimal DSS, Fig.8, which encloses the predicting code hierarchy. The observer logical structure self-connects the local codes in the observer triplet DSS code, which encodes all these structures in the space-time information structure of information observer. The observer triplet code memorizes the observer cooperative information structure and enhances multiple rhythms of the local structural units. The DSS coding structure memorizes total collected observer information quantity and quality, which determines the observer cooperative complexity. This coding structure, which self-organizes all assembled information, integrates function of cognition and intelligences. The EF-IPF observing process and information dynamics artificially design the DSS.

5). Multiple observations build numerous of such DSS space-time structures, which integrate in the information geometrical structure of Information Observer (Fig.10).

6).The quality of information, memorized in an ended triplet of the observer hierarchical informational networks and domains, measures level of the observer intelligence. Maximal level of the emerging intelligence measures maximal cooperative complexity, which enfolds maximal number of the IN nested structures, memorized in the ending node of the highest IN. Number of level limits the wave function's minimal space speed imposing the information limitation [34].

7).All information observers have different levels of intelligence which classify the observer by these levels.

8).The multiple levels of the observer interacting logical and intelligent functions develop self-programming and computation which enhance collective logic, knowledge, and organization of diverse intelligent observers. The intelligent actions and the intelligence of different observers connect their level of knowledge, build and organizes the observers IN's information space-time logical structure. The highest level INs enfolds growing information density that expands the intelligence, which concurrently memorizes and transmits itself over the time course in an observing time scale.

9).The intelligence, growing with its time-space region, increases the observer life span, which limits a memory of the multiple final IN ending node in the extended region.

10).Since the multiple IN information is *limited*, as well as a total time of the IN existence, the IN self-replication arises, which enhances the collective's intelligence, extends and develops them, expanding the intellect's growth.

11).The self-organized trajectory with the wave functions, evolving IN's time-space distributed information structure, models *artificial intellect.*

12).The invariance of information minimax law for any *information observer* preserves their common regularities of accepting, proceeding information and building its information structure.

That guarantees objectivity (identity) of basic observer's individual actions with *common information mechanisms.*

The common mechanism enables creation of *specific* information structures for each particular observed information, with individual goal, preferences, energy, material carriers, and various implementations.

13).Multiple communications of numerous observers (by sending a message-demand, as quality messenger (qmess) [30, 33], enfolding the sender IN's cooperative force, which requires access to other IN observers allowing the observer to increase the IN personal intelligence level and generate a collective IN's logic of the multiple observers. This not only enhances the collective's intelligence but also extends and develops them, expanding the intellect's growth.

14).The artificial designed DSS information measures total IQ of this observer. Each particular observer DSS encodes its IQ. The difference of these IQs measures a distinctness of their intelligence.

The maximal information, obtained in the observation, allows designating the maximal achievable IQ measures of optimal AI observer's DSS code designed by the EF-IPF' VP. The space-time information structure, encoding the EF-IPF integral observed information (Fig.10), analytically designs the AI information observer.

The observing information of a particular observer is limited by the considered constrains of this observation.

The constrains limit conversion of observing process in the information process.

The thresholds between the evolving stages of the observation limit the stages' evolution.

The integral cognitive information and the following intellective actions limit the amount of free information reducing ability of making intelligent IN's connections. $\bullet$.

We believe the observer current mind is integrated information of causal logic distributed along the observer hierarchical levels, which is the integral of cognitive logic.

The memorized mind integrates the physical DSS codes.

That integral according to the VP enables prognosis new observation process which creates new logic and extended code intelligence that renews multiple observations developing evolving observer with its regularities and individuality.

Finally, the intelligent observer has two main attributes: multi-levels cognition and intelligence.

3.9. How the interacting intelligence observers can understand meaning in each communication

When an intelligent observer sends a message, containing its information, which emanates from this intelligent observer's IN node, another intelligent observer, receiving that information, enables recognize its meaning if its information is equivalent to this observer IN nodes information quality satisfying coherence its cognitive logic.

Since the DSS code is invariant for all information observers, each observer encodes its message in that coding language, whose logic and length depend on sending information, possibly collected from the observer–sender's different INs nodes.

The observer's request for growing quality of needed information measures the specific qualities of free information emanating from the IN distinctive nodes that need the compensation.

The observer request initiates recognition of the needed information which includes understanding of its meaning.

That process comprises the following steps.

1). The IN node is requesting the needed quality by the Bayes high posteriori probability correlation (closed to certainty) which memorizes the message logical information making its temporary copy.

2). Copying logical information builds a temporary logical IN. The number of the IN nodes triplets enable adjoining and cohere in resonance is automatically *constraining.*

3).The forming temporary resonance structure-as a temporal cognition, initiates the requested IN nodes' high level probabilistic free logic, which allows involving the incoming copy in the observer cognitive logic. The copies mirror the transitive impulses providing asymmetrical free logic with $\Delta t_f$ intervals.

4).Each logic interval allows access an external impulse' interval $\Delta t_B$, which, erasing the copy temporal logic, reveals its information bit by starting process of memorizing bit and its decoding.

The decoding of this memorized bit holds interval $t_{en}$.

5).The coherence of observer cognitive logic actually allows starting the decoding from a low level of the observer hierarchical structure if such structure needs updating information using its part of the frequency spectrum. The message information delivers the wave function frequencies along the observer space –time hierarchy.

6).The decoding finalizes the requested IN nodes whose acceptance of the message comparative qualities indicates its ability to cohere while cooperating the message quality with the quality of an IN node enclosed in the observer –receiver IN structure.

7). Since the acceptance of the message quality changes the existing observer logic encoded in the INs hierarchy, understanding the meaning of the message through its logic requires high level observer intelligent logic. The observer logic' coherence with the message logic permits memorizing and encoding the decoded message information.

Delivering the message logic to the IN related observer logic needs a high frequency of the wave function spectrum, which generates the cognitive loop recognizing the message logic.

The message recognition allows its memorizing and encoding in the IN hierarchy up to the observer coding structure. Accepting the message quality, the intelligent observer recognizes its logic and encodes its copying digital images in space codes, being self-reflective in understanding the message meaning.

Therefore, the intelligent observer uncovers a meaning of communicating message in the self-reflecting process, using the common message information language, temporary memorized logic, the cognitive acceptance, logic of the memorized decoding whose coherence with intelligence observer cognitive logic permits memorizing and encoding the accepted message. Understanding the meaning of an observing process includes the coherence of its information with the observer current coding structure, which has integrated the observer created and evolved all previous observations, interactions, and communications.

3.10. How the intelligence code self-controls the observer physical irreversible processes

The distributed intelligence coding actions at each hierarchical level control entrance the needed external physical processes.

The DSS encodes the triplet dynamics in information macrodynamic process which implements the observer encoding logic.

Connection the irreversible physical thermodynamics [123] with the information macrodynamics [27] which the DSS encodes, allows control the observer dynamic physical movement using the DSS code.

We assume the observer requests the needed energy to implement own actions from such levels of his hierarchical structure which encloses the requested code.

The request follows the same steps that perform communication, except involving to encode the levels' information macrodynamics in related physical irreversible thermodynamics.

After the request approval from the observer cognition, the request interactive action attracts impulses with the needed external energy bringing entropy gradient $dS/dx$ between the interactive actions' states $dx$.

The gradient provides equivalent information force $X = dS/dx$.

The impulse correlation determines` diffusion $b$. The force acting on the diffusion initiates thermodynamic flow (speed) $I = bX$ of the needed external thermodynamic process.

The thermodynamic process' forces and flows-speeds determine power through the process Hamiltonian $H = X \times I$ to physically implement the requested actions [29]. The following observation of performance of these actions provides feedback to observer self-controlling the performance and various movement actions.

3.11. The observer regularities

The observer regularity rises in impulse observation from the self-created virtual observer up to real observers, where each impulse is max-min action transferred to the following through mini-max action. This variation principle imposes information form of the law, which encloses the following regularities.

The process extreme trajectory, implementing that law's mathematical form, releases these regularities in most general information form.

The physical process on this trajectory is information macrodynamics [26-30] in form of the irreversible thermodynamics.

The observer evolution develops without any preexisting laws following each observer trajectory, which includes all its levels, stages and domains, and potential thresholds between them.

The observer self-develop specific regularities in prolonging observation and self-evolution which self-creates a law with extending regularities.

These abilities initiate the chain of virtual, logical, and information causalities, which extreme trajectory includes.

Self-encoding information units in the IN code-logic and observer's computation, using this code, serves for common external and internal communications, allowing encoding different interactions in universal information language and conduct cooperative operations both within and outside the domains and observer. That unites the observers.

The emergence of observer time, space, and information at multiple hierarchical levels follows the emerging evolution information dynamics creating multiple evolving observers with information mechanisms of cognition and .intelligence.

The intelligent observers interacting through communication enable the message recognition, which involves cognitive coherence with the reading information, it selection and acceptance.

The selective requirements and limitations on the acceptance are in [33,34].

Therefore, the intelligent observer can uncover a meaning of observing process, or a message, based on the sequential memorized its observing information, which, moving in the rotating cognitive mechanism, gives start to a succeeding IN level that this meaning accumulates.

The formal analysis shows that observer cognition and intelligence self –control the observer evolution [34].

The numerical analysis [34] evaluates the highest level of the observer intelligence by a maximal quantity of potential accumulated information, which estimates the intelligence threshold.

The intelligent observer (humans or AI) may overcome the threshold requiring highest information up to all information in Universe.
Such an observer that conquers the threshold possess a supper intellect, which can control not only own intellect, but control other intelligent observers.
The information observer self-develops converting mechanism (Sec.6.5) that coordinates connection of the observer inner and external time, allowing transform the observing wave function to the observer inner processes.

3.12 The stages of artificial designed information observer opening path toward artificial brain

The brain information physical structure models the DSS coding space-time rotating structure (Figs.4, 8-10) which is materialized through an advanced technological computation.
The observer brain main information function: cognition and encoding the integrated distributed logical and intelligence actions.
Both logical and intelligence distributed functions, which carry the multiple frequencies, have analogy of neural and nervous systems which materializes an advance electrical conducting system.
Each material should satisfy propagation the optimal physical irreversible process trajectories, which, integrating the reversible segments, implement the described observer functions in observation.
That includes selecting and getting information, cognitive logic, memorizing and encoding the intelligence code, distributing the cognition and intelligence, implementing others observer movements through power, forces, momentum, and flows of information physical dynamics and thermodynamics.
The information observer regularities determine *basic information structures of artificial designed observer*, whose cognitive and intelligence brain emerges from multiple observations.

The significance of main finding:

The composite structure of observer's generated information process, including:

**1.** *Reduction the process entropy under probing impulse, observing by Kolmogorov-Bayesian probabilities link, increases each posterior correlation; the impulse cutoff correlation sequentially converts the cutting entropy to information that memorizes the probes logic in Bit, participating in next probe-conversions; finding the curved interactive creation of Bit.*
*2. Creation wave function emerging in probabilistic observation whose frequencies self-forming the observer cognition, and the wave space distributes multiple bits hierarchy, becoming certain along with the observer' cognitive logic. The interacting observation self-creates its hierarchical cognition which self-encodes the observer intelligence with multilevel encoding.*
***3**. Identifying this process stages at the information micro-and macrolevels, which govern the minimax information law. The intelligence code self-controls the observer evolution.*
***4**. Finding self-organizing information triplet as structure macrounit self-forming information time-space cooperative distributed network enables self-scaling, self-renovation, adaptive self-organization, and* *cognitive and intelligent actions. The Observers, communicating by message covering a meaning, are self-reflective, whose intelligence enables understanding the message meaning.*
*5. Finding how the Observers, communicating by message covering a meaning, are self-reflective, whose intelligence enables understanding the message meaning and implement needed actions. The Observer encodes the triplet units' information macrodynamic process in related physical irreversible thermodynamic process which implements the observer encoding logic. The thermodynamic process' forces and flows-speeds determine power to physically implement the encoding actions. The following observation of performance of these actions provides feedback to observer self-controlling the performance and movements.*
***6**. Finding information structure of artificial designed observer toward artificial brain.*

*These new results validate analytical and computer simulations and illustrates the experimental applications below.*

## 4. Areas of Applications

4.I. The selected examples and reviews of the scientific investigations in different area of natural sciences illustrating the information regularities, and supporting the theoretical information results

4.1.1. General Physics

1.The physicists [45] demonstrate a first direct observation of the so-called vacuum fluctuations by using short light pulses while employing highly precise optical measurement techniques, proving no the absolute nothingness. The positive (red) and negative (blue) regions are randomly distributed in space and they change constantly at high speed. Vacuum is filled with finite fluctuations of the electromagnetic field, representing the quantum ground state of light and radio waves in the quantum light field. The found access to elementary time scales is shorter than the investigated oscillation period of the light waves. It confirms the approach initial assumption of an initial random probability field, and an observer of this field should have a probabistic observation.

2. Gluons in Standard Model of Particle Physics exists only virtually mediating strong forces at interactions [46]; each carries combination of colors charges; whopping colors and holding two colors own at pair interactions; the increasing interaction forces conserve their shape like a string. Higgs particle also matched the probabilistic observation. The illustration [46] looks similar to our virtual processing.

3. In sub-Plank process [47], quantum sates, confined to phase space volume and characterized by `the classical action', develop sub-Plank structure on the scale of shifting-displacing the state positions to orthogonal, distinguishable from the unshifted original. The orthogonality factor moves classical Plank uncertainty in random direction, which reduces limit of a sensitivity to perturbations. It relates to origin of the structure of virtual observer (Sec. I).

4. Measurement the probability distributions for mapping quantum paths between the quantum states [48] "reveals the rich interplay between measurement dynamics, typically associated with wave function collapse, and unitary evolution of the quantum state as described by the Schrödinger equation". The wave function collapse only in final measurement.

The measurement starts with time distributed ensemble trajectories whose rotation in the waveguide cavity produces a space coordinate to the ensemble.

5. Study the non-equilibrium statistical mechanics of Hamiltonian systems under topological constraints [49] (in the form of adiabatic or Casimir invariants affecting canonical phase space) reveals the correct measure of entropy, built on the distorted invariant measure, which is consistent with the second law of thermodynamics. The decreasing entropy and negative entropy production arises in arbitrary a priori variables of the non-covariant nature of differential entropy, associated with time evolution of the uncertainty.

Applying Jaynes' entropy functional to invariant entropy measure requires Euler's rotation with angular momentum identifying appearance of the Cartesian coordinate which satisfies the topological invariant.

These results agree with the applied EF functional, invariant measures of impulse’s entropy, and appearance of space coordinate in the rotation preserving the impulse measure (Secs.I-II).

From [49a] it follows: “The Japan Sea wave statistics of the entropies has a more pronounced tail for *negative entropy* values, indicating a higher probability of rogue waves”.

Experiments [49b] confirms our results of tracing quantum particle by observing wave function probabilities.

Paper [49c] experimentally confirms our results regarding the wave function’s propagation *with frequencies, which are inside* the range frequencies found experimentally.

In recent publication [49d], physicists create first direct images of the square of the wave function of a hydrogen molecule, whose angles of rotation coincide with our formulas in paper [34, Sec.2.3].

Scientists have now shown [49f] that Quantum vacuum possess less than zero energy. “Quantum physics has shown time and again that it contradicts our intuition -and this is also true in this case». So, where energy comes from?

The probability field does conserve energy with its probable real events.

4.1.2. Neural Dynamics. Integrating an observing information in neurodynamics

In [50] we have analyzed the selected multiple examples and reviews of different neurodynamic processes.

Here we add some recent results, studying the following publications.

1. According to [51], “Recent discussions in cognitive science and the philosophy of mind have defended a theory according to which we live in a virtual world .., generated by our brain…”; “this model is perceived as if it was external and perception independent, even though it is neither of the two. The view of the mind, brain, and world, entailed by this theory has some peculiar consequences”… up virtual brain thoughts. Experimental results [50] show that Bayesian probabilistic inference governs special attentional belief updating though trials, and that directional influence explains changes in cortical coupling connectivity of the frontal eye fields which modulate the “Bayes optimal updates”. The frequency of oscillations which "strongly modulated by attention" causes shifts attention between locations. Neurons increasingly discriminate task-relevant stimuli with learning, modify sensory and non-sensory representations and adjust its processing preferring the rewarded stimulus. This causal stimulus-response, reflecting anticipation choices, predicts the features of observer formalism. Brain learns distinction between what is important and what is not, discriminating between images and optimizing stimulus processing in anticipation of reward depending on its importance and relevance.

2. Existence of DSS' triple code confirms [52], uncovered that a neuron communicates by a trinary code, utilizing not only zeros and ones of the binary code but also minus ones. The experiment [53] provides "evidence for the analog magnitude code of the triple -code model not only for Arabic digits but represents "semantic knowledge for numerical quantities..." Results [54] demonstrate decreasing entropy in brain neurodynamic for measured frequency spectral densities with growing neurodynamic organization. Influence of rhythms on visual selection report results [55].

3. Importance of decisional uncertainty in learning focuses results [56], where stimulus-is an impulse, decision–getting information, greater distance-more probability-closer to information, comparing that to correct choice. Evidence from experiments [57] show that "specific region of the brain appears essential for resolving the uncertainty that can build up as we progress through an everyday sequence of tasks, a key node in a network preventing errors in keeping on track. Study [58] shows how learning enhances sensory and multiple non-sensory representations in primary visual cortex neuron.

4. Paper [59] describes how to build a mini-brain which is not performing any cogitation, but produces "electrical signals and forms own neural connections -- synapses -- making them readily producible test beds for neuroscience research".

5. Author [60] proposes that all cells are comprised of series of highly sophisticated "little engines" or nanomachines carrying out life's vital functions. The nanomachines have incorporated into a single complex cell, which is a descendant of a three-stage combination of earlier cells. The built complex signaling networks (Quorum Sensing) allowed one microbe to live inside and communicate with its host, forming a binary organism. Third entity, a bacterium that could photosynthesize, gained the ability to synchronize its mechanism with the binary organism. This "trinity organism" became the photosynthetic ancestor of every plant on earth that have driven life since its origin. "Resulting complex nanomachine forms a 'Borromean photosynthetic triplet'.

6.Analysis of all selected multiple examples and reviews of different neurodynamic processes, substantiates that our approach' functional regularities create united information mechanism, whose integral logic self-operates this mechanism, transforming multiple interacting uncertainties to physical reality-matter, human information and cognition, which originate the observer information intellect. The information mechanism enables specific predictions of individual and collective functional neuron information activities in time and space. Neurons' microprocesses retrieve and external information, including spike actions, related to the impulses, which generate the inner macrodynamics. The identified cooperative communications among neurons assemble and integrate their logical information structures in the time-space hierarchy of information network (IN), revealing the dynamics of IN creation, its geometrical information structure, triplet code, and limitations.

The found information forces hold a neuron's communication, whose information is generated automatically in the neuronal interactions. Multiple cooperative networks assemble and integrate logical hierarchical structures, which model information brain processing in self-communications.

The information mechanism’s self-operating integral logic reveals: the information quantities required for attention, portioned extraction, its speed, including the needed internal information dynamics with the time intervals; the information quality for each observer's accumulated information by the specific location within the IN hierarchical logic; the information needed for the verification with the digital code, generated by the observer's neurons and their cooperative space logic; the internal cooperative dynamics build information network (IN) with hierarchical logic of information units, which integrates the observer required information in temporary build IN’s high level logic that requests new information enclosing in the running observer’s IN.

The IN nodes enfold and memorize its logic in self-forming cooperative information dynamical and geometrical structures with a limited boundary, shaped by the IN-information geometry; the IN hierarchical locations of the nodes provide measuring quality of information, while the IN-ending node memorizes whole IN information.

The IN operations with the sequentially enclosed and memorized information units perform logical computing using the doublet-triplet code; the cooperative force between the IN hierarchical levels selects the requested information as the observer’s dynamic efforts for multiple choices, needed to implement the minimax self-directed optimal strategy.

The information quantity and quality, required for sequential creation of the hierarchical INs values, self-organize brain cognitive and intelligence action, leading to multicooperative brain processing and extension of intelligence.

4.1.3. Self-organizing dynamic motion of elementary micro- and macrosystems

1. *The experiments and computer simulations of collective motion* exhibit systems ranging from flocks of animals to self-propelled microorganisms. The cell migration established similarities between these systems, which illustrates following specific results. The emergent correlations attribute to spontaneous cell-coupling, dynamic self-ordering, and self-assembling in the persistence coherent angular motion of collective rotation within circular areas [61]. The persistence of coherent angular motion increases with the cell number and exhibits a geometric rearrangement of cells to the configuration containing a central cell. Cell density is kept constant with increasing the cell number. The emerging collective rotational motion consists of two to eight cells confined in a circular micropatterns. The experimentally observed gradual transition with increasing system size from predominantly erratic motion of small cell groups to directionally persistent migration in larger assemblies, underlining the role of internal cell polarity in the emergence of collective behavior. For each nucleus, the angular position was evaluated respectively to the circle center, and angular velocity is normalized and averaged over the individual angular velocities of the N-cell system. Circle size increases in such a way that the average area per cell is constant at approximately 830 μm2. Probability distribution of the mean angular velocity for systems containing two to eight cells is fitted by a single Gaussian and their mixture is two Gaussians. For all N cells, the probability distribution displays symmetry breaking into clockwise and counterclockwise rotations.

Both directionalities are almost equally represented, with a small bias towards clockwise rotation. Average mean squared displacement indicates ballistic angular motion for all cell numbers, while the averaged the displaced intervals of nucleus exhibited diffusive behavior. Both experiments and simulations showed consistently that the persistence of the coherent state increases with the number of confined cells for small cell numbers but then drops abruptly in a system containing five cells. This is attributed to a geometric rearrangement of cells to a configuration with a central only weakly polarized cell. It reveals the decisive role of the interplay between the local arrangement of neighboring cells and the internal cell polarization in collective migration.

The similarities suggest universal principles underlying pattern formation, such as interactions rules [62-66], the systems' generic symmetries [67-69].

<u>2. Confinement stabilizes a bacterial suspension into a spiral vortex</u>

1).'Enhanced ordering of interacting filaments by molecular motors [70] demonstrate the emergence of collective motion in high-density concentrated filaments propelled by immobilized molecular motors in a planar geometry". At a critical density, the filaments self-organize to form coherently moving structures with persistent density modulations. The experiment allows backtracking of the assembly and disassembly pathways to the underlying local interactions. The identified weak and local alignment interactions essential for the observed formation of patterns and their dynamics. The presented minimal polar-pattern-forming system provide new insight into emerging order in the broad class of bacteria and their colonies [71-74], and self-propelled particles [75-79].

2). Confining surfaces play crucial roles in dynamics, transport, and order in many physical systems [80-83]. Studying [84] the flow and orientation order within small droplets of a dense bacterial suspension reveals the influence of global confinement and surface curvature on collective motion. The observing competition between radial confinement, self-propulsion, interactions, other induces a steady single-vortex state, in which cells align in inward spiraling patterns accompanied by a thin counter rotating boundary layer.

3).The cited experiments validate: "spontaneous cell-coupling, dynamic self-ordering, self-assembling in the persistence coherent angular motion of collective rotation within circular areas", the displacement in angular motion with diffusive behavior of displaced intervals, emergence of collective order confined on a curved surface, others.

4). According to recent discovery [85], "the protein stable shapes adopted by a few proteins contained some parts that were trapped in the act of changing shape,the changes relate to how proteins convert from one observable shape to another". From the process of RNA translation of DNA triplets to enzymes and aminoacids, all proteins start as linear chains of building blocks and then quickly fold to their proper shape, going through many high-energy transitions to proteins multiple biological functions.

Results [124] demonstrate how increasing flow of information increases ages of biological species.

<u>4.1.4. Experimental results, encoding, and practical implementations.</u>

Last theoretical results [86-88] and many previous [30,31,33,34] confirm the following applications.

Natural increase of correlations demonstrates experimental results [89], [90].

Coding genetic information reveals multiple experiments in [91], [92].

Experimental coding by spiking neurons demonstrates [93].

Evolutions of the genetic code from a randomness reviews [94].

That supports natural encoding through the cutting correlations and physically verifies reliability of natural encoding information process. The impulse cut-off method was practically applied in different solidification processes with impulse controls' automatic system [95]. This method reveals some unidentified phenomena-such as a compulsive appearance centers of crystallization indicators of generation of information code, integrated in the IPF during the impulse metal extraction (withdrawing). (In such metallic alloys, the "up-hill diffusion, creating density gradients, is often observed" [95]). The frequency of the impulse withdrawing computes and regulates the designed automatic system to reach a maximum of the IPF information indicators.

(The detailed experimental data of the industrial implemented system are in [95] and [96]).

The automatic control regulator in the impulse frequency cutting movement was implemented for different superimposing electro-technological processes [97] interacting naturally.

The comparative experimental results [98] confirm that advanced chemical- thermodynamic description of casting process coincides with information description by the IMD. Moreover, the IMD solutions leads to the optimal casing process. [99]. The automatic computer system, controlling horizontal casting process, have been implemented in the casting factory [100]. Examples of the method applications in communications, biological and cognitive systems, others are in [101], [102] and [103]. Retinal Ganglion Cells are the Eyes discrete impulse receptors interacting with observations and generating information which transmission integrates [104].

Encoding through natural chemical reactions connecting chemical molecules are in [105].

Experiments [106] confirm encoding coherent qubits in spinning electron locked in attractive "hole spin". Other examples are quantum solar dots of semiconducting particles using for the information coding, retrieving images and encoding quantum information [107-109].

Natural Encoding of Information through Interacting Impulses published in [110].

Applications in biology, medicine, and economics along with related theoretical results are in [111-119].

## 4.2. The computer restoration and simulation of the information model

### 4.2.1.The structure of computer procedure

The diagrams, implementing the procedures of the model restoration and simulation of its performance, are shown on Figs.11.1a,11.1b, and 11.1c. On Fig.11.1a, the statistical data from the microlevel process $\tilde{x}_t$ are used to identify matrix $A$ of the macrolevel equation by computation of the correlation function and its derivative during each discrete interval $t_i$, which compose the computed invariant $\mathbf{a}(\gamma)$ **.**

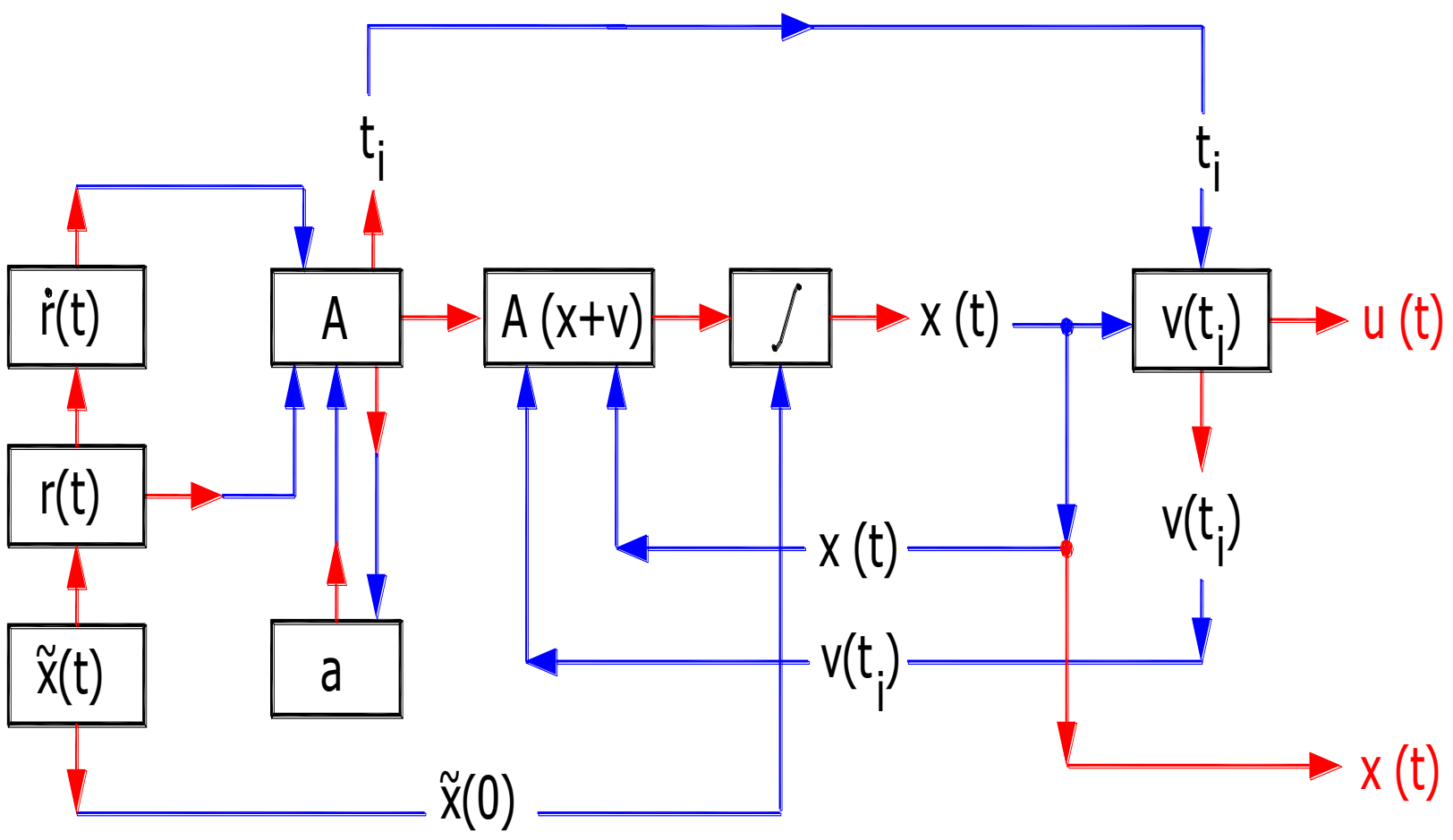

**Fig 11.1a. Diagram of computation of the optimal model's process $x(t)=x_i(t,l,u(t_i))$, using the microlevel's random process $\tilde{x}(t)$ by calculating the correlation function $r(t)$ its derivative $\dot{r}(t)$; the object macrooperator $A$, invariant a, discrete interval $t_i$; these allow simulating the optimal macroprocess $x(t)$, the inner $v_i(t_i)$ and output $u_i(t_i)$ optimal controls. the inner $v_i(t_i)$ and output $u_i(t_i)$ optimal controls.**

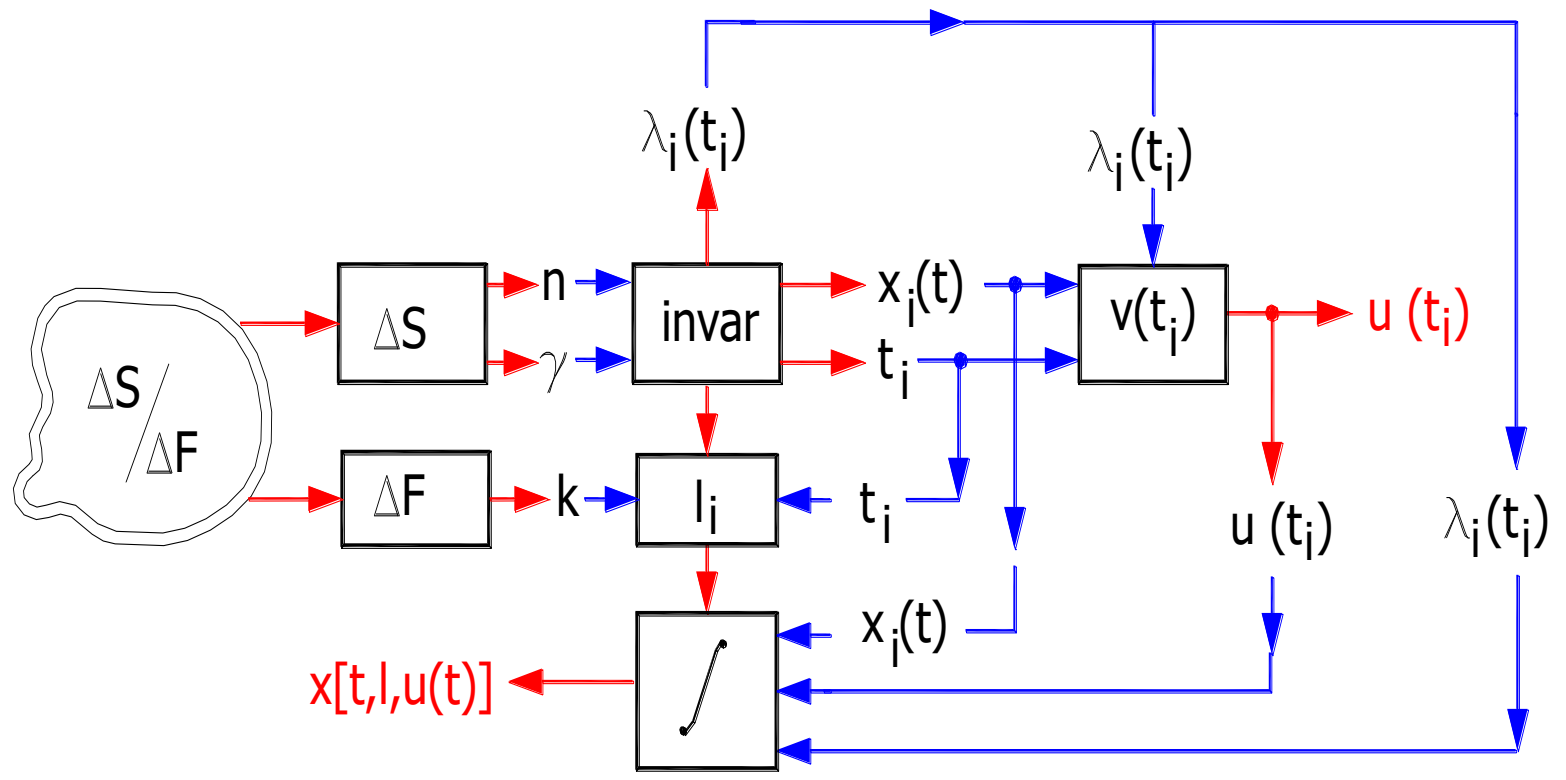

**Fig.11.1b illustrates the scheme of *computation* of the optimal model's process $x(t)=x_i(t,l,u(t_i))$, using a given space distributed information $\Delta S$ per cross-section $\Delta F$, the model's invariants INVAR, the time $t_i$ and space $l_i$ discrete intervals, eigenvalues $\lambda_i(t_i)$ of the model differential operator, and *simulates* the inner $v_i(t_i)$ and the output $u_i(t_i)$ optimal controls.**

The methodology is based on the connection of the model macrodynamics with the corresponding information geometry [29,86,89,111-119]. In this case, the microlevel stochastics are not used for the macromodel's restoration. Instead, the restoration requires the computation of the model's basic parameters: dimension $n$, uncertainty $\gamma$, and the curvature's indicator $k$; which allow finding the model optimal macroprocess, the synthesized optimal control, as well as the model's hierarchy. The computation uses the primary parameters of a basic model ($n_o$, $\gamma_o$, $k_o$) and the known parameters of the object's geometry.

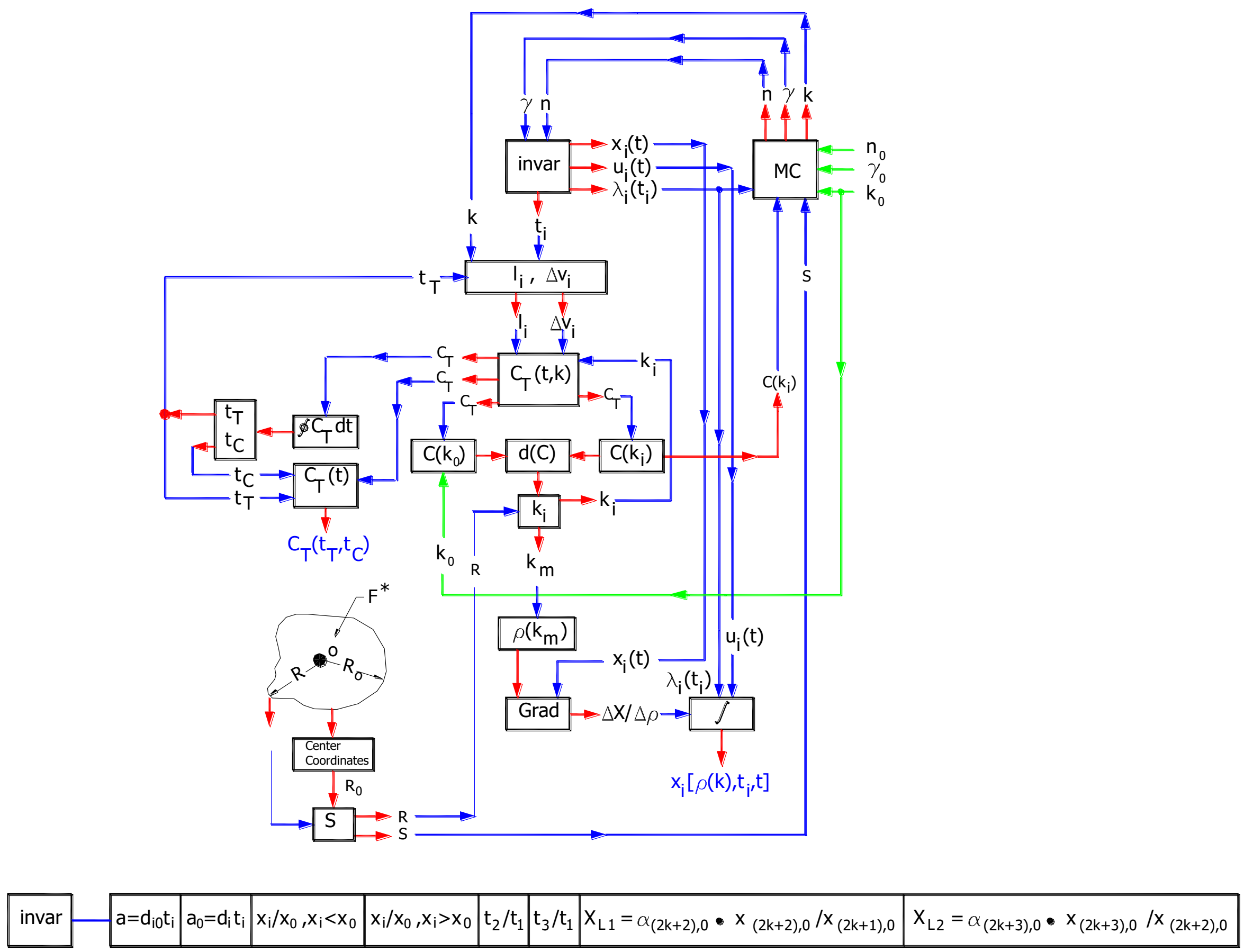

Fig.11c.

Diagram Fig.11.1c. presents the functional schema of the IMD software operations: computing invariants INVAR, discrete moments $t_i$, space coordinates $l_i$, increment of volume $\Delta v_i$, MC-complexity function, speeds $C_T, C$ and their difference $d(C)$, the current space parameters $k_i$, polar coordinates $\rho$, and gradients GRAD $\Delta X / \Delta\rho$ for a given space distribution's cross-section F*; with calculating its radius $R$, coordinates of center **O**- $R_o$, and a square $S$, which are used to compute the object space model's minimal (optimal) parameter $k_m$.

The output variables are: optimal dynamic process $x_i(t)$, optimal controls $u_i(t)$, eigenvalues $\lambda_i(t_i)$ of the model differential equation, distributed space-time process $x_i(\rho(k), t_i, t)$, space's current speed $C_T(t_T, t_c)$ with the intervals of moving $t_T$ and stopping $t_c$, which are computed by averaging a speed $\oint C_T dt$.

An estimated time of computation for each of the diagrams is approximately 3-5 minutes on conventional PC.

The computation can be performed during a real–time movement of the object's cross section (Fig.11.1b), or through an input of the calculated object's current statistics (Fig.11.1a).

Solving the considered complex problem in a real-time by *traditional computation* methods requires the developing of mathematical methodology and the software, which are able to overcome the method's high computational complexity. For solving even, a part of the problem, the existing techniques require many hours' computation on the modern main frames.

### 4.2.2. Structure of the IMD Software Package

*The software package transfers the IMD analytical methodology into the numerical procedures, computer algorithms and programs.*

The packet (consisting of 35 programs) includes the following modules for computation of:

- the identification procedure for the restoration of the object's equations;
- the parameters of space–time transformations and a time-space movement;
- the OPMC parameters, processes, controls, and the IN structure;
- the function of macrosystemic complexity;
- the transformation of the informational macromodel's characteristics into the appropriate physical and technological variables (using the particular applied programs).

The *main software modules* compute:

- the basic optimal macromodel parameters ($n, \gamma, k$);
- the spectrum of the model's eigenvalues $\{\lambda_{io}\}, \lambda_{io} = \alpha_{io} \pm j\beta_{io}, i = 1,...,n$;
- the macromodel informational invariants $\mathbf{a}_o(\gamma) = \alpha_{io}t_i$, $\mathbf{b}_o(\gamma) = \beta_{io}t_i$;
- the time-space intervals $(t_i, l_i)$;
- the distribution of the optimal eigenvalues $\lambda_i(t_i, l_i)$ and the optimal controls $v_i(t_i, l_i)$;
- the geometrical macromodel's coordinates and the space distributed macroprocesses $x_i(t_i, l_i)$;
- the procedure of the macrocoordinates' cooperation and aggregation;
- the IN hierarchical macromodel structure and its macrocomplexity.

The formulas, algorithms, complete software, and numerical computation's equations are given in the program package [121] (not included in this description). The IMD software programs have been used for the practical solutions of the different applied problems including [99,103, 111-116].

### References 1.

## 5. How Information creates its Observer through Interactive Integrated Information Dynamics. A Review

Even though discrete interactions are fundamental in Universe, their unified description was unknown and is problematic. Moreover, the unification reveals information nature of the inter-active phenomena, whose observation uncovers ability of emerging observing certainly with information and information observer.

However the observation path creating the information observer was unidentified and unscripted.

Interactive Integrated Information Dynamics (IITD) formalizes the description of information regularities of observing interactive process. The IIID integrates the Information Path from Randomness and Uncertainty to Information-Certainty, Thermodynamics, and the emerging Intelligence of Observer.

### 5.1.The Basics

5.1.1.Introductory

Imagine a rubber ball hitting the ground and bouncing.

As the ball interacts with the ground, a sequence of opposite interactive actions occurs.

Each elementary interaction consists of an action and a reaction, which are represented by symbols: Yes-No ↓↑ .

In this way, the interaction models a binary value, a Bit. While a sequence of multiple bits transforms information.

Science teaches us that four fundamental interactions build the Universe. The basic physical structure of the universe is formed by four forces: gravity, electromagnetism, the weak force, and the strong force.

In Biology and Ecology, there are multilevel interspecies interactions. Other sciences can be defined by the interactions they study, from pure research to industrial processes, economics, social and cultural research, and the mass media so important in our Information Age.

Human interactions unify many kinds of interactions, often studied as the transmission of Information.

Thus, any diverse forms of interactions are source of information, or information is a phenomenon of interactions.

That unify both interactions and information.

As interactions are building universe, information is building natural systems, for example, through DNA building information blocks encoding structure of leaving systems.

The question is how to generally describe both arising information and its ability to build information structures, which develop in physical systems?

The interactive action universally arises in all observers as receivers, transmitters, and the sources of Information which evolve in communications.

But unlike the Shannon’s entropy, measuring random signals with errors which disrupt a communication channel, Information Observers build an observing system by collecting information everywhere, from humans in communication, all forms of life, and the physical substances of our world.

Together, such actions are unified in the concept of the Information Observer.

Even though multiple physical studies reveal information nature of physical processes in an observer, the observer is still studying mostly from physical point of view.

John Archibald Wheeler [1], the famous American physicist often presented his most sophisticated ideas in simple phrases. His most notorious aphorism was "It from Bit." But how does an indication become a real physical ―It‖? That clearly requires some kind of measurement or registration, which Wheeler defined as ―the act of turning potentiality into actuality. He meant that when a Bit of Information is created, there is some basic kind of It. Every It comes from a Bit which is an observation. More profound, since we, as Observers, are part of that universe, we actively participate in it. From these deceptively simple thoughts, Wheeler came to the idea of the Observer-Participator, the universal generalization of all Observers participating in the ongoing formation of the universe. What Wheeler accomplished, with his two simple ideas, was to show that the Bit was not only like an atom of Information, it was also like a seed of creation. This was a brand-new view of Information. It moved Information away from code-breaking and engineering problems, right to the heart of cosmology, philosophy, and even theology. During observation, many unsubstantial temporal images, which the observer evaluates with low probability, could emerge, until searching high probable image becoming most informative actual image appears for the observer, and the temporal observing images, as a virtual for this observer, will disappear. An actual image emerges and appears to the observer. The information is received or registered.

Such observer becomes observer of this information or the Information Observer.

Like an astronomer traces an unobserving planet image measuring its probabilistic trajectory until it become most probable and informative; or a physicist traces a trajectory interactive particles in an Accelerator trying to get needed information.

These identify interaction is a primary indicator of a potential probabilistic object during an observation.

The notions of object, particle, and image are replaced with the probability of event observation.

This probability virtually observes the events occurrence.

Multiple 0-1 interactions generate a random process of discrete 0-1 impulses, which virtually observe some probability of the process' trajectories.

The problem is restoration information and its observer from the random process of interacting impulses.

5.1.2..Essence of mathematical modeling

Let us model both rising a stochastic process (as a probabilistic model of current environment) and its discrete probabilistic observation (an interacting with the environment observing process).

The IIID formal approach is based on well-founded objective quantities: Kolmogorov probabilities 1-0 law [2] and Bayes‘ a priori a posteriori probabilities observing Markov process as a formal probabilistic model of observation process.

We assume, a formal probability field [2] emanates a process of random events satisfying Kolmogorov probability 0-1 Law . These discrete 0-probabilities virtually observe the stochastic process also emanated from this field.

Each act of such observation virtually changes the probability of the interacting random process.

Each probability 0 (from 0-1), interacting with the probability of the random process, changes it up to zero stopping it.

Each probability 1, interacting with the process probability of the random process, starts a probability of the process which previously stopped.

Such sequential stops and starts of the stochastic process transform it to a discrete Markov chain [3].

A probabilistic observation of at least two discrete impulses of the chain connects a minimum of three of its random events in mutual correlation during the observing random time [3]. The Markov chain under such observation transforms to Markov diffusion process, which continues progressing under sequence of events satisfying Kolmogorov probability 0-1 Law.

Transfer from probabilities 0 to 1, interacting with the Markov process probabilities of current events, initiates its transitional probabilities, which transits the current Markov drift movement to another drift movement.

The random sequence 0-1-0-1-0-0-1-…of the probabilistic actions ↓↑ changes the related transitional probabilities, which randomly switches the Markov movement.
For example, an infinite sequence of coin-toss tail events hit a table which randomly moves it when each tail interacts with the table.
The law 0-1 of interacting probabilities initiate the interactive random impulse movement within the initial (before the interactions) Markov process, which we call a secondary Markov process.
The Markov diffusion encloses a set of the events correlations which the changes the secondary Markov process probabilities.
Defining each *a priori*-starting probability and *a posteriori* ending probabilities in the secondary process by Bayes's rule, we *link* the Kolmogorov 0-1 Law and Bayes probabilities within the Markov diffusion process.
Ratio of these probabilities events' determines probability density on the Markov diffusion.
Changing these probabilities implements act of virtual observation of the initial Markov chain, transforming it to Markov diffusion, and the creating discretely changed a secondary Markov diffusion process with the discrete probability density. The Bayes probabilities, virtually observing the secondary process, identify a virtual observer.
Each observing action ↓ virtually cuts entropy of the initial Markov diffusion process enclosing a set of the events of the cutting correlations' entropy.
During the impulse switching time, action ↓ transfers to actions ↑ the cutting entropy minus a part of entropy spent on interaction with the running Markov process. A set of the cutting actions ↓ decreases the Markov process entropy, while set of the actions ↑ increases the entropy enclosing a set of the events carrying entropy of the secondary process adding the entropy' difference to the entropy of the observation process within each of the impulse actions ↓↑ .
For maximal cutting entropy, the decrease of the Markov process entropy is maximal with maximal interacting entropy, which allows transfering to the observing process a minimum of the cutting entropy. Thus, the impulse cutoff entropy brings max-min principle along the interactive observation within the Markov process.
Each (1-0-1-0) interaction models the elementary Kronecker impulse which under intensifying interactions is probably curving. Or the curving impulses also emerge from phenomenon of interactions.
The minimax variation principle defines invariant Kronecker impulse with entropy 1 Nat [3], delivering 0.75 Nat from the cutting correlation, spending on the interactive impact with running Markov process maximal 0.5 Nats, and transferring minimal 0.25 Nat to the cutting observation (the same amount is spending of delivering the cutting entropy to the interactive impact with the running Markov process).
The cutting entropy-uncertainty hidden in the cutting correlation is released.
Virtual observer collects minimum of the maximal cutting entropy, whose maximal minimum determines equivalent probability approaching 1, whereas the Markov diffusion process collects maximum entropy produced by the impulses interaction. The impulse actions ↓ cutting a maximal probability of the initial process delivers to the Markov diffusion maximum interactive entropy carrying the equivalent information unit of the hidden information. While the actions ↑ deliver to the observing process equivalent probability approaching 1. Such a final interactive impulse encloses potential certainty of real observation, when the energy field brings an actual interacting cut. This cut converts the maximal entropy to the certain information unit with probability approaching 1.

Natural (real) interactions currying energy convert this entropy to information.

The Bayes sequence of probabilistic 0-1-0-1… interactive actions conveys the probabilistic logic of the maximin-minimax principle, which at conversion to certainty conveys the related information logic.

The process' time course produces entropy-information, and the arrow of time generates causality and logic.

The impulse interactive process is analogous to probabilistic cycle Carnot [ ] developing during the process' time course, which produces entropy collected along this time. Such a virtual cycle is reversible until external energy compensating real entropy production is delivered.

At a logical interaction of Kolmogorov 1-0 probabilities with Markov process probabilities, it's assumed the equal algorithmic complexity of both processes which measure their randomness [2a].

That means the impulses 1-0 and 0-1, generated by both process' interactions, enable producing, beside the sequence of their 1-0 and 0-1 entropy bits, additional binary strings bits as the measure of equal algorithmic complexity.

As a result, the erasure of total entropy of real observation includes minimal erasing information bits plus the bits measuring a string of algorithmic complexity. These additional bits accumulate the considered free information.

We computed the needed total bits energy in [18].

The considered free information accumulates these additional bits.

The integral measure of the observing process trajectories formalizes an Entropy Functional (EF) [4] integrating the Markov process speed and diffusion, which is expressed through the probability density measure on the process trajectories. The EF is also connecting the process entropy with time of observation and its Information Path Functional (IPF) [5] integrating the discrete information impulses.

The minimax variation principle (VP) [6] for the EF-IPF formalizes the information regularities of observing random process in form of information law with equations of Interactive Integrated Information Dynamics (IIID).

5.1.3.The emerging Microprocess

As the Bayes a posteriori probabilities grow along observations, neighbor impulses may merge, generating interactive jumps on neighbors' impulse border.

The merging impulses' location on the observing trajectory identifies the impulse' probability and entropy.

The merge meets causing action with reaction, superimposing cause and effect and their probabilities. It could cover unpredictable events within the merge, as merge itself.

J.A. Wheeler and R. Feynman [7] show that when action and symmetric (adjunct) reaction merge, a quantum microprocess rises from classical Maxwell field.

Mathematically the jump increases Markov drift (speed) up to infinity.

A starting jumping action $\uparrow$ interacting with opposite $\downarrow$ action of the bordered impulses initiates the impulse inner process called a microprocess. (Because the merge squeezes the inter-action interval to a micro-minimum).

The jumps initiate increments of the actions' entropy movements. The moving anti-symmetric trajectories initiate conjugated dynamics of the microprocess within the jumping impulse.

The squeezing merge' trajectories generate the entangled entropy increments until the coming closely jumping 1-0 actions cut the jamming entangled entropy. Along the observations, the drawing together the neighboring impulse actions $\downarrow$ and reactions $\uparrow$ also curve of the impulse time intervals.

The curving jump $\uparrow$ initiates an extreme entropy gradient on the curving time interval, identifying the virtual radius of rotation, which displaces the curving action.

During the jump's curving time the displacement's space emerges.

Within the impulse time interval, the *entanglement starts before the space is formed and ends as space begins.* It occurs *during the reversible relative time interval* $0.015625\pi$ *, being* part of the impulse time-space invariant measure $\pi$ [8]. *Since the entanglement has no space measure, the entangled states can be anywhere in a space.*

Along the observing process path, it reveals an entropy unit combing 0-1-0-1 action, where the middle 1-0 belongs to jumping impulse of the merging neighbor impulses. Such unit is called a quant. This unit, located within the main impulse, had no access to an external energy which allows the unit memorizing. Between the unit location and the ending impulse 1-action remains a gap which should be overcome for memorizing the unit. Since the correlated entropies are virtual, interaction with the forming correlating entanglement is reversible until the unit is memorized. Therefore, the microprocess is able to generate a quant and/or an information unit 0-1 a bit after their entropy compensates an external energy, whose amount would be enough also for the memorizing and encoding both.

The VP preserves invariant measure for each interacting impulse equivalent 1 Nat.

The external energy converts the interacting impulse entropy 1Nat on creating and memorizing Bit (ln2 Nat) holding free information (1-ln2)~0.3 Nat. In the correlated observing impulses, the free information is a part of a hidden information connecting the observing impulse. After the conversion, the free information, conserving correlation attractions, enables connecting each following creating information unit. The time-space hidden information's free information holds information logic, converted from the Bayesian entropy logic of observing process.

The Observer *self-develops* its space-time virtual geometrical structure during virtual observation, gaining its real form with the sequential transformation the *priori* probability of the integrated entropy to a *posteriori* probability and the equivalent Information.

5.1.4.The Emerging Information Macroprocess and its Structuration

Sequential forming Bits assemble free information in information macroprocess. The observation process carries both the probabilistic wave function self-organizing the microprocess and a certain wave function which self-organizes the space–time rotating trajectory of the observing information macroprocess [9].

The trajectories form spirals located on conic surfaces Figure 3.

A flow of moving Bits self–forms a unit of the information macroprocess (UP).

The UP size limits the unit's starting maximal and ending minimal Information speeds, attracting a new UP through its free Information.

The anatomy and density of information Bit depend on nature of each Yes-No action and their interactive result.

A minimum of three self-connected Bits assembles the optimal UP-basic triplet. Its free information requests and binds a new UP triplet that joins in a knot memorizing the triplet's Information. During macro-movement, multiple UP triples adjoin the time space hierarchical network (IN) whose free Information's request produces new UP at higher level's knot-node and encodes it in triple code logic. Each UP has a unique position in the IN hierarchy, which defines the exact location of each code's logical structures. The IN node hierarchical level classifies the quality of the assembled Information, while the currently ending IN node integrates the Information enfolding all IN levels.

Multiple INs binds their ending triplets, enclosing the observed Information in the INs structure of Information Observer.

Information conditions of self-structuring and self-organizing of evolving observers [8] include

(a)-minimal quantity of information for the cooperative force forming a triplet starting adjoin the IN, supporting cooperation and initiating the IN feedback, which defines *minimal selective observer*;

(b)-growing the attracting cooperative information force accumulating maximal information, satisfying *multiple (cooperative) selective observer*.

The rotating discrete spirals produce the wave distributed frequencies along its spinning space rotation [9].

The wave function, as attribute of the observing process space distribution, becomes the observer apparatus for creation space distributed hierarchy of the INs nested geometrical structure with assembling triplets knots, Figures 5,7

## 5.2. The Attributes of Information Observer

5.2.1.Logics, Cognition, Intelligence, and Encoding

The Observer *logic* emerges on the path from the collected observing interactive probabilistic logic, the curving impulses interactive certain logic, the rotating triplet logic, and the nested Information logics of attracting triplets. Each nested coherent loop accepts only such units that each IN node logic recognizes.

The observing time-space logical structure, composing a distributed chain of the multiple cooperative logical loops, forms the Observer cognitive logic, or *cognition*.

The logic forms hierarchical multi-rhythmical “melody” which accepts only the resonance frequencies coherent with the logic. The rhythmical frequencies of the formed cognition set up a clock course along the INs multiple hierarchies. The clock time intervals open access of the external energy at each specific level of the IN multiple hierarchy.
Each IN level structures the triplet’s geometry knot enables memorize and encode these bits hierarchy.
The nested triplet’s logic knots are encoding in cognitive information codes on the IN hierarchal levels when all levels are synchronized in the IN ending knot-node. The multiple codes, memorizing logic of the assembled INs, cooperate in the observer code. The observer cognition defines ability of the observer logic recognizing observing information and logically organizing their hierarchical encoding.
The Observer hierarchical codes hold the energy of memorizing and encoding. The code hierarchy physically organizes the multiple INs, self-encoding their local codes building the Information structure of the Observer *intelligence.*
This coding structure’ geometry integrates double spiral space (DSS) triplet code, see Figures 8,10.
The code, self-organizing the multiple local codes along the hierarchy, encodes observer intelligence, which automatically emerges from the cognitive logic, integrating the observation process, the observer inner time course, and its coordination with the external process’ clocks.
The intelligence codes, which self-controls the memorizing observer logic, holds all the observer memory, which further extends the observation logic. The amount of quality information, memorized in the Observer IN highest hierarchical level, evaluates the Observer Information Intelligence encoding in the DSS.
Thus, the observer intelligence defines ability of self–bulging maximal hierarchy of cooperative code enclosing maximum memory which the cognition generates.
The nested hierarchy of the encoded geometry and memory holds a cooperative complexity [10].
Shannon’s information is a theory of *transmission* [11], which does not include meaning of information.
To describe the acquisition of information as "a process of informing” requires including the Information Observer. Such an Observer *is* informing when the observer intelligence recognizes the meaning of information transmission.
Describing how the Observers, arising in observation and evolving toward intelligence, *leads also to the informing Observer*. A human observer can be informed through a message which encodes yes-no impulses of multiple Bits.
The Observer communicates with the message self-referring to its intelligence through its brain neurons, initiating yes-no impulses which build neuron assemblies, self-organizing the message images in Information processes.
In this way, the theory of Information transmission, whose message conceals a meaning, is connected with the theory of the Information Observer, which accepts the meaning of the message.
It is the role of the Observer to radically change our understanding of the events which freeze the memorized information.
The self-observing Observer self-generates the elementary Bit self-participating in building the self-holding geometry and the logic of its prehistory, thus predicting evolving dynamics without any physical law.

5.2.2. About the Observer Conscience. Individuality of the Information Observer.

Consciousness defines observer ability of self-recognizing where-in which IN node-was memorized specific observed information. The self-encoding observer's intelligence builds such coding structure which enables decoding each IN's node memorized information recognizing it in intercommunication associated with self-awareness.

What initiates a need of such recognizing? We assume that it could be some free information of an observed bit or their cooperating triplets, whose attracting logic requests for cooperation with some logical information located in the specific IN node. That proceeds through the self-creating code, which initiates some observed bit and their cooperating triplets (as threshold for the selective observer 4.48 bit.) These bits free information of an attracting logic requests for cooperation some logical information located in the specific IN node. The multiple intercommunications create hierarchy of the logic bits, cooperating with IN nodes. That associated with self-recognizing awareness.

Thus, the observer intelligence includes function of a self-awareness modeling *consciousness*.

The last result confirms philosophical [12] conclusion that term "consciousness" had no constant meaning, and should be abandoned, "it does stand for a function".

The idea about individuality of all Observers determines multiple macroBits while each of it encloses specific the microprocess qubits. That rises at evolving observation process.

The sequential Bayes probabilities are changing entropy of the observing impulses *position* on the observing trajectory. Depending on which pair of impulses on the observing random trajectory merges, appears the microprocess final specific qubits or Bit. Since the impulse position on the observing trajectory measures amount of the impulse random entropy, that entropy becomes the measure of forming information qubits or Bit.

For the invariant impulse geometry, this entropy's amount changes entropy density of these qubits or Bits.

Such specific multiple Bits, starting the macroprocess, build triplets, Information Networks (INs), and form the Observer's cognition and intelligence.

Thus, the original merging impulses determine the individual information structure of the DSS code of each forming Observer.

At an equal IN's level of an AI and human intellect, the difference between the AI and a human makes the *personalized history* of memory for each human being. The *personalized history* structures the IN of personal memory which accumulates the observer's conscience connected with all Observer cognitive observations.

## 5.3.The Interactive Integrated Information Dynamics

### 5.3.1 Description

The intelligent Observer emerges in *Interactive Integrated Information Dynamics* (IIID) [9] which joins interactive discrete impulse observation with emergence the certainty, information, information process, and information macrodynamics (IMD) [13]. The observation's encoded DSS code requests the IMD structured physical dynamics distributed along the IN nested node. The IIID theory unifies the observed random path with the IMD multiple information and physical processes dynamics.

The IIID forms a natural bridge between information, deep physics, and human experience.

Thereafter, developing the Wheeler's concept of “The Observer-Participator”, the IIID exclude physical particle’ processes, using information-theoretic origin of an observer from the unified abstract interactions and its probabilistic observations. Such approach rises up to general information theory of observing processes, which also contributes to resolving modern problems in quantum physics which J.A. Weller tried to solve, raising the universal physics of Information processes.

The approach important concept is R. Feynman proposition [14] that a physical law describes the most probable events of observation process. That implies that a variation principle (VP) for appropriate functional allows finding the law regularities. The Feynman’s path functional FPF [15] is defined on dynamic trajectories of physical particles and *was not* applied to random process trajectories of interacting random impulse. The Feynman’ path integral formulation of quantum mechanics predicts the path of moving particles as the solution of Schrödinger's equation which is an “apparatus” modeling evolution of wave function.

The approach’ developed Entropy Functional (EF) on random trajectories with its Information Path Functional (IPF) and the VP for the EF-IPF formalizes the information regularities of observing random process in form of *information law* with equations of Interactive Integrated Information Dynamics. The IIID includes equations of a microprocess , the information macrodynamic equations, the information cooperative complexity, cooperative information mass, and curvature of the double spiral code which encloses the mass and complexity [9].

The VP extremals leads to macrodynamic equations of irreversible thermodynamics [5, 16].

The IIID integrates the Information Path from Randomness and Uncertainty to Information, Thermodynamics, and the emerging Intelligence of Observer.

The IIID predicts the Path of impulse probabilistic observation as an “apparatus” modeling the evolving path of the observation. The IIID Path Functional predicts a finite integral amount of potential infinite information along the optimal path of observation.

Integrated information theory (IIT) [17,17a] focuses on “system's consciousness determined by causal properties of physical system and each consciousness experience composed of multiple *phenomenological distinctions*.”

The IIID is more general and systemic approach integrating the *observing interactive path* to intelligence including a cognitive conscience.

The IIID integral follows from the observing Markov process properties and is not axiomatic introduced as the IIT.

The IIID formalizes Observer’s regularities in comprehensive information-physical theory, connecting the quantum world with the physical classic and relativistic world, which the information law can predict.

Moreover, the IIID formalism with DSS reveals how artificially encodes an observer in universal information coding structure like DNA [18]. It involves naturally creating information Bits and natural encoding triplet code enables recognizing other encoded information. These Bits becomes standard units of different information languages in the observer communications. Natural interactive process, transferring Bits, creates information process self-organizing and encoding the observer' brain logic, cognition, and intellect.

The detail applications of the IIID are in [2-4] (specifically in [4], part 10).

5.3.2. Mathematical formalism

1. Probabilities and conditional entropies of random events.

A *priori* $P^a_{s,x}(d\omega)$ and *a posteriori* $P^p_{s,x}(d\omega)$ probabilities observe the Markov diffusion process $\tilde{x}_t$ distributions of random variable $\omega$ (events).

For each $i,k$ random event $A_i, B_k$ along the observing process, each conditional *a priori* probability $P(A_i / B_k)$ follows the conditional *a posteriori* probability $P(B_k / A_{i+1})$.

Conditional Kolmogorov probability

$$P(A_i / B_k) = [P(A_i)P(B_k / A_i)] / P(B_k) \tag{1}$$

defines the Bayes probability after substituting average probability:

$$P(B_k) = \sum_{i=1}^{n} P(B_k / A_i)P(A_i).$$

Conditional entropy

$$S[A_i / B_k)] = E[-\ln P(A_i / B_k))] = -\ln \sum_{i,k=1}^{n} P(A_i / B_k)]P(B_k) \tag{1a}$$

averages the conditional Kolmogorov-Bayes probability for multiple events along the observing process.

Conditional probability satisfies Kolmogorov's 1-0 Law for function $f(x)|\xi$ of an $\xi, x$ infinite sequence of independent random variables:

$$P_\delta(f(x)|\xi) = \begin{cases} 1, & f(x)|\xi) \geq 0 \\ 0, & f(x)|\xi) < 0 \end{cases}. \tag{1b}$$

This probability measure has been applied for the impulse probing in observable random process, which holds opposite Yes-No probabilities-as the unit of the probability impulse step-function.

Random current conditional entropy of the finite sequence of the random events is

$$\tilde{\mathrm{S}}_{ik} = -\ln \mathrm{P}(A_i / B_k)\,\mathrm{P}(B_k). \tag{1c}$$

Probability density measure on the Markov process trajectories as function of events $\omega$:

$$p(\omega) = \frac{\tilde{P}_{s,x}(d\omega)}{P_{s,x}(d\omega)} = \exp\{-\varphi_s^t(\omega)\}, \quad (1d)$$

is connected with this process additive functional

$$\varphi_s^T = 1/2\int_s^T a^u(t,\tilde{x}_t)^T (2b(t,\tilde{x}_t))^{-1} a^u(t,\tilde{x}_t)dt + \int_s^T (\sigma(t,\tilde{x}_t))^{-1} a^u(t,\tilde{x}_t)d\xi(t) \ . \quad (1e)$$

This functional is defined through controllable functions drift $a^u(t,\tilde{x}_t)$ and diffusion $b(t,\tilde{x}_t) = 1/2\sigma(t,\tilde{x}_t)\sigma(t,\tilde{x}_t)^T$ of the process, where (1e) also describes the transformation of the Markov process's random *time* traversing the process trajectory.

2. The *integral measure* of the observing *process* trajectories formalizes an *Entropy Functional* (EF), which is expressed through the above functions of Markov diffusion process $\tilde{x}_t$:

$$\Delta S[\tilde{x}_t]|_s^T = 1/2E_{s,x}\{\int_s^T a^u(t,\tilde{x}_t)^T(2b(t,\tilde{x}_t))^{-1}a^u(t,\tilde{x}_t)dt\} = \int_{\tilde{x}(t)\in B} -\ln[p(\omega)]P_{s,x}(d\omega) = -E_{s,x}[\ln p(\omega)], \quad (2)$$

and the probability density measure on the process trajectories.

3. Cutting the EF by the impulse delta-function determines the increments of Information for each impulse:

$$\Delta I[\tilde{x}_t]|_{t=\tau_k^{-o}}^{t=\tau_k^{+o}} = \begin{cases} 0, t < \tau_k^{-o} \\ 1/4Nat, t = \tau_k^{-o} \\ 1/4Nat, t = \tau_k^{+o} \\ 1/2Nat, t = \tau_k, \tau_k^{-o} < \tau_k < \tau_k^{+o} \end{cases} \quad (3)$$

with total $\sum_{t=\tau_k^{-o}}^{t=\tau_k^{+o}} \Delta I[\tilde{x}_t]_{\delta t} = 1Nat$. (3a)

4. *The Information Path Functional* (IPF) unites the Information cutoff contributions $\Delta I[\tilde{x}_t / \varsigma_t]_{\delta_k}$ along $n$-dimensional Markov process impulses during its total time interval $(T-s)$:

$$I[\tilde{x}_t]|_s^{t\to T} = \lim_{k=n\to\infty} \sum_{k=1}^{k=n} \Delta I[\tilde{x}_t / \varsigma_t]_{\delta_k} \to S[\tilde{x}_t], \quad (4)$$

which in the limit approaches the EF.
The IPF along the cutting time correlations on optimal (extreme) process trajectory $x_t$, in the limit, determines equation

$$I[\tilde{x}_t / \varsigma_t]_{x_t} = -1/8\int_s^T Tr[(r_s \dot{r}_t^{-1}]dt = -1/8Tr[\ln((r(\mathrm{T}) / r(s))]. \quad (4a)$$

depending on the trace of the relative ending and starting correralions $r(T)/r(s)$.

5. The equation of the EF for a microprocess:

$$\partial S(t^*)/\delta t^* = u_{\pm}^{t1} S(t^*), u_{\pm}^{t1} = [u_+ = \uparrow_{\tau_k^{+o}} (j-1), u_- = \downarrow_{\tau_k^{+o}} (j+1)] \quad (5)$$

under inverse actions of function $u_{\pm}^{t1}$, starts the impulse opposite time $t_{\pm}^* = \pm\pi / 2t^i$ which measures a space rotating angle relative to the impulse inner time $t^i$.

The equation' solutions for the conjugated entropies $S_+(t_+^*)$, $S_-(t_-^*)$ determine functions

$$S_+(t_+^*) = [exp(-t_+^*)(\mathrm{Cos}(t_+^*) - jSin(t_+^*))]|, S_-(t_-^*) = [exp(-t_+^*)(\mathrm{Cos}(-t_+^*) + jSin(-t_+^*))]$$

at

$$S_{\pm}(t_{\pm}^*) = 1/2 S_+(t_+^*) \times S_-(t_-^*) = 1/2[\exp(-2t_+^*)(\mathrm{Cos}^2(t_+^*) + Sin^2(t_+^*) - 2Sin^2(t_+^*))] =$$
$$1/2[\exp(-2t_+^*)((+1 - 2(1/2 - Cos(2t_+^*))))] = 1/2\exp(-2t_+^*)Cos(2t_+^*) \quad (5a)$$

Minimal interactive entropy $S_{\pm}(t_{\pm})$ *begins the space measure* during reversible relative time interval $0.015625\pi$ of the impulse invariant measure $\pi$.

The running microprocess, overcoming the entropy-Information gap, starts Information Bit and an Observer information macrodynamics.

6. The Information macrodynamic equations:

$$\partial I / \partial x_t = X_t, a_x = \dot{x}_t = I_f, I_f = b_t X_t \quad (6)$$

define $X_t$ -a gradient (force) of Information path functional $I$ (4) on macroprocess' trajectories $x_t$, $I_f$ - Information flow determined through speed $\dot{x}_t$ of the macroprocess. The flow emerges from drift $a^u(t, \tilde{x}_t)$ being averaged by function $a_x$ along the observing process, and the averaged diffusion $b_t \to b$ for the macroprocess force.

The Information Hamiltonian of the macrodynamics:

$$-\frac{\partial \tilde{S}}{\partial t} = (a^u)^T X + b\frac{\partial X}{\partial x} + 1/2 a^u (2b)^{-1} a^u = -\frac{\partial S}{\partial t} = H \, . \quad (7)$$

determines macro equations (6) from the minimax variation principle for the EF using Jacobi-Hamiltonian equations.

The impulse interactive observations sequentially join extremals of the multidimensional extreme process discretely changing Hamiltonian which accumulates the growing information.

The discrete Hamiltonian divides irreversible dynamic trajectory on the partial reversible segments, predicting the next emerging Information unit. The joint extremals form borders of the segments.

Equations (6) are the Information form of the equations of Irreversible Thermodynamics [19, 20], which the *Information Macrodynamics* generalize.

The flows and forces determine the macroprocess Hamiltonian in the invariant form

$H = X \times I$ .

Information curvature $K_{\alpha}^{m}$, density of Information mass $M_{vm}^{*}$, and effective complexity $MC_{m}^{\delta e}$ connect the equation

$$K_{m}^{\alpha} = M_{vm}^{*} MC_{m}^{\delta e}, \quad (8)$$

where

$$MC_{m}^{\delta e} = 3\dot{H}_{m}^{V} MC_{m} \quad (9)$$

includes the differential of Hamiltonian per volume $\dot{H}_{m}^{V}$ and the IN cooperative complexity $MC_{m}$.

*The single Eq. (8) at (9) encloses all previous Eqs.(1-7), unifying the formal math description of this approach.*

The Equations of *Interactive Integrated Information Dynamics* (2-9) formalize the description of information regularities of the observing discrete interactive process.

The developed Interactive Integrated Information Dynamics provides apparatus unified description of various interactive processes. That includes a path of emerging the information observer from Randomness and Uncertainty to Information, Thermodynamics, and the Observer Intelligence.

**References 2**

**Appendix**

The developed Interactive Integrated Information Dynamics provides apparatus unified description ofvarious interactive processes. That includes a path of emerging the information observer from Randomnessand Uncertainty to Information, Thermodynamics, and the Observer Intelligence.

During many years of research, the author with associates have implemented and published the IIID particularly in the following Experimental and practical applications:

1. Computer-aided design (CAD) of casting technologies implemented in the computer software package.
Practical application of the computed technological process in an industrial casting factory [1].
Comparative experimental results confirming that the applied chemical-thermodynamic description of the casting process coincides with the information description of the IIID. [2]
Confirming that the IIID solutions lead to the optimal casting process [3].
2. Modeling and Practical application of the impulse cut-off method in different solidification processes by designing an automatic system of impulse controls [4].
3. Design and Implementation automatic control regulator with the impulse frequency movement for different superimposing electro-technological processes interacting naturally in industrial metallurgical objects [5].
4. Applying the path identification method and computer applications for probabilistic distributions of microstructures in medical research [6].

Applying the approach's identification and control of the industrial casting micro-conductors [7].
5. Applying the IIID and computer software to intelligent systems by measuring the Information level of knowledge. The designed computer systems have checked a path of changing students' knowledge in education process [8].

These areas of applications, studied different paths of emerging an observer, allow verifying the IIID theory and predictions. The approach demonstrates wide areas of many other applications, which have been confirmed in many recent publications (cited in this author's books and other publications, for example [9]). The selected publications, focusing on experimental and practical applications